\documentclass[a4paper,12pt]{article}
\usepackage{amsfonts,amsthm,amsmath,amssymb,graphicx,hyperref} 
\newcommand{\tr}{\operatorname{tr}}

\newcommand{\Sym}{\operatorname{Sym}}

\def\Sym{\textrm{Sym}}

\def\a{\alpha}
\def\b{\beta}

\def\m{\mu}

\def\n{\nu}

\def\r{\rho}

\def\L{\Lambda}

\def\C{\mathbb{C}}

\def\Z{\mathbb{Z}}

\newcommand{\cA}{\mathcal A}

\newcommand{\cF}{\mathcal F}

\newcommand{\cL}{\mathcal L}

\newcommand{\cN}{\mathcal N}
\newcommand{\cO}{\mathcal O}

\newcommand{\cW}{\mathcal W}

\newcommand{\be}{\begin{equation}}
\newcommand{\bea}{\begin{eqnarray}}
\newcommand{\ee}{\end{equation}}
\newcommand{\eea}{\end{eqnarray}}

\newcommand{\nn}{\nonumber}

\newcommand\bC{ {\bf{C}} }

\def\mQ{ { \mathbb{Q} } }

\def\mC{ \mathbb{C}}
\def\cW{ \mathcal{W} }

\def\mR{ \mathbb{R}} 
 
\def\cF{ \mathcal{F}} 

\begin{document} 

\makeatletter
\@addtoreset{equation}{section}
\makeatother
\renewcommand{\theequation}{\thesection.\arabic{equation}}

\rightline{QMUL-PH-14-03}
\rightline{WITS-CTP-131}
\vspace{1.8truecm}

\vspace{15pt}

%%%%%%%%%%%%%%%%%

{\LARGE{ 
\centerline{\bf  CFT$_4$ as $SO(4,2)$-invariant  TFT$_2$} 
}}  

\vskip.5cm 

\thispagestyle{empty} \centerline{
    {\large \bf Robert de Mello Koch
${}^{a,} $\footnote{ {\tt robert@neo.phys.wits.ac.za}}}
   {\large \bf and Sanjaye Ramgoolam
               ${}^{b,}$\footnote{ {\tt s.ramgoolam@qmul.ac.uk}}   }
                                                       }

\vspace{.4cm}
\centerline{{\it ${}^a$ National Institute for Theoretical Physics ,}}
\centerline{{\it School of Physics and Centre for Theoretical Physics }}
\centerline{{\it University of the Witwatersrand, Wits, 2050, } }
\centerline{{\it South Africa } }

\vspace{.4cm}
\centerline{{\it ${}^b$ Centre for Research in String Theory, School of Physics and Astronomy},}
\centerline{{ \it Queen Mary University of London},} \centerline{{\it
    Mile End Road, London E1 4NS, UK}}

\vspace{1truecm}

%%%%%%%%%%%%%%%%%
\thispagestyle{empty}

\centerline{\bf ABSTRACT}

\vskip.2cm 

% The color combinatorics of a class of  four-dimensional matrix conformal field theory (CFT$_4$)
% correlators has recently been shown  to be described by two dimensional topological field theories (TFT$_2$)
% based on permutation groups. Here we approach the space-time dependence of correlators from the TFT$_2$ point of view.

We show that correlators of local  operators in  four dimensional free scalar field theory can be expressed in terms of amplitudes in 
a two dimensional topological field theory  (TFT$_2$). We describe the state space of the TFT$_2$, which has $SO(4,2)$ as a global symmetry, 
and includes both positive and negative energy  representations. Invariant amplitudes in the TFT$_2$ 
correspond to surfaces interpolating  from multiple circles to the vacuum. They  are constructed from 
$SO(4,2)$ invariant linear maps from the tensor product of the state spaces to complex numbers. 
When appropriate states labeled by 4D-spacetime coordinates are inserted at  the circles, the TFT$_2$ amplitudes 
become correlators of the four-dimensional CFT$_4$. The TFT$_2$ structure includes an associative algebra, related to 
crossing in the 4D-CFT, with a non-degenerate pairing related to the CFT inner product in the CFT$_4$. 
In the free-field case, the TFT$_2$/CFT$_4$ correspondence can largely be understood as realization of free quantum field theory as a 
categorified form of  classical invariant theory for appropriate $SO(4,2)$ representations. We discuss the prospects 
of going beyond free fields in this framework.

\setcounter{page}{0}
\setcounter{tocdepth}{2}

\newpage

\tableofcontents

\setcounter{footnote}{0}

\linespread{1.1}
\parskip 4pt

{}~
{}~

\section{Introduction}

In this paper we develop a new perspective on  correlators of four dimensional conformal field 
theory  using two dimensional topological field theory.  We are primarily concerned with 
the correlators of local operators in   CFT$_4$ on $\mR^{3,1}$ (or the Wick-rotated $\mR^4$), where the space of states 
forms representations of the conformal group $ SO(4,2)$. The explicit calculations in the paper will 
start from  the simplest CFT$_4$, namely the free scalar field. We will outline how elements 
of  the discussion generalize in various related free theories, and we will also see  that many  
key elements in the discussion would be sensible for interacting  theories. 
The symmetry $SO(4,2)$  plays the   role of a global symmetry in 
 the TFT$_2$. 

The space of local operators of the  CFT$_4$ will determine the state space of
the TFT$_2$. In a TFT$_2$, we associate a state space $ \cW $ to 
a circle, and tensor products $ \cW^{ \otimes k } $ to a disjoint union of $k$ circles. 
In $G$-invariant TFT$_2$, the state space is a linear representation of $G$. 
Here the state space $ \cW$ is a linear representation of $G = SO(4,2)$.  
TFT$_2$ associates, to an interpolating surface (cobordism)  from $k$ circles to the vacuum, 
 an $SO(4,2)$ invariant multi-linear map from $ \cW^{ \otimes k }$ to complex numbers.  
For the case of the free scalar field, we will specify this map explicitly. 
There are two basic ingredients that go into this map.
One is the fact that the 2-point function of the elementary field $\varphi $ of scalar field theory 
can be viewed as a generator for the matrix elements of the $SO(4,2)$ invariant map 
from $ V_+ \otimes V_- \rightarrow \mC$. Here $ V_+$ is the basic 
positive energy  (lowest weight) representation of $ SO(4,2)$ formed by $ \varphi$ and its derivatives,
while $ V_-$ is the dual negative energy (highest weight) representation. 
The other ingredient that goes in the construction of the map 
from $ \cW^{ \otimes k } \rightarrow \mC$ is  the combinatorics of Wick contractions. 
Using  this invariant map, we can compute the $k$-point correlation functions 
of arbitrary composite operators.  For the general background on TFT$_2$, 
including careful definition of orientations, of  ingoing versus outgoing boundaries
and of cobordisms and their equivalences, we found   \cite{Kock} to be a very useful reference.

Section 2 starts with the motivations from AdS/CFT leading to this work.
It has recently been found that the combinatoric part of extremal correlators, notably three-point functions, 
of half-BPS operators can be expressed in terms of 2d TFTs built from lattice gauge theories 
where permutation groups play the role of gauge groups \cite{MatxBelyi,Tom,CJR,Feyncount,quivcalc}.
These belong to the class of theories considered by Dijkgraaf-Witten \cite{DW}, and form examples
of TFTs obeying the  axioms stated by Atiyah \cite{Atiyah}. This naturally raises the 
question of whether the full spacetime dependent correlators, not just the combinatoric part,
can be understood using an appropriate TFT$_2$. Since the space-time dependences 
of three-point correlators are completely determined by the invariance of the theory 
under the conformal group $ SO(4,2)$, the right TFT$_2$ has to have a global 
$SO(4,2)$ invariance. The notion of TFT$_2$ with a global $G$-invariance has been explained 
in \cite{segal-moore}.  One subtlety we have to deal with is that standard TFT$_2$'s 
have finite dimensional state spaces. This requirement of finite dimensionality 
follows from the way the axioms are set up, and physically relates to the fact that 
the TFT$_2$'s have amplitudes corresponding to surfaces of arbitrary genus.  To allow 
infinite dimensional state spaces, we restrict to genus zero surfaces, so we have a genus zero TFT$_2$, 
which involve a genus zero subset of the equations and algebraic structures entering finite TFT$_2$'s. 
We describe this genus zero subset of equations and corresponding geometry of two dimensional 
cobordisms. Key among these properties are  the existence of a product, corresponding to 
the 3-holed sphere with two ingoing and one outgoing boundary. This product is associative.
Another key property is the existence of a non-degenerate pairing $ \cW \otimes \cW \rightarrow \mC$, 
which has  to be $ SO(4,2)$ invariant. The student of perturbative QFT is familiar with the 
fact that zero-dimensional Gaussian integrals provide a brilliant toy model to learn about 
QFT. Here the Gaussian integration model is used to provide what is arguably the simplest 
example of TFT$_2$, having infinite dimensional state space, associativity and non-degeneracy.

Section \ref{sec:basic} describes how the basic two-point function of the elementary scalar in  free 
scalar quantum  field theory is understood 
in terms of $ SO(4,2)$ invariants. Let $V_+$ be the irreducible representation of $ SO(4,2)$, with states of 
positive scaling dimensions (positive energy in radial quantization), consisting of the field $\varphi$ and its derivatives, with the equations of motion 
set to zero. $V_-$ is the conjugate representation. There is no $ SO(4,2)$ invariant in the tensor product 
$ V_+ \otimes V_+ $, but there is an $SO(4,2)$ invariant bilinear map $\hat\eta : V_+ \otimes V_- \rightarrow \mC$. 
$\hat\eta$ is extended to an invariant map from $\hat\eta :V \otimes V \rightarrow\mC$, where $V = V_+ \oplus V_-$. 
This bilinear invariant plays a crucial role, so we study some of its properties.
We describe the invariant explicitly by exploiting   $SU(2)\times SU(2)$ and $SL(2)$ subalgebras of $SO(4,2)$. 
 The  $SO(4,2)$ invariance is neatly expressed as a set of partial differential equations obeyed by the generator of matrix elements of $\hat\eta$.
Some additional representation theoretic constructions are described, notably a map $ \rho : V_+ \rightarrow V_-$, 
which is related to an automorphism of the $so(4,2)$ Lie algebra. By using $ \hat\eta, \rho$, we can construct an
inner product $g$ on $V_+$ and on $V_-$. The positivity of $g$ is related to unitarity of the CFT. 
The invariance of $\hat\eta $ is important in giving a TFT$_2$ interpretation of the 2-point function 
in terms of an $SO(4,2)$ invariant map.  These ideas can be understood in a simple toy model. 
Consider the spin half representation of $SU(2)$, denoted $V_{1 \over 2} $.  A problem in classical invariant theory 
is to count the number of times the trivial (one-dimensional) representation $ \mC$ appears in the Clebsch-Gordan decomposition of various tensor powers.  If we take $ V_{1 \over 2} \otimes V_{1 \over 2}$ it appears once. 
A more refined question is to describe the form of this state, which is 
\bea\label{invtsu2} 
| { 1\over 2}  , { 1 \over 2 } > \otimes | { 1 \over 2}  , - { 1 \over 2 }  >  ~ -  ~ |{ 1 \over 2 }  , -{ 1 \over 2 }  >  
\otimes   | { 1 \over 2 }  , {  1 \over 2}  > 
\eea
We are using the usual notation for $SU(2)$ reps where states in an irrep are labeled by $ |j ,m > $ 
with $ J_3 = m $ and quadratic Casimir equal to $ j ( j +1)$. 
Now if we let  $ x_1^{2  J_3} \otimes x_2^{ 2 J_3 } $   act on this invariant state, we get back the state 
with a factor. 
\bea\label{invtsu22}  
{ x_1 \over x_2 }   - { x_2 \over x_1 } 
\eea
The equations (\ref{invtsu2}) (\ref{invtsu22}) are equivalent ways 
to  describe the form of the invariant state in $ V_{1 \over 2} \otimes V_{ 1 \over 2 } $. 
In the application to the TFT$_2$ construction of CFT$_4$ correlators, these  $ x_1 , x_2 $ 
are replaced by 2 spacetime coordinates, and we are describing the invariant representation $ \mC $ 
in $ V_+ \otimes V_- $. The precise description of the invariant state in this way is a refinement of 
the counting of invariants. In this sense, this is a categorification of invariant theory and the TFT$_2$ 
construction of free field correlators involves  a categorification of invariant theory for certain 
representations of $ SO(4,2)$. 

Section \ref{sec:corrs} describes the state space  $ \cW $ of the TFT$_2$ corresponding to the free field CFT$_4$.
Loosely speaking $ \cW $ contains states corresponding to all the composite operators 
in free field theory. The slight surprise is that it is $V = V_+ \oplus V_- $, rather than $V_+ $ or $V_-$ alone which enters the construction of  $ \cW $.  
This is related to the fact mentioned above that there is an $SO(4,2)$ invariant in $ V_+ \otimes V_- $ but 
not in $ V_+  \otimes V_+ $, so a construction of CFT$_4$ correlators from $SO(4,2)$ invariants in TFT$_2$ 
has to involve both $ V_+ $ and $V_-$ in the construction of  the TFT$_2$ state space. 
 We have 
\bea 
&& \cW = \bigoplus_{ n=0}^{ \infty } \cW_n \cr 
&& \cW_{n} = Sym ( V^{ \otimes n } ) 
\eea
The $ n=0$ subspace $ \cW_0 = \mC$. 
The  $n$-fold symmetric product arises because of the bosonic statistics of the free scalar.
TFT$_2$ involves assigning $ SO(4,2)$ invariant maps to interpolating   surfaces (cobordisms) from   disjoint unions of 
circles to the vacuum. We describe such an invariant map from $\cW^{ \otimes k  }\rightarrow \mC$
for any $k \ge 0$. It is constructed from the basic invariant $\hat\eta$, several copies of which are tensored 
according to Wick contraction combinatorics of QFT. We identify the basic field as a linear superposition of 
states, labeled by position $x \in \mR^4 $, living in $V_+$ and $V_-$, 
\bea 
\Phi ( x  )  = \Phi^{+} ( x ) + \Phi^- ( x^{\prime} ) 
\eea
with $x'$ related to $x$ by inversion. Using tensor products of this field, we have states corresponding to 
composite fields living in $ \cW_n$ for all $n$. Choosing coordinates  $x_1, x_2 \cdots x_k $ for the composite fields
thus defined, using tensor products and applying  the invariant map, we get arbitrary correlators of 
composite fields at non-coincident points. 

In section \ref{sec:nondegen} we focus attention on the 2-point functions of arbitrary composite operators, viewed 
from the TFT$_2$ perspective. This is the amplitude for two circles going to the vacuum, denoted $ \eta : \cW \otimes \cW \rightarrow \mC $. 
We show that the non-degeneracy equation is satisfied, i.e. there is an 
inverse $\tilde \eta $ of $ \eta $.  This equation corresponds to the fact that we can  
glue a cylinder with two incoming boundaries to one with two outgoing boundaries, along one boundary from each, 
to give a cylinder with one in and one out boundary (see Figure \ref{fig:nondeg}).  There is no gluing along two boundaries 
to produce a torus, which would give infinity because of the infinite dimensionality of the state spaces. 
This restriction is clear at the level of equations, but subtle at the level of rigorous category theoretic  axiomatics.
These subtleties are discussed in Section \ref{sec:discuss}.  The approach we take in the bulk of the paper 
is to define TFT$_2$ in terms of this restricted set of genus zero equations. 

In section \ref{sec:3ptOPE}, we discuss 3-point functions and the operator product expansion. 
The relation between the two is provided by the inverse of $\eta$ discussed in Section \ref{sec:nondegen}. 
The amplitude for 3 circles to vacuum is the 3-point function. The amplitude for 
2-circles to one circle is the OPE. The amplitude for one circle to two is the co-product. 

In section \ref{sec:crossass}, we discuss crossing and associativity. We explicitly 
prove the crossing property of the 4-point  amplitude of TFT$_2$. 
This is related, using the non-degeneracy condition, to associativity, 
and also to what is sometimes called the Frobenius equation or the 
nabla-delta equation.

Section \ref{sec:counting} looks at a basic problem in free scalar field theory, 
which is the { \it enumeration}  of primary fields according to $SO(4,2)$ representation 
and multiplicity. We find that TFT$_2$'s with infinite dimensional state spaces, 
of the kind described in Section 2, play a role in the counting and lead to explicit 
new results for the case of three primary fields. This shows that the notion of genus zero 
TFT$_2$'s, with infinite dimensional state spaces which  we have identified, 
is integral to the architecture of CFTs - not just to the whole CFT, but also to how the whole 
CFT is assembled from simpler parts. We hope to return to this theme by considering 
the { \it construction} of primary fields in the future. This would be  another application of  the counting to construction philosophy
which finds various applications in the study of BPS states, integrability of giant graviton fluctuations and quiver combinatorics
\cite{countcons,doubcos,quivcalc}. 
It is worth elaborating on counting to construction in this free scalar field setting.
In the context of a free $O(N)$ vector model, the decomposition of the tensor product of the singleton representation (associated
with the free scalar field) with itself into irreducible representations, as decsribed in the Flato-Fronsdal theorem\cite{Flato:1978qz}, 
is the kinematics underlying the higher spin holography\cite{Sezgin:2002rt,Klebanov:2002ja,Sezgin:2003pt}.
The relevance of the tensor product of the singleton with itself follows from the fact that to form $O(N)$ singlets in the
vector model, one has to contract a product of two scalars.
In the case of the matrix model, since we can take the product of an arbitrary number of matrices and trace to get a scalar,
we need a more general version of the Flato-Fronsdal theorem which considers the tensor product of an arbitrary
number of singleton representations.
The generalized theorem will play a central role in the kinematics underlying the holography of the free CFT.
Section \ref{sec:counting} represents a concrete framework within which this generalized Flato Fronsdal theorem can be tackled.

Section \ref{sec:discuss} discusses outstanding problems and future directions. 
In particular, Section \ref{sec:genfree} considers generalized free fields, where the 
irrep $V_+ $ is replaced by a more general irrep of $ SO(4,2)$. In this more general 
set-up, we can still construct an $ SO(4,2)$ invariant TFT$_2$ as defined in section \ref{sec:tft2eqs}. 
However, there is an additional condition related to having a local stress tensor 
that is not satisfied in the case of generalized free fields. We outline how this additional stress tensor condition 
can be expressed in terms of the $SO(4,2)$-invariant TFT$_2$ data.

The TFT$_2$ construction we have developed with $SO(4,2)$ can be repeated after replacing $SO(4,2)$ with other 
groups. If we consider $ SO(d,2)$, we can relate CFTs in $ d $ dimensions to TFT$_2$. We can also consider a  
compact group $G$. This will give $G$-invariant TFT$_2$'s.
A unique invariant map $\hat\eta :V\otimes V\rightarrow\mC$ can be defined for any finite dimensional 
representation $V$ of $G$ which contains, with unit multiplicity,  the trivial irrep in the tensor product decomposition $V\otimes V$. $V$ can be an irreducible representation if that irrep is self-dual, 
or it can be a direct sum of some irreducible representation $V_+ $ with its dual $V_-$. 
We can define the state space 
\bea 
  \cW = \bigoplus_{ n = 0}^{\infty} Sym ( V^{ \otimes n } ) 
\eea
The amplitudes $C_{A_1 ,\cdots ,A_k}$ can be defined using tensor products of the elementary $\hat \eta$ as in Section \ref{sec:corrs}.
Since the proofs of non-degeneracy in Section \ref{sec:nondegen} and of associativity in Section \ref{sec:crossass} are 
purely combinatoric, they will continue to hold in this more general set-up.  It would be  interesting to investigate 
applications of this general construction and to find
path integral constructions which give rise to these TFT$_2$'s.

It is worth noting here that connections between 4D quantum field theories 
and two dimensional topological field theories, in diverse incarnations, have been 
a fruitful area of research. Superconformal indices of  4D theories have been related to 2D TFT \cite{GPRR}. In such applications 
the 2D surface has a physical origin as the surface that a 6D theory has to be compactified on
to arrive at the 4D theory. Another way to related 4D QFT to 2D TFT is to twist the 
4D theory so it becomes topological and then consider the 4D  theory on a product of Riemann surfaces 
\cite{KapWit}. It is instructive to compare the  present construction with these precedents. 
We have a 4D theory, and we are looking at local correlators, with non-trivial spacetime dependences. There is no dimensional reduction and the 2D surface arises as a geometrical device to encode, via its cobordism equivalences, the crossing and non-degeneracy properties of  the 4D conformal field theory. 
The spacetime coordinates of local operators  have become labels 
of states in the state space 
$\cW$ associated boundary circles of the 2D surface. This is somewhat like string theory where 
spacetime momenta become labels of vertex operators inserted at points on the worldsheet.
The observables do not depend on worldsheet metric,  because we integrate out the worldsheet metrics, hence the topological nature.   
In this sense, our construction  has some analogies to the twistor string proposal for $\cN=4$ SYM \cite{witten-twistor}.  

It is also useful to consider the results of the present work in  light of the general phenomenon 
of dualities in string and field theory. For instance T-duality in string theory is constructive and, for toroidal backgrounds,  technically very simple : it exchanges the  momentum and winding modes of string excitations. At the same time it has a conceptually very important aspect : it exchanges 
small and large sizes. Strong-weak dualities allow the computation of the strongly coupled theory in terms of its weakly coupled dual, but in most cases the explicit construction of the map is not  known. 
The present CFT$_4$/TFT$_2$ correspondence is constructive, so in this sense, more like T-duality than $S$-duality. The construction  encodes both 
the structure of the local operator and the space-time coordinates in the choice of  boundary data. 
The conceptually intriguing part is that the four space-time coordinates of local operators are simply labels of states  at the boundaries of  the surfaces in two dimensions. So space-time as a stage  for propagating fields has disappeared in the TFT${}_2$ picture. This can be viewed as a form of space-time emergence, admittedly only in  the context of free CFT${}_4$ 
at this stage, but this  is a proof of principle that space-time emergence (as opposed to just emergence of space) is possible in the world of dualities.

\section{ Genus zero TFT$_2$ equations   } \label{sec:tft2eqs}

\subsection{ Motivations and strategy } 

The approach to correlators of CFT$_4$ we develop here, is motivated 
by studies  of extremal correlators of half-BPS operators 
 in $ \cN =4$ super Yang Mills theory with $U(N)$ gauge group.
The half-BPS states correspond to multi-traces of an $ N \times N $   complex matrix $ Z $. 
For every positive integer $n$, these are
 gauge invariant observables  which can be constructed using  permutations $ \sigma  \in S_n $ 
\bea 
\cO_{ \sigma } (x)   \equiv  Z^{i_1}_{ i_{ \sigma ( 1 ) } } (x)  \cdots    Z^{i_n }_{ i_{ \sigma ( n ) } } (x) 
\eea
The correlators can be written as \cite{MatxBelyi,Tom,CJR,Feyncount,quivcalc}
\bea\label{halfbpscorr}  
\langle \cO_{ \sigma_1} ( x_1) \cO_{ \sigma_2 } ( x_2 )\rangle  = { 1 \over |x_1 - x_2 |^{ 2n } } \sum_{ \sigma_1' \in T_1 } \sum_{ \sigma_2' \in T_2 } \sum_{ \sigma_3 \in S_n } \delta ( \sigma_1' \sigma_2' \sigma_3 ) N^{ C_{ \sigma_3} }  
\eea
$T_1$ is the conjugacy class of the permutation $ \sigma_1$. $T_2$ is the conjugacy class of the 
permutation $ \sigma_2$. $C_{\sigma_3}$ is  the number of cycles in  the permutation $ \sigma_3$. 
The combinatoric part of the correlator is constructed from a quantity $ Z_{{\rm{TFT}}_2 ( S_n ) }  ( T_1 , T_2 , T_3 )$
which is a function of 3 conjugacy classes 
\bea 
Z_{{\rm{TFT}}_2 ( S_n ) }  ( T_1 , T_2 , T_3 ) = \sum_{ \sigma_1' \in T_2 } \sum_{ \sigma_2' \in T_2 } \sum_{ \sigma_3' \in T_3 }
\delta ( \sigma_1' \sigma_2' \sigma_3' )  
\eea
This 2D topological field theory is an example from the class of TFT$_2$'s associated with finite groups $H$ (here $S_n$), which 
were first discussed by Dijkgraaf and Witten \cite{DW}. For closed Riemann surfaces, this sums over homomorphisms 
from the fundamental group of the surface  to the group $H$. For manifolds with boundary, we sum over homomorphisms 
subject to a condition that the boundary group elements are restricted to some conjugacy classes. In the above case, 
we have the partition function on a 3-holed sphere, with $T_1, T_2, T_3$ being  the three specified conjugacy classes 
at the boundaries. 

Given that the combinatoric part has an elegant TFT$_2$ description, it is natural to ask if the same is 
true for the space-time dependent part of the correlator \ref{halfbpscorr}. This is known to be determined 
by the conformal group $SO(4,2)$.  The simplest set-up to investigate  this question is to consider ordinary
(non-matrix) free scalar field theory. The main result of  this paper is to describe this  as an $SO(4,2)$-invariant  TFT$_2$. 
 We return to  matrix scalar field theories briefly  in Section \ref{sec:furtherexs}   and outline 
  how the $SO(4,2)$  and  $S_n$ appear in the TFT$_2$ description in that case. 

A TFT$_2$  associates a vector space $\cW$ to a circle and tensor products of $\cW$ to disjoint unions of 
circles. Cobordisms are associated to homomorphisms between tensor products of the vector spaces. 
From a physical point of view, once we have chosen a basis, there is discrete data   : structure constants 
$ C_{AB C}   $ and a bilinear pairing $ \eta_{AB}$, which obey some consistency conditions. 
These consistency conditions correspond to equivalences between different ways to 
construct cobordisms. They include, importantly, a non-degeneracy condition 
and an associativity condition.

\begin{figure}[ht]%
\begin{center}
\includegraphics[width=0.3\columnwidth]{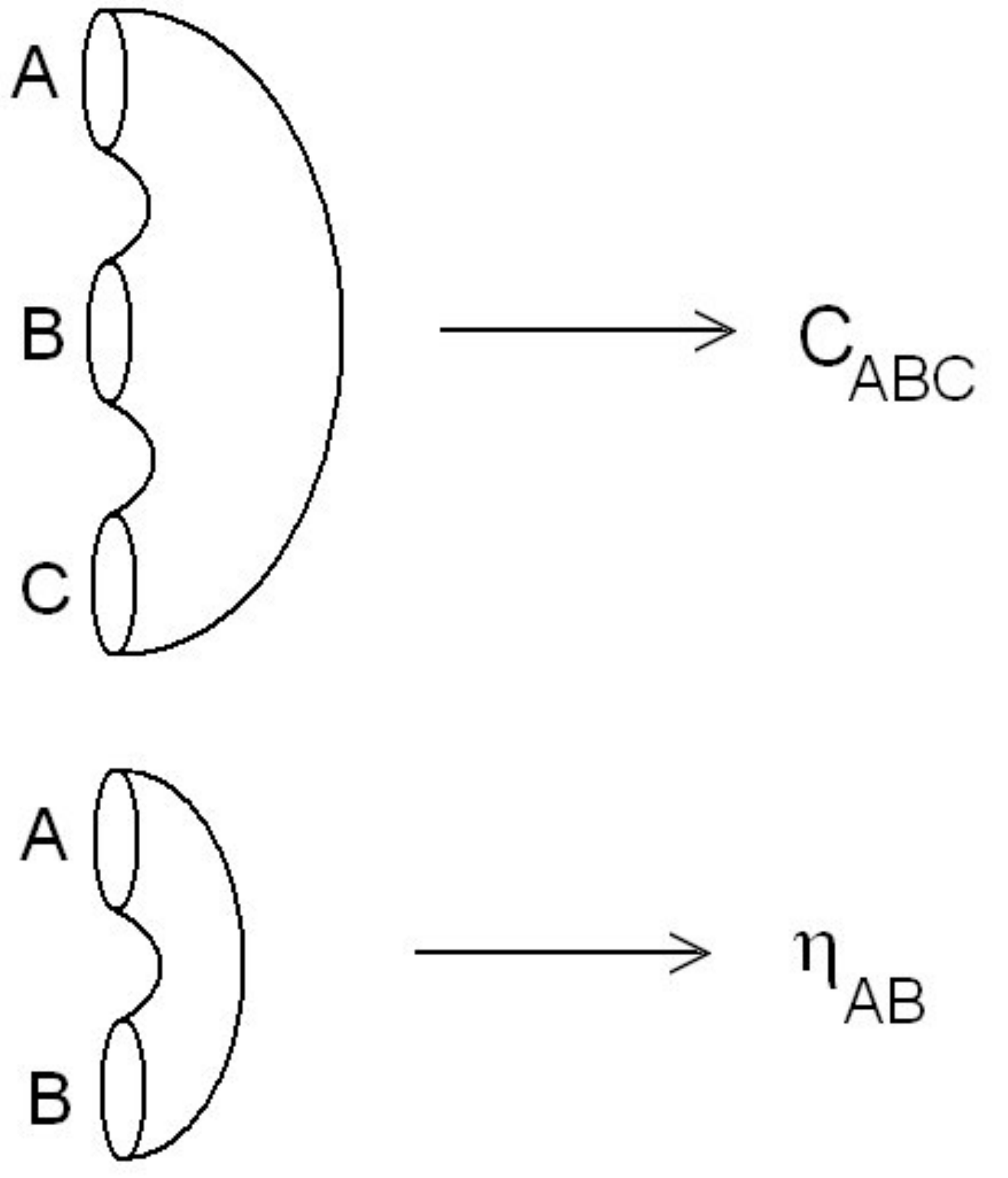}%
\caption{ Basic building blocks and corresponding cobordisms   }%
\label{fig:CETApics}%
\end{center}
\end{figure}

A variation on the above is associated to theories with global symmetry group $G$. 
Then $\cW$ is a representation of $G$. The homomorphisms $ \rho_{ k_1 , k_2 } : \cW^{ \otimes k_1 } \rightarrow 
\cW^{ \otimes k_2 } $ are $G$-equivariant, that  is for any $g \in G $ the homomorphism 
and $g$ action commute  $ \rho_{ k_1 , k_2  } \circ g = g \circ  \rho_{ k_1 , k_2 } $. So the vector space of states $\cW $, whose basis states are labeled by 
$A, B , \cdots $,  form a representation of a group $G$ (or its Lie algebra, when $G$ is a Lie group). The data 
$ \eta_{ AB} $   and  $C_{AB C } $, are equivariant maps to  the trivial representation $ \mC$, equivalently they are $G$-invariant  maps. This notion of  TFT$_2$ with global symmetry group $G$ is mentioned in \cite{segal-moore} prior to developing  TFT$_2$ with local $G$- symmetry, where the geometrical category 
involves circles with $G$-bundles and the cobordisms involve surfaces equipped with $G$-bundles.

\subsection{ Genus zero restriction and infinite dimensional state spaces} 

The standard axiomatic approach to TFT2 \cite{Kock} requires the state spaces to be finite dimensional
and includes finite amplitudes for surfaces of arbitrary genus.  The first observation is that there is a well-defined subset, which we may call {\it the  genus zero subset of the TFT$_2$ equations}, which do not involve summing over  states in a (stringy) loop. These genus zero equations consist of a rich algebraic system including an associative product, non-degeneracy, unit, co-unit and co-product.  These equations allow solutions involving infinite dimensional state spaces. So the $ \eta_{AB} , C_{ABC} $ are infinite dimensional arrays of numbers, with $A , B , \cdots $ running over an infinite discrete set of states.  
We will first write down some of  these genus zero equations, and then show that there 
are simple non-trivial solutions, which we will call toy model solutions. The first toy model can be viewed as quantum field theory 
reduced to zero dimension and consists of Gaussian integration. A second toy model is related to $SU(2)$ tensor product multiplicities, 
which has applications in counting primary fields in CFT$_4$, as we will see in Section \ref{sec:counting}.
 
In subsequent sections we will show how these equations - with an appropriate 
choice of state space -  provide a realization of 
free scalar field CFT$_4$ as a TFT$_2$.  The equations obeyed by these structure constants
 have geometrical  analogues in terms of equivalences of cobordisms (see \cite{Kock} for the 
 geometrical definitions and the equations, and a physics presentation in \cite{FHK}). We write  the key genus zero equations.

\begin{itemize} 

\item 
Non-degeneracy : The 2-point  function $ \eta_{AB}$ has an inverse $ \tilde \eta^{AB}  $
\bea   \label{non-degen-eqn}
\eta_{AB} \tilde \eta^{BC} = \delta_A^C 
\eea
$ \tilde \eta^{AB}$ corresponds to the cobordism from vacuum to two circles, 
while the identity on the RHS corresponds to the cylinder. 
Corresponding to (\ref{non-degen-eqn}) is the  relation between cobordisms  in Figure \ref{fig:nondeg}. 
In the case of finite dimensional state spaces, we also have $ \eta_{AB}\tilde \eta^{BA} = {\rm Dimension}$, 
which is closely related to non-degeneracy. This is a genus one cobordism from vacuum to vacuum which
is excluded from our genus zero subset of equations.

\item 
$G$  invariance : There invariance under a group  $G$, which can be finite or a Lie group. 
In case of a Lie group, the invariance can be expressed in terms of the Lie algebra. 
 For an element $\cL$ in  the Lie algebra of $G$, we have
\bea\label{invce-eqs} 
 \cL_{A}^{A'} C_{A'BC} +  \cL_{B }^{B'} C_{AB'C} + \cL_{C}^{C'} C_{ABC'} &=&  0 \cr 
 \cL_{A}^{A' }\eta_{A' B} + \cL_{B}^{B'} \eta_{A B'} &=&  0
\eea
In our application to  CFT$_4$, $G$ is $SO(4,2)$. In the toy model of  Section \ref{sec:modelGaussian}, 
$G$ is trivial.

\item 
Using the inverse $ \tilde \eta $ of $ \eta $, 
the  3-to-vacuum amplitude  $ C_{ABC} $  can be related to a 
2-to-1 amplitude, which is the structure constant of an algebra. 
\bea\label{2to1}  
C_{AB}^{~~~D}  = C_{ABC} \tilde \eta^{ CD } 
\eea
In the applications to free field theory, this structure constant
will be related to the operator product expansion, while $ C_{ABC}$ will be related
to 3-point correlators. As a relation between cobordisms, this is shown in Figure \ref{fig:rel-ope-corr}.
Note that we might imagine associating one-dimensional pictures 
to such data, i.e. in this case a trivalent graph, and describing the equations 
in terms of relations between graphs. However  a trivalent graph is not a manifold. 
Indeed in one dimension, cobordisms exist from one set of points to another 
only if the numbers of points are both even or both odd \cite{Kock}. 
 Here we keep as closely as possible to the standard topological  field theory framework of cobordisms,
 hence two dimensions are naturally selected as the right geometrization of the equations. 

\item Using $ \tilde \eta$ , we can also relate the 3-to-vacuum amplitude $C_{ABC}$ 
to a 1-to-2 amplitude, which is called a co-product. 
\bea 
  C_{A}^{~~DE }   = C_{ABC} \tilde \eta^{ B D} \tilde \eta^{ CE } 
\eea
The figure corresponding to this is Figure \ref{fig:rel-ope-corr}. 

\item Symmetry Relations
\bea 
C_{ABC} = C_{BAC} = C_{ACB} 
\eea

\item Associativity  : 
\bea\label{asseqs}  
C_{ AB }^{~~~ E} C_{E C}^{~~~D}  = C_{BC}^{~~~E} C_{EA}^{~~~D} 
\eea
This corresponds to the fact that the two different gluings shown in  Figure \ref{fig:cross-ass} 
give equivalent cobordisms.

\item Crossing : 
\bea 
C_{AB}^{ ~~~ E} C_{CD}^{~~~ F  } \eta_{ EF }   = C_{BC}^{~~~E}  C_{ AD }^{~~~ F} \eta_{ EF}  
\eea
Using the non-degeneracy equation, this is equivalent to associativity, which we elaborate on 
in Section \ref{sec:asscross}.  We also show there the  equivalence to the Frobenius relation.

\item  $G$-invariance conditions for $ \tilde \eta^{AB}  $  and $ C_{AB}^{ ~~~ C}  $ follow from 
the previous equations (\ref{invce-eqs}) 
\bea 
&&  \cL^A_{A'} \tilde \eta^{ A' B } + \cL^B_{B'} \tilde \eta^{ A B'} = 0 \cr 
&& \cL_A^{ A'} C_{ A' B }^{ ~~~~  C} + \cL_{B}^{B'} C_{ A B'}^{ ~~~~  C}  - \cL^C_{ C'} C_{AB}^{~~~~  C'}  = 0 
\eea

\item In the context of the TFT$_2$/CFT$_4$ construction in Section   \ref{sec:basic},
we will use an automorphism $ \rho $  of $ G$ to define an inner product $g$ on the state space. 
The relation is of the form  $ g ( \cdot , \cdot ) = \eta ( \rho (\cdot ) , \cdot ) $. We can impose a 
unitarity constraint on the TFT$_2$ by requiring positivity of  this inner product.

\begin{figure}[ht]%
\begin{center}
\includegraphics[width=0.7\columnwidth]{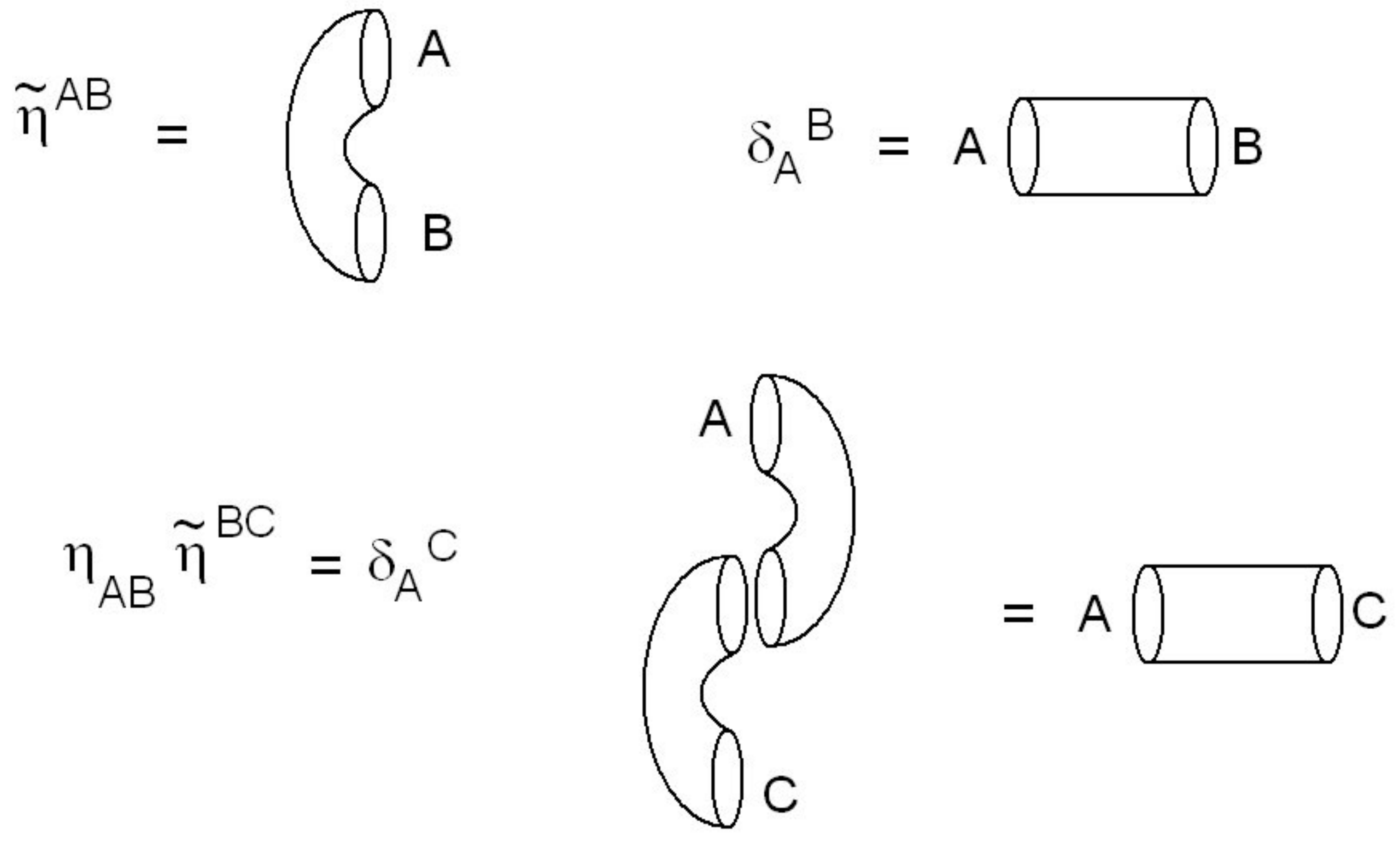}%
\caption{ The non-degeneracy equation and cobordisms.   }%
\label{fig:nondeg}%
\end{center}
\end{figure}

\begin{figure}[ht]%
\begin{center}
\includegraphics[width=0.5\columnwidth]{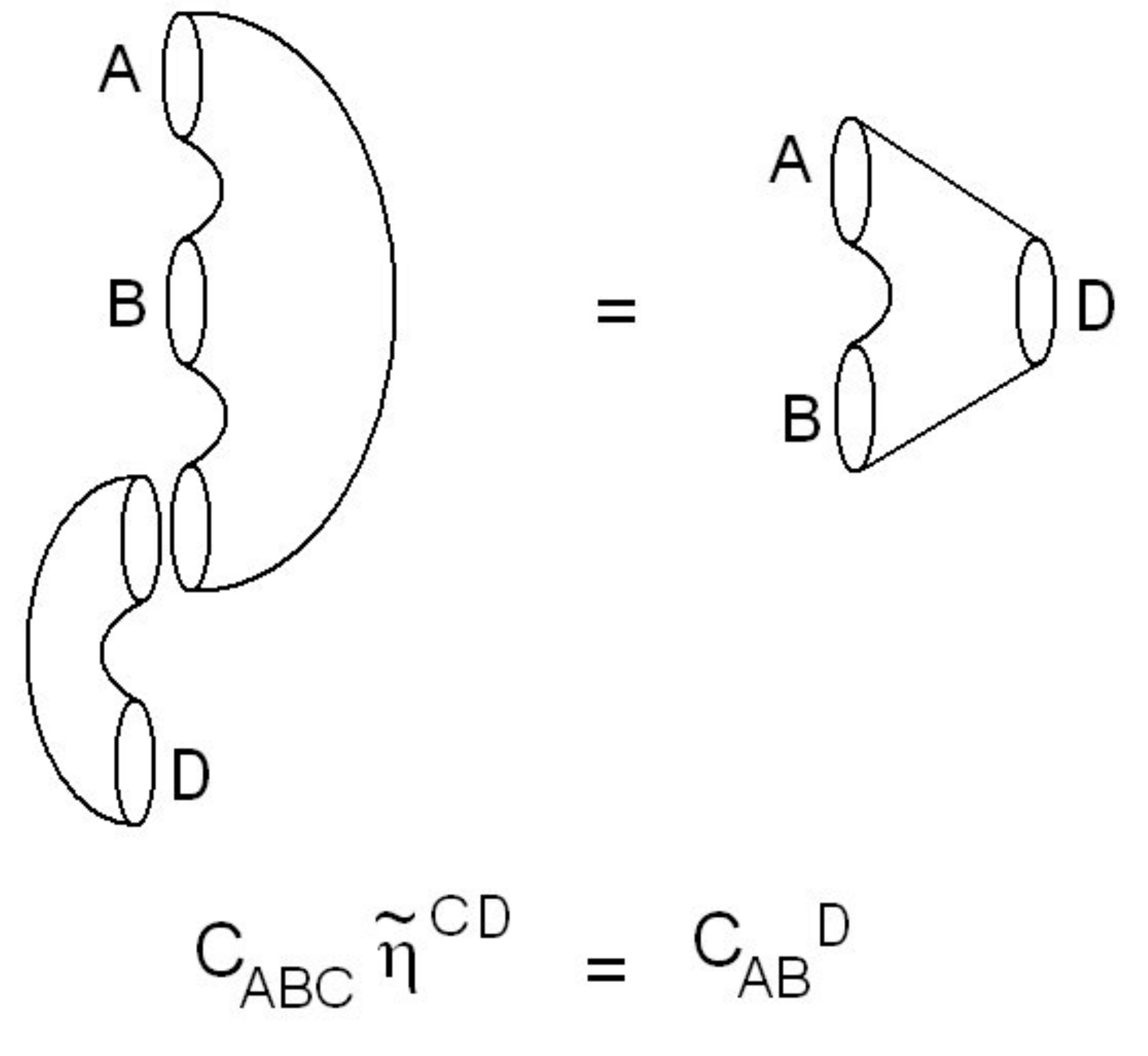}%
\caption{ Relating correlator to product   }%
\label{fig:rel-ope-corr}%
\end{center}
\end{figure}

\begin{figure}[ht]%
\begin{center}
\includegraphics[width=0.5\columnwidth]{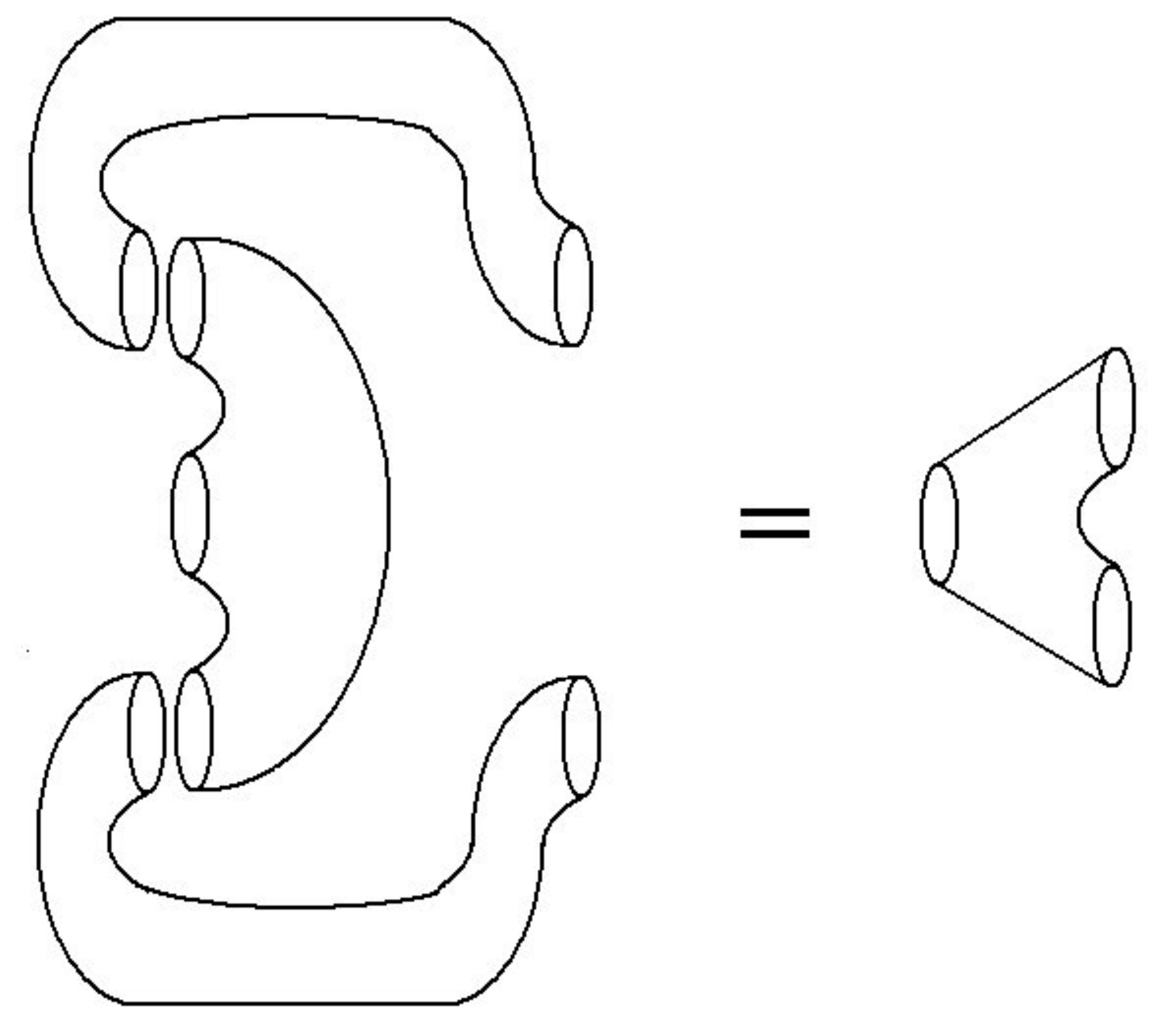}%
\caption{ Relating correlator to co-product  }%
\label{fig:rel-corr-cop}%
\end{center}
\end{figure}

\begin{figure}[ht]%
\begin{center}
\includegraphics[width=0.5\columnwidth]{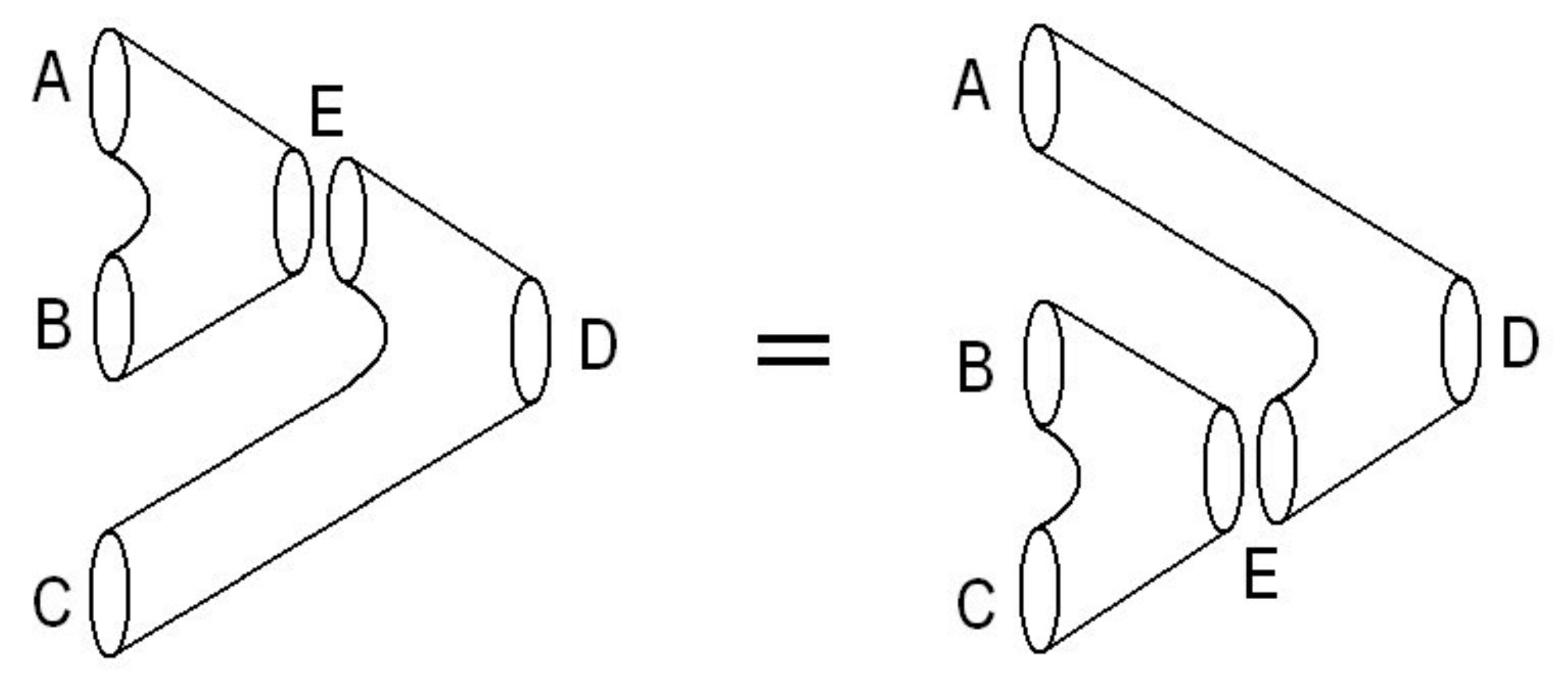}%
\caption{Associativity and Crossing  }%
\label{fig:cross-ass}%
\end{center}
\end{figure}

\item Higher point correlators can be constructed from 
three-point correlators, e.g 
\bea 
C_{ABCD} = C_{AB}^{~~~E} C_{EC}^{~~~F} \eta_{FD} 
\eea
There is a similar construction for $n$-point correlators. 

\end{itemize} 

We will show that all these equations have realizations in the context 
of discrete data underlying correlators of CFT$_4$. These same equations are also realized by 
simpler toy-models. One of them is Gaussian integration. 
Another is related to $SU(2)$ fusions and has applications 
in the counting of primary fields. 
Both of these toy models have infinite dimensional state spaces.
TFT$_2$ defined by these equations thus  contains the discrete structure of 
CFT$_4$  as well as the related toy models. We will take these equations for  
$C_{ABC},\eta_{AB}$ as our definition of TFT$_2$ - they are essentially 
genus zero restrictions of standard TFT$_2$ equations. 
 We have not given an axiomatic definition of the kind that exists 
 for the case of usual TFT$_2$ (corresponding to Frobenius algebras) 
  where all the possible gluings  of the basic $ \eta , C , \tilde \eta $ are allowed, higher genus surfaces 
 are included and state spaces are constrained to be finite dimensional. 
 Finding the right axiomatic framework for the equations presented here 
  is an interesting problem, which we discuss in section \ref{sec:eqsrax}.

\subsection{ Equations related to the properties of the vacuum state }

As mentioned in the introduction and described in detail in section 4, the state space 
$\cW $ in the case of the TFT$_2$ construction of CFT$_4$
 is graded by the degree of the symmetric  tensors
\bea
 \cW = \oplus_{n=0}  \cW_n \qquad \cW_n = Sym (V^{\otimes n}) 
\eea
The pairing $\eta :\cW \otimes\cW \rightarrow \bC$ is diagonal in the grading in the sense that 
\bea 
  \eta (\cW_n,\cW_m) \propto \delta_{n,m} 
\eea
The case $ n = m =1$ is the case of the basic pairing, denoted $\hat\eta$. 
The case $n = m = 2$ and higher corresponds to sums over all possible Wick contractions 
between composite fields, which are quadratic in $\varphi$ etc. 
The explicit formulae for $n \ge 1$ are in later sections. 
To describe the degree zero or vacuum sector introduce the state $e^{(0)}$ 
so that $\cW_0 = \mC e^{(0)}$.  That is, a general state in $\cW_0$ is 
$a e^{(0)}$ for some complex number $a$.  This is the one-dimensional representation of 
$ SO(4,2)$. We define 
\bea 
&& \eta ( e^{(0)}  , e^{(0)}  ) = \eta_{00}  = 1 
\eea
Then bilinearity requires that 
\bea 
&& \eta (a_1 e^{(0)}, a_2 e^{(0)}) = a_1 a_2 
\eea

Let us define the co-unit. This a homomorphism   $ \epsilon : \cW \rightarrow \mC $.  It corresponds  
to the cobordism from circle to vacuum. 
We define 
\bea 
\epsilon  ( a e^{(0)} + e^{(1)} + e^{(2)} + \cdots )  = a 
\eea
It just picks up the coefficient of  the vector in the trivial representation. Since this is a projector to the 
trivial irrep of $SO(4,2)$, it is $SO(4,2)$ equivariant. 
If we denote $ e_A $ the general basis vectors of $ \cW $  (say a basis  that diagonalizes the 
CFT inner product   - which can be constructed by  group theory) and we denote $A =0$ the vector $e^{(0)}$, then 
we have 
\bea 
\epsilon_{ A } = \delta_{A , 0 } 
\eea
Then we have the unit, which is a map $ \tilde \epsilon : \mC \rightarrow \cW $.  Pictorially 
it is the map from vacuum to circle. 
\bea 
\tilde \epsilon ( 1  ) = e^{(0)} 
\eea
We define $ \tilde \epsilon^{ A } $ as the coefficient of  the $ A$'th basis vector in $ \tilde \epsilon ( 1 ) $. 
Then we can write 
\bea 
\tilde \epsilon^A = \delta_{ A , 0 } 
\eea
And 
\bea 
\epsilon_A \tilde \epsilon^A =  1 
\eea
This means that the $S^2$ partition function of the TFT$_2$, which is obtained by gluing the vacuum-to-circle 
amplitude, with the circle-to-vacuum amplitude is $1$.

The definition of $ C_{ABD} $ for general degree states is given 
later in terms of Wick-contractions. Letting $D$ be a degree zero state
amounts to only having contractions between $ A , B$. It follows that we will have 
\bea 
C_{AB0}  = \eta_{AB} 
\eea
When both $ B , D  $ are degree zero states, 
\bea 
C_{A00}  \equiv \epsilon_A = \delta_{A0} 
\eea
With these definitions, the equations corresponding to 
capping off an incoming circle or an outgoing circle hold. 

\begin{figure}[ht]%
\begin{center}
\includegraphics[width=0.5\columnwidth]{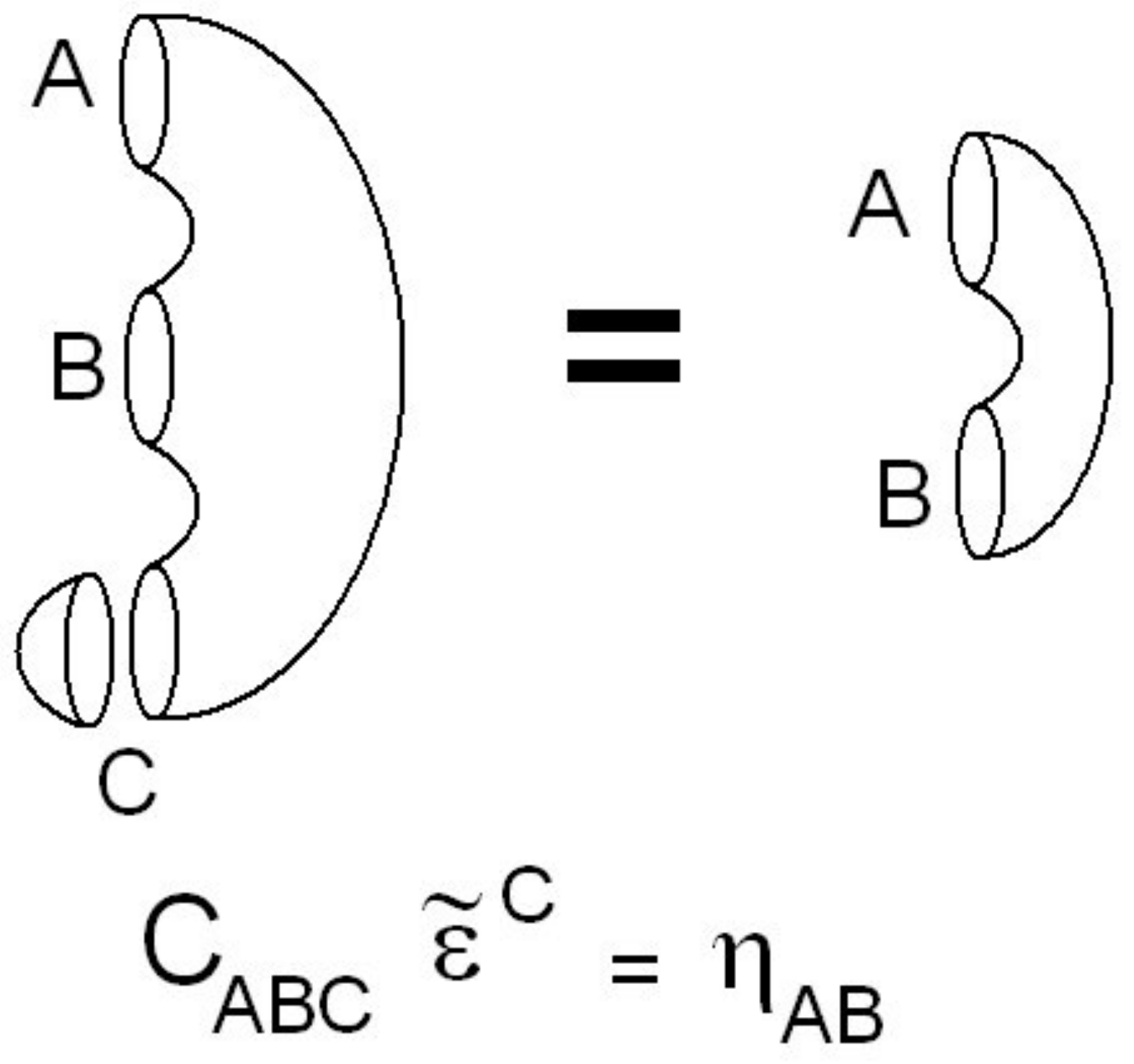}%
\caption{Capping 3-point correlator  }%
\label{fig:cap-3pt}%
\end{center}
\end{figure}

If we have 3-circles going to vacuum, and cap off one circle, then
we get just  the amplitude for 2 circles going to vacuum. 
\bea 
C_{ABC} \tilde \epsilon^C = C_{ ABC } \delta_{ C , 0 } = C_{ AB 0 } = \eta_{ A B } 
\eea
The figure for this equation is Figure \ref{fig:cap-3pt}.

\begin{figure}[ht]%
\begin{center}
\includegraphics[width=0.3\columnwidth]{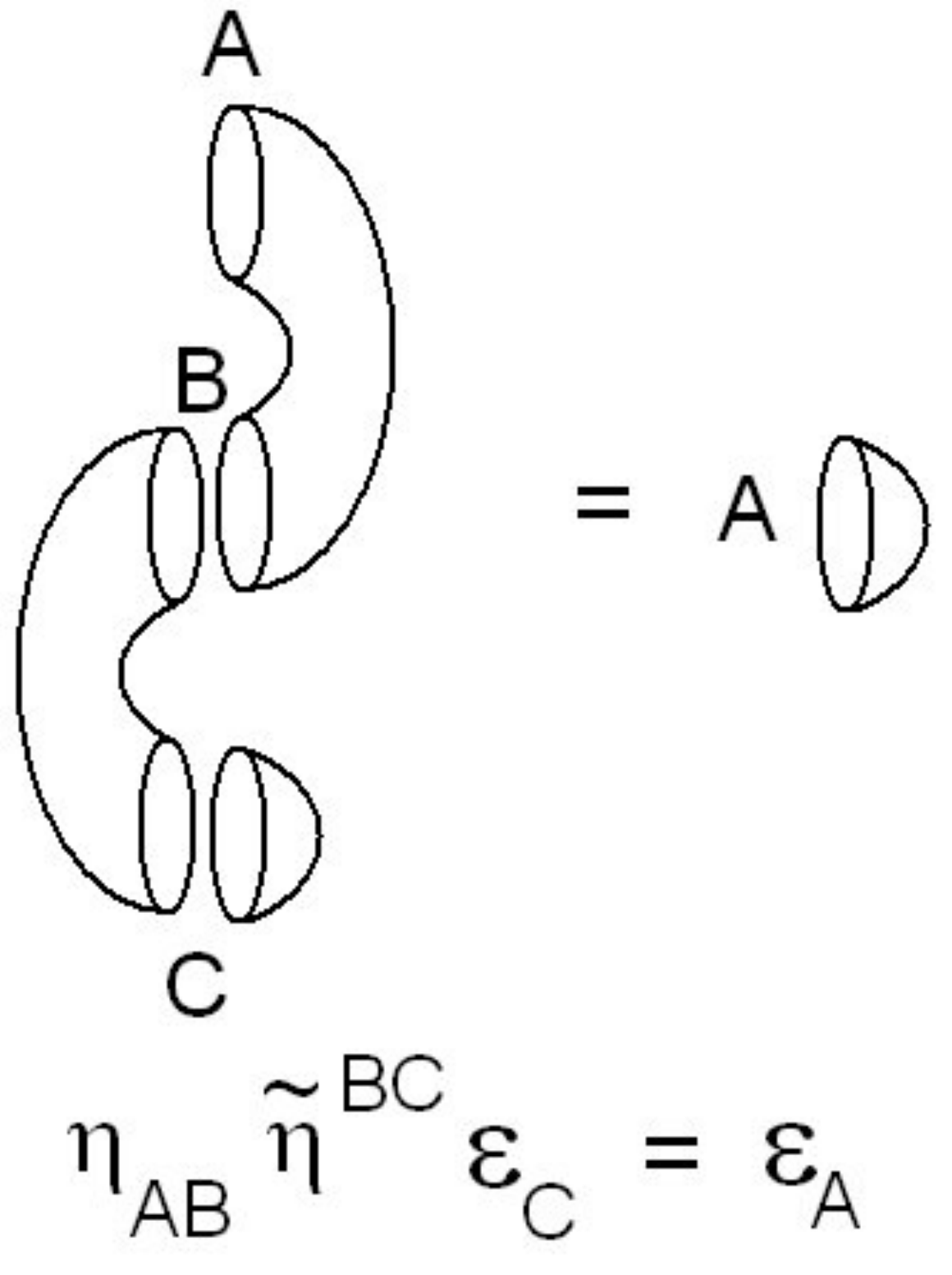}%
\caption{Capping the snake cylinder equation }%
\label{fig:snake-cap-cylinder}%
\end{center}
\end{figure}

For $\tilde\eta $, it makes sense to define 
\bea\label{tildebet0} 
\tilde \eta^{ A 0 } = \delta_{ A 0 }  
\eea
Then the equation 
\bea 
\eta_{AB} \tilde\eta^{BC} = \delta_{A}^C 
\eea
becomes in the case $ C =0$, 
\bea 
\eta_{AB} \tilde  \eta^{B0} = \eta_{AB} \delta_{ B , 0 } = \eta_{A,0} = \delta_{A,0} 
\eea
which is consistent with the definition (\ref{tildebet0}).  
We can write this as 
\bea 
\eta_{AB} \tilde  \eta^{BC} \epsilon_C = \epsilon_A  
\eea
The cobordism equivalence  corresponding to this equation is  Figure \ref{fig:snake-cap-cylinder}.

\subsection{ Associativity and crossing equations }\label{sec:asscross} 

\begin{figure}[ht]%
\begin{center}
\includegraphics[width=0.5\columnwidth]{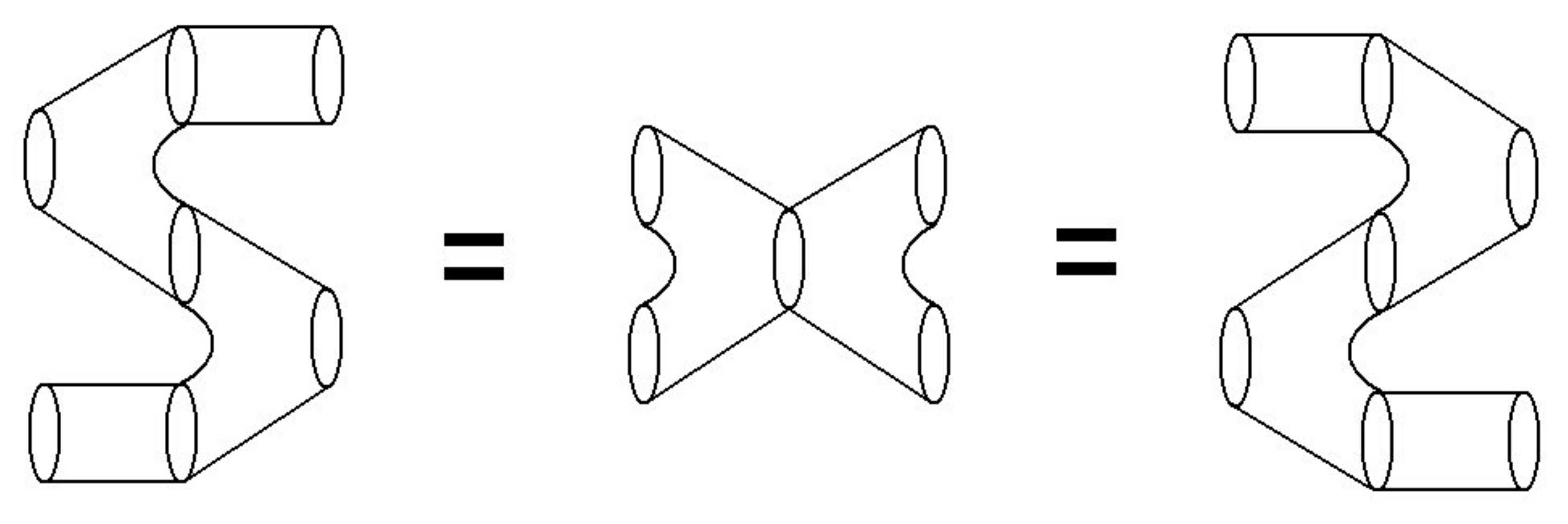}%
\caption{The Frobenius relation}%
\label{fig:frob-rel}%
\end{center}
\end{figure}

Here we use the non-degeneracy property  (\ref{non-degen-eqn}) to show that associativity, crossing 
and Frobenius equations are equivalent. The Frobenius relation is given in
Figure \ref{fig:frob-rel}.

The crossing equation is 
\bea 
C_{AB}^{ ~~~ E} C_{CD}^{~~~ F  } \eta_{ EF }   = C_{BC}^{~~~E}  C_{ AD }^{~~~ F} \eta_{ EF}  
\eea
Using the $ \eta $ to lower indices 
\bea 
C_{ C D}^{ ~~~F} \eta_{EF}  = C_{CDE} = C_{ ECD} = \eta_{EF} C^{ F}_{~~ CD } 
\eea
The symmetry of $C$ follows because we have a CFT of bosons and  
is an algebraic property of the Wick contraction map acting on $\cW\otimes\cW\otimes\cW$. 
We can do the same steps on the RHS and arrive at 
\bea 
C_{ BC}^{ ~~~E}  C_{EAD } 
\eea
Now raise the $D$ index on both side using $ \tilde \eta$ and 
we get the associativity equation is 
\bea 
C_{ AB }^{~~~ E} C_{E C}^{~~~D}  = C_{BC}^{~~~E} C_{EA}^{~~~D} 
\eea

We also have the Frobenius equation \cite{Kock}  ( sometimes called the  nabla-delta equals delta-nabla relation), which can demonstrated using the existence of the inverse $\tilde \eta$ . 
Start from the nabla-delta equation
\bea 
C_{ AB }^{ ~~~E } C_{ E}^{ ~~ CD } = C_{A}^{~~ED} C_{ E B }^{~~~C } 
\eea
and lower the $C,D$ indices to obtain
\bea 
C_{AB}^{~~~ E } C_{E CD } = C_{ AED} C^{ E}_{ ~~~ BC } 
\eea
The RHS can be rearranged using the symmetry of $C_{ ... }$ as follows
\bea 
C_{AED} C^{E}_{ ~~~ BC  } =  C_{ ADE } C^{E}_{ ~~~ BC  }  =  C_{ AD }^{ ~~~ E }  C_{E BC  } 
\eea
This proves that the crossing equation implies both associativity and the equality of nabla-delta 
and delta-nabla. This last equation is called the Frobenius relation.
These manipulations use the inverse of $\eta$, without 
ever encountering $ \eta_{AB}\tilde  \eta^{AB}$ which diverges.

\subsection{Toy Model : Gaussian Integration}\label{sec:modelGaussian} 

Our toy model employs the ring of polynomials in one variable.
The states of the model are defined by
\bea 
&& \phi_n =  : x^n : = e^{ - { 1 \over 2 }  \partial_x^2 } x^n
\eea
It is straight forward to introduce an inner product on this space of states
\bea
&& g_{nm} = \langle\phi_n | \phi_m \rangle = {\int dx e^{-x^2/2}  : x^n : ~ :x^m : \over \int dx e^{ -x^2/2}}
\eea  
This inner product will play the role of the bilinear pairing of the TFT.
Carrying out the integral above, we find 
\bea 
g_{nm} = \langle\phi_n | \phi_m \rangle = \delta_{nm} n! 
\eea
Since $g_{nm}$ is clearly invertible, we have proved that this model has a non-degeneracy equation.
Now, define the TFT structure constants by
\bea
   C_{n_1,n_2,n_3}= \langle\phi_{n_1 }\phi_{n_2}\phi_{n_3}\rangle
\eea
where
\bea
&& \langle \phi_{n_1 }\phi_{n_2} \phi_{n_3} \cdots \phi_{n_p} \rangle 
=  { \int dx e^{-x^2/2}  : x^{n_1}  : ~ :x^{n_2}  : ~ : x^{n_3}  : ~\cdots ~ :x^{n_p}  :  \over     \int dx e^{ -x^2/2}  }
\label{intforsc}
\eea
Again, carrying out the integral we find
\bea 
   C_{n_1 , n_2 , n_3 } = \sum_{ k =0} \delta ( n_3 , n_1 + n_2 - 2k ) k! { n_1 \choose k } { n_2 \choose k } n_3!
   \label{structure-gauss}
\eea
The structure constants $C_{ab}^{~~c}$ give a product and we can get a co-product from $C_{a}^{~bc}$
\bea 
&& \nabla ( \phi_a \otimes \phi_b ) = C_{ab}^{~c} \phi_c \cr  
&& \Delta ( \phi_a )  = C_{a}^{~bc} \phi_b \otimes \phi_c 
\eea
The integral expression for $C_{abc}$ prove that the structure constants are symmetric.
The above structure constants and bilinear pairing will define a TFT$_2$  provided the crossing equation
\bea\label{assoc} 
\langle \phi_{n_1}\phi_{n_2}\phi_{n_3}\phi_{n_4}\rangle  = \sum_{m}C_{n_1,n_2,m}C_{n_3,n_4,m}{1\over m!}=\sum_{m}C_{n_1,n_3,m}C_{n_4,n_2,m}{1\over m!} 
\eea
holds. Since the numerical values of the structure constants are given in (\ref{structure-gauss}), it is straightforward
to verify (\ref{assoc}) by plugging numbers in (e.g. with the help of Mathematica). More generally, it is clear that the crossing equation 
 follows from the fact 
that the two expressions in (\ref{assoc}) are two ways of calculating the integral (\ref{intforsc}). 
Thus, this model does indeed define a TFT.

A simple generalization of the above model, is to consider a Gaussian Hermitian matrix model.  
The natural analog of the observables above are parametrized by $[n,\sigma]$, where $n$ is a positive integer and 
$\sigma $ is a permutation  in $S_n$
\bea
   \phi_{[n,\sigma ]}&&= :X^{i_1}_{i_{\sigma(1)}}\cdots X^{i_n}_{i_{\sigma(n)}}:\cr
                     &&= e^{-{1\over 2}{\rm Tr}\left({d\over dX}{d\over dX}\right)} X^{i_1}_{i_{\sigma(1)}}\cdots X^{i_n}_{i_{\sigma(n)}}\cr
                     &&\equiv e^{-{1\over 2}{\rm Tr}\left({d\over dX}{d\over dX}\right)} {\rm Tr}(\sigma X^{\otimes\, n})
\eea
Again, a natural inner product on this set of states ($dX$ is the usual $U(N)$ invariant  measure for Hermitian matrices)
\bea
   \langle \phi_{[n,\sigma ]} |\phi_{[m,\tau]}\rangle =
   {\int dX \, e^{-{1\over 2}{\rm Tr}(X^2)}\, :{\rm Tr}(\sigma X^{\otimes \,n}):\, :{\rm Tr}(\tau X^{\otimes \,m}):\over
    \int dX \, e^{-{1\over 2}{\rm Tr}(X^2)}}
\eea
gives us the bilinear pairing of the TFT. Carrying out the integral above, we find 
\bea
\langle \phi_{[n,\sigma ]} |\phi_{[m,\tau]}\rangle =\delta_{nm}
\sum_{ \sigma_1' \in T_1 } \sum_{ \sigma_2' \in T_2 } \sum_{ \sigma_3 \in S_n } \delta ( \sigma_1' \sigma_2' \sigma_3 ) N^{ C_{ \sigma_3} }  
\label{pairperm}
\eea
$T_1$ is the conjugacy class of the permutation $\sigma$. $T_2$ is the conjugacy class of the 
permutation $\tau$. $C_{\sigma_3}$ is  the number of cycles in  the permutation $ \sigma_3$. 
This model again has a non-degeneracy equation.
Note however, the pairing (\ref{pairperm}) is not diagonal in this permutation basis.
This will obscure the associativity of the model, although it could be verified by explicit computations. 
A simpler description is obtained by changing basis with a Fourier transform on the symmetric group 
\bea
\phi_{ [ n , R ] } &&= { 1 \over n! } \sum_{ \sigma \in S_n  } \chi_R ( \sigma )    \phi_{ [ n  , \sigma  ] } \cr 
&& = {1\over n!}\sum_{\sigma\in S_n}\chi_R (\sigma):X^{i_1}_{i_{\sigma(1)}}\cdots X^{i_n}_{i_{\sigma(n)}}:
\eea
Above $R\vdash n$ is a Young diagram with $n$ boxes and $\chi_R (\sigma)$ is a character of the symmetric group.
Our states are now the Schur polynomials for which the pairing is diagonal
\bea
  \langle \phi_{[n,R]} |\phi_{[m,S]}\rangle =\delta_{RS}f_R
\eea
with $f_R$ the product of the factors of Young diagram $R$ \cite{CJR}.
The TFT structure constants are now
\bea
  C_{[n,R],[m,S],[p,T]}=\langle \phi_{[n,R]}\phi_{[m,S]}\phi_{[p,T]}\rangle
\eea
where
$$
 \langle \phi_{[n_1,R_1] } \cdots \phi_{[n_p,R_p]} \rangle 
=    {\int dM \, e^{-{1\over 2}{\rm Tr}(M^2)}\, \phi_{[n_1,R_1]}\,\cdots\, \phi_{[n_p,R_p]}\over
    \int dM \, e^{-{1\over 2}{\rm Tr}(M^2)}}
$$
The crossing  equation
\bea
\langle \phi_{[n_1,R_1]}\phi_{[n_2,R_2]}\phi_{[n_3,R_3]}\phi_{[n_4,R_4]}\rangle  
&& = \sum_{p=|n_1-n_2|}^{n_1+n_2}\sum_{R}C_{[n_1,R_1],[n_2,R_2],[p,R]}C_{[n_3,R_3],[n_4,R_4],[p,R]}{1\over f_R} \cr 
&& = \sum_{p=|n_1-n_3|}^{n_1+n_3}\sum_{R}C_{[n_1,R_1],[n_3,R_3],[p,R]}C_{[n_4,R_4],[n_2,R_2],[p,R]}{1\over f_R} \cr 
&& ~~ 
\eea 
will hold because, again, the two expressions appearing above are two ways of calculating the integral needed to evaluate the right hand side. 
The product takes the form
\bea 
  \phi_{[n_1,R_1]}*\phi_{[n_2,R_2]} = \sum_k C (n_1,n_2,k,R_1,R_2,R_1 *_k R_2) \phi_{[n_1+n_2-2k,R_1 *_k R_2 ]} 
\eea
where $C (n_1,n_2,k,R_1,R_2,R_1 *_k R_2)\in\Z$ is a combinatoric factor. 
We have defined a  star-product labeled by $k$. 
In this Fourier basis of Young diagrams, the $k=0$ product is 
given by Littlewood-Richardson coefficients. In the original permutation basis 
this is the outer product which takes $ \sigma_1 \in S_{n_1}, \sigma_2 \in S_{n_2} $ 
to give $ (  \sigma_1 \circ \sigma_2 ) \in S_{ n_1} \times S_{n_2} \in  S_{n_1+ n_2}$. 
For the case, $n_1 = n_2 $ and $ k = n_1 = n_2$, the product is 
the ordinary product of permutations. The intermediate cases correspond to products 
where one permutation acts on a set of $n_1 $ integers, the second on a subset of 
$n_2$ integers, where the sets overlap over $k$ elements. This structure is readily derived 
using diagrammatic tensor space techniques \cite{cr,BDS}.

\subsection{ Zero area YM2  :   An example of infinite dimensional state spaces and restricted amplitudes } 

In the above discussions and in most of the paper, we focus on examples where TFT$_2$ with  
infinite dimensional state spaces arise, with a restriction to genus zero. 
There are also examples where infinite dimensional state spaces arise, and 
infinities are avoided by restricting to surfaces of  genus greater than one. 
The partition function of YM2 on a surface of genus $G$ with area $A$ \cite{Migdal}, with $ g_{YM}^2 $ set to $1$,
is 
\bea 
Z_{YM2} ( G , A  ) = \sum_{ R } ( Dim R )^{ 2 - 2G }  e^{ - A C_2 ( R ) } 
\eea
where the sum is over irreps  $R$ of the gauge group, $ Dim R $ is the dimension of the 
representation $R$, $C_2(R)$ is the quadratic Casimir. In the zero area limit, we
have 
\bea
Z_{YM2} ( G , A=0  ) = \sum_{ R } ( Dim R )^{ 2 - 2G }  
\eea
This diverges for $ G =0 , 1$ for a Lie group, e.g.  $ SU(N)$,  since there are infinitely 
many irreps with arbitrarily large dimensions.   However the partition function is well defined for $ G >1$. 
The $A=0$ limit is interesting as a topological limit, from the point of view of 
the moduli space of flat $SU(N)$ connections \cite{Witten2dYM}  and also from the large $N $ 
expansion and gauge-string duality \cite{GrTa,CMR}.

\section{Basic CFT$_4$   2-point function as an invariant map in TFT$_2$  } \label{sec:basic}

In free massless scalar field  theory in four dimensions, all the correlators 
of composite local  operators inserted at distinct points, can be obtained 
from the basic 2-point function of the elementary field
\bea 
\langle \varphi ( x_1 ) \varphi ( x_2  ) \rangle 
= { 1 \over ( x_1 - x_2 )^2 }  \equiv G ( x_1 , x_2 ) 
\eea
where $ x_1^{\mu}  , x_2^{ \mu }   $ are points in $\mR^4$. 
If we transform $ \varphi ( x_2 ) \rightarrow \varphi' ( x_2') $ 
where $ x_2' ={  x_2 \over x_2^{ 2 } } $ , we encounter 
\bea 
\langle \varphi ( x_1) \varphi' ( x_2' ) \rangle = ( x_2 )^2 \langle \varphi ( x_1 ) \varphi ( x_2 ) \rangle =  { 1 \over ( 1 - 2 x_1 \cdot x_2' + (x_2')^2  x_1^2 )  } \equiv F ( x_1 , x_2' ) 
\eea
In this section we will show that this quantity encodes  the way 
the one-dimensional representation $ \mC $ appears inside the 
tensor product $ V_+ \otimes V_-$. The representation $ V_+$ is 
the irrep of $ SO(4,2) $ where the lowest energy state has dimension $  D=1$, 
corresponding to the state $ \varphi $ of CFT$_4$, and the other states correspond to derivatives of $ \varphi$, or equivalently to strings of $ P_{\mu}$ acting on 
the lowest weight state.  We use energy/dimension/weight  interchangeably in 
this paper, since we are working with radial quantization. Its conjugate 
is $V_-$ which is the representation with highest weight (energy)  having 
$D=-1$ and where states are generated by $ K_{\mu}$. 

We start by describing the properties of the bilinear  $SO(4,2)$ invariant map $\hat\eta : V_+ \otimes V_- \rightarrow \mC $. 
A Lie algebra element $ \cL \in so(4,2)$ acts on the tensor product as 
\bea 
\cL \otimes 1 + 1 \otimes \cL \equiv \Delta ( \cL ) 
\eea
The complex number field is the one-dimensional representation where the Lie algebra acts as zero. 
An equivariant map $ \hat \eta $ to $ \mC$ obeys 
\bea 
\Delta  ( \cL ) \circ  \hat \eta  = \hat \eta \circ \Delta ( \cL )  
\eea
The LHS is  zero because $ \mC$ is the trivial irrep, and the expanding the RHS gives 
\bea 
\hat \eta ( \cL v_1 , v_2 ) + \hat \eta ( v_1, \cL v_2 )    = 0 
\eea
This equivariance property exactly fits the definition of what is required in TFT$_2$ with $SO(4,2)$ 
as global symmetry. 

The requirement that this map is $SO(4,2)$ invariant fixes it up to an overall constant. Equivalently the decomposition of  the tensor product 
$V_+ \otimes V_-$ 
in terms of irreducible representations contains a unique copy of the 
trivial one-dimensional  representation $\mC$, where the Lie algebra elements 
act as zero.
The states in $V_+$ are of the form 
\bea 
C^I_{ \mu_1 \cdots \mu_n } P_{ \mu_1 } \cdots P_{\mu_n } v^+
\eea
where the $C^I$ are symmetric traceless tensors. 
We will  introduce (Euclidean) spacetime coordinates $x^{ \mu } $ 
to keep track of  these states. We can think of $ x^{ \mu}$ as a 
way to describe the states in $V_+ $ via a continuous variable as
opposed to a discrete variable.  Analogously for $ V_- $ we have 
states obtained by acting with $ K_{\mu} $ and  the variable $ x' $ 
is  the continuous variable.  We will show that 
the invariant map $\hat\eta$, described 
 in this spacetime basis is  the 2-point function  $ F ( x_1 , x_2' ) $.
Further, in the spacetime basis, there are a simple set of differential equations expressing the $SO(4,2)$ invariance of $\hat\eta$.
Finally, using $\hat\eta$ and a map $\rho:V_+\to V_-$ we are able to define an inner product  on $ V_+ $ and on $ V_-$. 
This is a map from $ V_{ \pm }  \times V_{ \pm}  $ to $ \mC$ which is 
{ \it sesquilinear}. 
The inner product obtained in this way is the usual one which is used, for example, to study the bounds unitarity places on
operator dimensions\cite{Mack:1975je}.

\subsection{Flat space quantization and radial quantization}

Start from the $so(4,2)$ algebra (we use $g ={\rm diag}(-,-,+,...,+)\,\,\,)$
\bea
   \left[S_{MN},S_{PQ}\right]=g_{NP}S_{MQ}+g_{MQ}S_{NP}-g_{MP}S_{NQ}-g_{NQ}S_{MP}
\eea
The indices $ M , N , \cdots $  run over $ \{ - 1 , 0 , 1 , \cdots d \} $. 
Note that the structure constants are all real and the generators are all antihermittian $S_{MN}^\dagger =-S_{MN}$.
We can write this algebra in two different ways, which make different subgroups manifest.
For useful background material see \cite{Minwalla:1997ka} and section 2.1.2 of \cite{Aharony:1999ti}.

\subsubsection{Manifest $so(3,1) \times so(1,1)$ subgroup}

This rewriting is relevant for quantization of the theory on $R^4$ or $R^{3,1}$. Each equal time slice  is a copy of  3-dimensional Euclidean space, $R^3$.
Identify
\bea
  M_{\m\n}'&=&S_{\m\n}\qquad D'=S_{-1,d}\cr
  P_\m'&=&S_{\m,-1}+S_{\m,d}\qquad K'_\m=S_{\m,-1}-S_{\m,d}
\eea
with $ \mu \in \{ 0 , 1 , \cdots , d-1 \} $.
From the anti-hermiticity of the $S_{MN}$s we find
\bea
  (M_{\m\n}')^\dagger &=& -M_{\m\n}'\qquad (D')^\dagger =-D'\cr
  (P'_\m)^\dagger &=&-P'_\mu         \qquad (K'_\m)^\dagger =-K'_\m
\eea
i.e. all of the generators have pure imaginary eigenvalues. 
The algebra obeyed by these generators is
\bea
\big[M_{\m\n}',M'_{\a\b}\big] &=& g_{\m\b}M'_{\n\a}+g_{\n\a}M'_{\m\b}-g_{\m\a}M'_{\n\b}-g_{\n\b}M'_{\m\a}\cr
\big[M_{\m\n}',D'\big]&=& 0\qquad \big[D',P'_\r\big]=-P'_\r\qquad \big[D',K'_\r\big]=K'_\r\cr
\big[M_{\m\n}',P'_\r\big]&=& g_{\n\r}P'_\m-g_{\m\r}P'_\n \qquad
\big[M_{\m\n}',K'_\r\big]=g_{\n\r}K'_\m-g_{\m\r}K'_\n\cr
\big[K_\m',P_\n'\big]&=&2M'_{\m\n}-2g_{\m\n}D'
\eea
Notice that the $M_{\m\n}'$ generate $SO(3,1)$ and $D'$ generates $SO(1,1)$.

\subsubsection{Manifest $so(4) \times so(2)$ subgroup}\label{radialalgebra}

This rewriting is relevant for the radial quantization. Equal ``time slices'' are three-spheres, $S^3$.
Identify
\bea\label{Hconj} 
  M_{pq}&=&S_{pq}\qquad D=iS_{-1,0}\cr
  P_p&=&S_{p,-1}+iS_{p,0}\qquad K_p= S_{p,-1} - iS_{p,0}
\eea
with $ p \in \{ 1, \cdots , d \} $. 
Notice that from the anti-hermitticity of the $S_{MN}$'s we find
\bea\label{hermconj1} 
(M_{pq})^\dagger &=& -M_{pq},\qquad (D)^\dagger = D\cr
(P_p)^\dagger    &=& -K_p,\qquad (K_p)^\dagger = -P_p
\label{daggeraction}
\eea
Thus, $D$ will have real eigenvalues and $M_{pq}$ will have purely imaginary eigenvalues.
The algebra obeyed by these generators is
\bea
  \big[M_{pq},M_{rs}\big]&=&\delta_{qr}M_{ps}+\delta_{ps}M_{qr}-\delta_{qs}M_{pr}-\delta_{pr}M_{qs}\cr
  \big[M_{pq},D\big]&=& 0\qquad \big[D,P_p\big]=P_p\qquad \big[D,K_p\big]=-K_p\cr
  \big[M_{pq},P_r\big]&=&\delta_{qr}P_p-\delta_{pr}P_q \qquad
  \big[M_{pq},K_r\big]=\delta_{qr}K_p-\delta_{pr}K_q\cr
  \big[K_p,P_q\big]&=&2M_{pq}-2\delta_{pq}D
  \label{SO4SO2algebra}
\eea
Clearly then, $M_{pq}$ generate the $SO(4)$ subgroup, while $D$ generates the $SO(2)$ subgroup.

\subsection{ Invariant pairing } 

We will use the writing of $so(4,2)$ which makes the $so(4)\times so(2)$ subalgebra manifest. 
We want to consider two different $so(4,2)$ representations $V_+$ and $V_-$. 
$V_+$ is built on the lowest weight state $v^+$ which obeys
\bea
   D v^+ = d v^+ \qquad M_{pq} v^+  =0
\eea
The remaining states in this irrep are constructed by acting with traceless combinations of $P_p$s on $v^+$.
A convenient way to describe this is to write
\bea
  v^{+,I,l} =C^{I}_{p_1\cdots p_l}P_{p_1}\cdots P_{p_l}v^+
\eea
where the tensor  $C^{I}_{p_1\cdots p_l}$ is symmetric traceless in the $p_1,\cdots,p_l$ indices. 
The index $I$ runs over the states in the $SO(4)$ irrep $({l\over 2},{l\over 2})$.
We will often trade $I$ for $SU(2)\times SU(2)$ state labels $(m_L,m_R)$.

$V^-$ is built on the highest weight state $v^-$ which obeys
\bea
   D v^- =- d v^- \qquad M_{pq} v^- =0
\eea
The remaining states in this irrep are constructed by acting with traceless combinations of $K_p$'s on $v^-$.
The representations that we consider most of the time are relevant for the description of a free massless bosonic 
scalar field in 4 dimensions, in which case we set $d=1$.
For the remainder of this section, we will set $d=1$.

The invariant pairing is $\hat\eta : V_+ \otimes V_- \rightarrow \mC$. Concretely
\bea
   \hat\eta \left( C^I_{p_1\cdots p_l}P_{p_1}\cdots P_{p_l}v^+ , C^{J}_{q_1\cdots q_{l'}}K_{q_1}\cdots K_{q_{l'}}v^- \right) 
    = f_{l,m_L,m_R;l'm_L',m_R'}
\eea
On the right hand side we have traded $I$ for $(m_L,m_R)$ and $J$ for $(m_L',m_R')$.
In the next section we will prove that the requirement of $so(4,2)$ invariance determines $f_{l,m_L,m_R;l'm_L',m_R'}$ up to an overall constant. 
A convenient way of summarizing the action of $\hat\eta$ is in terms of the tensors
\bea
   T_{p_1 p_2\cdots p_n,q_1 q_2\cdots q_n}=\hat\eta\left( P_{p_1}P_{p_2}\cdots P_{p_n} v^+,K_{q_1}K_{q_2}\cdots K_{q_n} v^-\right)
   \label{etaonstates}
\eea
which are themselves nicely summarized as
\bea
T_{p_1\cdots p_n, q_1\cdots q_n}\prod_{a=1}^n y^{\prime q_a}x^{p_a}
=2^n n! \sum_{l=0}^{\big[{n\over 2}\big]}(-1)^l (x\cdot y')^{n-2l}|x|^{2l}|y'|^{2l}{(n-l)!n!\over (n-2l)!2^{2l}l!}
\label{tensorev}
\eea
To evaluate (\ref{etaonstates}), we can use the $so(4,2)$ invariance 
of the pairing to shift (say) $P_p$'s from the left slot to the right slot.
The action of $P_p$ on the state in the right slot is then easily computed 
by using the $so(4,2)$ algebra as well as the fact that $P_p$ annihilates $v^-$.
In the next section, this logic will be applied also to  $SU(2)\times SU(2) $ and $ SL(2)$ sub-algebras, to derive
an explicit formula for $f_{l,m_L,m_R;l'm_L',m_R'}$.
A straight forward computation now gives
\bea
   F(x,y')&&\equiv \hat\eta\left( e^{-ix\cdot P'}v^+,e^{iy'\cdot K'}v^-\right)\cr
&&=\sum_{n=0}^\infty T_{p_1\cdots p_n, q_1\cdots q_n}\prod_{a=1}^n y^{\prime q_a}x^{p_a}{1\over (n!)^2}\cr
&&=\sum_{n=0}^\infty 2^n \sum_{l=0}^{\big[{n\over 2}\big]}(-1)^l (x\cdot y')^{n-2l}|x|^{2l}|y'|^{2l}{(n-l)!\over (n-2l)!2^{2l}l!}\cr
&&={1\over 1-2x\cdot y' +x^2 y^{\prime 2}}
\label{explicitF}
\eea
For a closely related discussion, see section 3 of \cite{Pappadopulo:2012jk}.

\subsubsection{ Description of pairing in terms of $ SU(2) \times SU(2)$ and $SL(2)$ subalgebras }\label{fsection}

The requirement of $so(4,2)$ invariance determines $f_{l,m_L,m_R;l'm_L',m_R'}$ up to an overall constant.
This is most easily demonstrated by requiring that $\hat\eta$ is invariant under $sl(2)$ and $su(2)\times su(2)$ subalgebras.
We will prove that the $\hat\eta$ obtained in this way enjoys the full $so(4,2)$ invariance.

 The complete set of states of $V_+$ can be obtained
by acting with elements of the $sl(2)$ and $su(2)\times su(2)$ subalgebras on $v^+$, and the complete set of states of 
$V_-$ can be obtained by acting with elements of the $sl(2)$ and $su(2)\times su(2)$ algebras on $v^-$.
This can be seen  by considering the $SO(4,2)$ character in equation (\ref{nisonehere}) for $V_+$.
The coefficient of $s^{q+1}$ in this character
\bea
   \chi_{V_+}(s,x,y)=\sum_{q=0}^\infty s^{q+1}\chi_{q\over 2}(x) \chi_{q\over 2}(y)
\eea
is $\chi_{q\over 2}(x)\chi_{q\over 2}(y)$. 
This implies that the states of scaling dimension $q+1 $ fill out a multiplet of spin $(j_L,j_R)=({q\over 2},{q\over 2})$. 
This complete multiplet can be generated by applying $SU(2)\times SU(2)$ rotations to $(H_+)^q v^+$, which shows that we do indeed generate
the complete set of states in $V_+$ by acting with the $sl(2)$ and $su(2)\times su(2)$ subalgebras on $v^+$.
A similar argument shows that we generate the complete set of states of dimension $-q-1$ in $V_-$ by acting with elements 
of $su(2)\times su(2)$ on $(H_-)^q v^-$.

Invariance under $sl(2)$ and $su(2)\times su(2)$ leads to 
\bea 
   f_{l,m_L,m_R;l',m_L',m_R'}=f_{l}\left(SL(2)\right)f_{m_L}\left(SU(2)\right)f_{m_R}\left(SU(2)\right)
                              \delta_{l,l'}\delta_{m_L,- m_L'}\delta_{m_R,- m_R'} 
\eea
We will demonstrate how invariance under $sl(2)$ fixes $f_{l}\left(SL(2)\right)$.
The demonstrations for $f_{m_L}\left(SU(2)\right)$ and $f_{m_R}\left(SU(2)\right)$ are very similar.

The subalgebra that we study is described in detail in Appendix \ref{sec:subgroupapp}.
The two $su(2)$ algebras have raising and lowering operators given by $J_{\pm}^R$ and $J_{\pm}^L$, while the raising and lowering operators
of $sl(2)$ are $H_\pm$.
In terms of these generators, the pairing $\hat\eta$ is
\bea\label{genpair}  
\hat\eta( (J_+^R )^{p_{R}} (J_+^L)^{p_{L}} (H_+)^l v^+ , (J_-^R )^{p_{R}'}(J_-^L )^{p_{L}'}(H_-)^{l'} v^-) 
= \delta_{p_{L},p_{ L}'}\delta_{p_{R},p_{R}'}\delta_{l,l'}f_{l,p_L }^{su(2)}f_{l,p_R }^{su(2)} f_{d,l }^{sl(2)}   
\eea

To demonstrate how $sl(2)$ invariance fixes $f_{d,l}^{sl(2)}$, consider the  positive discrete series irrep $\tilde{V}_+$  of $ sl(2)$ and the  negative 
discrete series irrep $\tilde{V}_-$ 
\bea 
&& \tilde{V}_+  = Span \{ H_+^l v^+  ~~ : ~~ H_- v^+ = 0  ~~ , ~~  H_3 v^+ =  d v^+  \} \cr 
&& \tilde{V}_- = Span \{ H_-^l v^-   ~~ : ~~ H_+ v^- = 0  ~~ , ~~  H_3 v^- = -d v^-  \} 
\eea
Note that $\tilde{V}_{\pm}$ are subspaces of the irreducible $SO(4,2)$ representations  $V_{\pm}$ that we introduced above.
Invariance of the pairing $ \hat\eta( H_+^l  v^+ , H_-^{l '}   v^- ) $ under $ H_3$ gives 
\bea
\Delta (H_3) \,\,\hat\eta ( H_+^l  v^+ , H_-^{l '}v^-) &=& \hat\eta( H_3 H_+^l  v^+ ,  H_-^{l '} v^- ) +  \hat\eta( H_+^l  v^+ ,  H_3  H_-^{l '}   v^- ) \cr  
&=&((d + l) - (d + l'))   \hat\eta ( H_+^l  v^+ , H_-^{l '}v^-)  
\eea
which shows it vanishes unless $l=l'$, so that 
\bea 
\hat\eta( H_+^l  v^+ , H_-^{l '}   v^- ) = f_{d,l}^{sl(2)} \delta_{ l l' }  
\eea
Then we have 
\bea\label{Hminrec}  
\Delta(H_- )\,\,\hat\eta(H_+^l v^+,H_-^{l'}v^-)&=&\hat\eta(H_- H_+^{l}v^+, H_-^{l'} v^-) +  \hat\eta(H_+^{l}v^+, H_-^{l'+1}v^-)\cr 
&=&  \delta_{l,l'+1} (\sum_{i =0}^{l-1} (-2i + d)f_{d,l-1}^{sl(2)} + f_{d,l}^{sl(2)})  
\eea
which gives 
\bea 
f_{l,d}^{sl(2)} = l(l+d-1) f_{l-1,d}^{sl(2)} 
\eea
This is solved by 
\bea\label{fsl2}  
f_{ l , d }^{sl(2)}  = l! { ( d + l -1 ) ! \over ( d -1) ! } 
\eea
A very similar argument requiring invariance under $su(2)$, shows that
\bea\label{fsu2}  
f^{su(2)}_{l,p} = (-1)^p {l! p!  \over (l-p)!}
\eea

\subsubsection{$so(4,2)$ invariance of the pairing}\label{invpair}

To obtain the pairing $\hat\eta$ we have required invariance under the $sl(2)$ and $su(2)\times su(2)$ algebras.
In this section we will show that the pairing we have obtained enjoys the bigger $so(4,2)$ invariance.
The requirement of $so(4,2)$ invariance translates into a set of partial differential equations for $F(x,y')$.
The demonstration then amounts to showing the $F(x,y')$ does indeed obey these partial differential equations.

Recall that
\bea
   F(x,y')=\hat\eta\left( e^{-ix\cdot P}v^+,e^{iy'\cdot K}v^-\right)
   \label{ForF}
\eea
To start, we will consider invariance under dilatations. 
Towards this end, note that
\bea
  D e^{-ix\cdot P}v^+ =\sum_{n=0}^\infty (n+1){(-ix\cdot P)^n\over n!}v^+
                      =\left( x\cdot {\partial\over\partial x}+1\right)e^{-ix\cdot P}v^+
\eea
and
\bea
  D e^{iy'\cdot K}v^-=-\sum_{n=0}^\infty (n+1){(iy'\cdot K)^n\over n!}v^-
                      =-\left( y'\cdot {\partial\over\partial y'}+1\right)e^{iy'\cdot K}v^-
\eea
Thus, the statement of invariance under dilatations
\bea
  \Delta (D)\,\,\hat\eta\left( e^{-ix\cdot P}v^+,e^{iy'\cdot K}v^-\right) = 0
\eea
is equivalent to the differential equation
\bea
   \left( x\cdot {\partial\over\partial x}-y'\cdot {\partial\over\partial y'}\right) F(x,y')=0
\eea
It is straightforward to check that the function given in (\ref{explicitF}) obeys this equation.

Next, some algebra shows that
\bea
P_{p}(y'\cdot K)^n v^- =&&
  n(n-1)[(y'\cdot K) y_p' -(y')^2 K_p](-iy'\cdot K)^{n-2} v^-\cr
&&+iy_p' n(n+1)(-iy'\cdot K)^{n-1} v^-
\eea
which implies
\bea
P_p e^{iy'\cdot K}v^- = -
\big[
2iy_p' y'\cdot {\partial\over\partial y'}-i(y')^2 {\partial\over\partial y_p'}+2iy_p'
\big]e^{iy'\cdot K}v^-
\eea
It is also clear that
\bea
  P_p e^{-ix\cdot P}v^+ = i{\partial\over\partial x^p}e^{-ix\cdot P}v^+
\eea
Consequently, the statement
\bea
  \Delta (P_p)\,\,\hat\eta\left( e^{-ix\cdot P}v^+,e^{iy'\cdot K}v^-\right) = 0
\eea
is equivalent to the differential equation
\bea
\left[
i{\partial\over\partial x^p}
-2iy_p' y'\cdot {\partial\over\partial y'}+i(y')^2 {\partial\over\partial y_p'}-2iy_p'
\right] F(x,y')=0
\eea
Again, this equation is obeyed by (\ref{explicitF}).

Finally, note that
\bea
  M_{pq}(iy'\cdot K)^n v^- = i n (y_p' K_q-y_q'K_p)(i y'\cdot K)^{n-1} v^-
\eea
and 
\bea
  M_{pq}(-ix\cdot P)^n v^+ = -i n (x_p P_q-x_q P_p)(-i x\cdot P)^{n-1} v^+
\eea
Consequently, the statement
\bea
  \Delta (M_{pq})\,\,\hat\eta\left( e^{-ix\cdot P}v^+,e^{iy'\cdot K}v^-\right) = 0
\eea
is equivalent to the differential equation
\bea
\left[
x_p{\partial\over \partial x_q}-x_q{\partial\over \partial x_p} +
y'_p{\partial\over \partial y'_q}-y'_q{\partial\over \partial y'_p}
\right] F(x,y')=0
\eea
This equation is again obeyed by (\ref{explicitF}), which completes the demonstration of $so(4,2)$ invariance.

\subsection{Invariant pairing to Inner product via twist map $ \rho$} 
\label{sec:etarogee} 

In this section we define an inner product $g:V_{ \pm } \times V_{ \pm } \to \C$. 
Since we have the invariant map $\hat\eta:V_+\times V_-\to\C$, if we introduce a map $\rho :V_+ \rightarrow V_-$ 
we can construct an inner product by composing $\rho$ and $\hat\eta$.
We will choose the map $\rho$ so that the inner product obtained is the usual inner product of the CFT used to test unitarity.
The positivity of  this inner product is what puts constraints on dimensions of fields. 
For example, a scalar should not have dimension lower than $1$, as first proved by Mack \cite{Mack:1975je}.
$\hat\eta $ is the building block of $ SO(4,2)$ invariant maps, which can be used to construct correlators. 
The $\rho$ map gives the relation between $\hat\eta$ and $g$, and is related to an automorphism. 

The map $ \rho : V_+ \rightarrow V_-$ is given by
\bea 
\rho (P_{\mu_1} P_{\mu_2} \cdots P_{\mu_n} v^+ ) = K_{\mu_1} \cdots K_{\mu_n}v^-  
\eea
This map obeys the conditions
\bea 
\rho ( \cL_a v_1 ) = -  \cL_a^{ \dagger }  \rho ( v_1 ) 
\eea
for any $v_1$. 
The dagger of the generators is given in (\ref{daggeraction}).
We also define 
\bea
 \rho ( \lambda v_1 ) = \lambda^* \rho ( v_1) 
\eea
for complex scalars $\lambda$. The inner product is now given by
\bea 
g ( v_1 , v_2 ) = \hat\eta ( \rho ( v_1 ) , v_2 ) 
\eea
This construction of a sesquilinear inner product from 
a bilinear pairing appears in the context of 3-dimensional TFT in \cite{Tur}.  
Consistency is guaranteed by checking that $g(v_1 , v_2)$ has the properties of the inner product
\bea 
&& g ( v_1 , v_2 ) = ( g ( v_2 , v_1 ))^* \cr 
&& g ( v_1 , \lambda v_2 ) = \lambda g ( v_1 , v_2 ) \cr 
&& g ( \lambda v_1 , v_2 ) = \lambda^* g ( v_1 , v_2 )
\eea
To see the last two of these, 
\bea 
&& g ( v_1 , \lambda v_2 ) = \hat\eta ( \rho ( v_1 ) , \lambda v_2 ) = \lambda \hat\eta ( \rho ( v_1 ) , v_2 ) = \lambda g ( v_1 , v_2 ) \cr 
&&  g ( \lambda v_1 , v_2 ) = \hat\eta ( \rho ( \lambda v_1 ) , v_2 ) = \hat\eta ( \lambda^* \rho ( v_1 ) , v_2 ) = \lambda^* g ( v_1 , v_2 ) 
\eea
We have used the bilinearity of $ \hat\eta $ and the definition $ \rho ( \lambda v ) = \lambda^* \rho ( v ) $. 
To see the symmetry consider 
\bea 
&& g ( \lambda_1 P_{\mu_1} \cdots P_{\mu_n } v^+ , \lambda_2 P_{\nu_1} \cdots P_{\nu_n } v^+ ) \cr 
&&  = \lambda_1^* \lambda_2 \hat\eta ( K_{\mu_1} \cdots K_{\mu_n  } v^- , P_{\nu_1} \cdots P_{\nu_n } v^+ ) \cr 
&& = \lambda_1^* \lambda_2 T_{\vec \mu , \vec \nu } 
\eea
and 
\bea 
&& g (  \lambda_2 P_{\nu_1} \cdots P_{\nu_n } v^+ , \lambda_1 P_{\mu_1} \cdots P_{\mu_n } v^+ ) \cr 
&& = \lambda_2^* \lambda_1 \hat\eta ( K_{\nu_1} \cdots K_{\nu_n } v^- , P_{\mu_1} \cdots P_{\mu_n } v^+ \cr 
&& = \lambda_2^* \lambda_1 T_{\vec \nu  , \vec \mu } 
\eea
The explicit formulae (\ref{tensorev}) show that $ T_{\vec \mu , \vec \nu } = T_{\vec \nu , \vec \mu } $
and $ T_{\vec \mu , \vec \nu }^* = T_{\vec \mu , \vec \nu } $. 
This proves $ g ( v_1 , v_2 ) =  ( g ( v_2 , v_1 ) )^* $.  

It is also useful to note that 
\bea 
 g ( \cL_a v_1 , v_2 )  = \eta ( \rho ( \cL_a v_1 ) , v_2 ) = - \eta ( \cL_a^{ \dagger} \rho ( v_1 ) , v_2 ) 
 = \eta ( \rho ( v_1 ) , \cL_a^{ \dagger} v_2 ) = g ( v_1 , \cL_a^{ \dagger}  v_2 )  
\eea 
This explains why this inner product has the usual hermiticity property. 
Finally we observe that the map $ \alpha(\cL_ a)\equiv -\cL_a^{ \dagger }$ used above is
an automorphism of the Lie algebra 
\bea 
[ \alpha ( \cL_a ) , \alpha ( \cL_b ) ] =  \alpha (  [ \cL_a , \cL_b ] ) 
\eea

\section{ State space, amplitudes, and  correlators in the TFT$_2$   }\label{sec:corrs} 

In section \ref{sec:basic} we have made use of two representations $V_+$ and $V_-$.
This has allowed us to describe the two point function of the basic field of free scalar field theory 
in four dimensions in terms of the invariant irrep $ \mC$ of $SO(4,2)$ appearing in the tensor 
product $ V_+ \otimes V_- $. 
In this section we will consider a larger state space $\cW$, which will allow us to extend our discussion to arbitrary
correlation functions in CFT$_4$.

The state space $\cW$ is a vector space 
\bea\label{statespace} 
\cW = \cW_{0} \oplus V \oplus {\rm Sym}(V^{\otimes 2})\oplus {\rm Sym}(V^{\otimes 3})\oplus\cdots 
\eea
where 
\bea 
&& \cW_{0} = \mC \cr 
&& V  =   V_+  \oplus V_-  
\eea
$V_+$ is again the lowest weight representation obtained by acting with some $\cL_ a^+$ in the enveloping algebra of $so(4,2)$
on the lowest weight state $v^+$.
The plus in  $\cL_ a^+$ indicates that this is the subalgebra generated by the $P_{\mu}$'s. 
$V_-$ is a lowest weight representation obtained by acting with $ \cL_a^-  $ (products of $K_{\mu}$'s) on the highest weight state $v^-$. 
Our field is represented by the following sum of states in $V_+ \oplus V_-$ 
\bea\label{Foundational} 
   \Phi (x) && = {1\over\sqrt{2}}\left( e^{iP\cdot x}v^+ + (x')^2 \rho (e^{iP\cdot x'}v^+) \right) 
 = { 1 \over \sqrt { 2 } } ( e^{ i P \cdot x } v^+ + (x')^2 e^{ - i K \cdot x' } v^- )  \cr 
               && \equiv \Phi^+ ( x ) + \Phi^- ( x ) 
\eea
where a primed coordinate is always related to the unprimed coordinate by inversion
\bea
   x^\mu = {x^{\prime\,\mu}\over x^{\prime\, 2}}
\eea
This is a key equation translating between fields in CFT$_4$ and the states in TFT$_2$.
The explicit $(x')^2$ multiplying the second term is needed to ensure that both terms have the same scaling dimension.
If $v^+$ has dimension $d$ we would have
\bea
   \Phi_d (x) = {1\over\sqrt{2}}\left( e^{iP\cdot x}v^+ + (x')^{2 d} \rho (e^{iP\cdot x'}v^+) \right)
\eea

We will demonstrate that the TFT$_2$ correctly computes arbitrary CFT$_4$ correlation functions.
Our demonstration will build up from the basic two point function in a series of steps. 
Let us denote by $ < \varphi ( x_1) \varphi  ( x_2)  ... \varphi ( x_{2k}  ) >_{CFT_4} $ the usual 4D CFT free field correlator. 
We are using the hatted notation for the usual quantum field to distinguish from the sum of states in $V_+ \oplus V_-$ which we use as  
the foundational equation in our TFT$_2$ approach (\ref{Foundational}).
We want to show that 
\bea 
 < \varphi ( x_1) \varphi  ( x_2)  ... \varphi ( x_{2k}  ) >_{CFT_4} = <  \Phi ( x_1 ) \otimes \cdots \otimes  \Phi ( x_{2k}  )  >_{TFT_2} 
\eea

We have already described the two point function as the  $SO(4,2)$ invariant.
To extend this idea to arbitrary correlation functions, we need a bilinear $so(4,2)$-invariant map
\bea 
\eta : W \otimes W \rightarrow \mC 
\eea
The basic building block for this map will be 
\bea 
   \hat\eta_{+-}  : V_+ \otimes V_- \rightarrow \mC 
\eea
which is the map $\hat\eta$, described in section \ref{sec:basic}.
It is used to define 
\bea 
   \hat\eta_{- + } : V_- \otimes V_+ \rightarrow \mC 
\eea
by symmetry as 
\bea 
   \hat\eta_{ -+ } ( v_1^- , v_2^+ ) = \hat \eta_{ + - } ( v_2^+ , v_1^- ) 
\eea
We may write 
\bea 
   \hat\eta_{ - + } = \hat\eta_{ + - } \circ \sigma 
\eea
where $\sigma : V_+ \otimes V_- \rightarrow V_- \otimes V_+ $ is the twist map. 
On $V$ we can define
\bea 
   \hat\eta : V \otimes V \rightarrow \mC 
\eea
by the direct sum 
\bea 
   \hat\eta = \hat\eta_{ + - } \oplus \hat\eta_{ - +} 
   \label{twopointeta}
\eea
It is useful to think of this $\hat\eta $ as a block off-diagonal matrix. 
% The twist symmetry can be written as follows
% 
% \bea 
 %   \hat\eta_{ +- ; a^+ b^- } = \hat\eta_{ - + ; b^- a^+ } = \hat\eta_{ a^+ b^- } = \hat\eta_{ b^- a^+ } 
% \eea
%
% and it implies the symmetry of $\hat\eta$.
In section \ref{sec:basic} we have given a formula for $\hat\eta_{a^+ b^- }$, in the basis where $a^+=l,p_L,p_R$ and $b^- =l',q_L,q_R$, as
\bea 
   \hat\eta_{l,p_L,p_R ;l',q_L,q_R} = f_{l}^{SL(2)} f_{p_L}^{SU(2)} f_{p_R}^{SU(2)} \delta_{l,l'}\delta_{p_L p_R}\delta_{q_L q_R} 
\eea

Next, suppose we have a 4-point function. The TFT$_2$ computes this correlator by mapping the tensor product
\bea 
\Phi  ( x_1) \otimes \Phi (x_2 )  \otimes \Phi (x_3 )  \otimes \Phi ( x_4 ) 
\eea
to a number.
This is accomplished with the map 
\bea
   \eta : V\otimes V\otimes V\otimes V \rightarrow \mC 
\eea
defined by $ \eta = \hat\eta^{12}\hat\eta^{34} + \hat\eta^{13}\hat\eta^{24} + \hat\eta^{14}\hat\eta^{23}$.
We are using the products of the basic map (\ref{twopointeta}) on $V\otimes V$ to produce a map on  $V\otimes V \otimes V\otimes V$.
The details of how we do this are fixed so that we reproduce the combinatorics of Wick's theorem.
This action gives
\bea 
&& <  \Phi ( x_1 ) \otimes \cdots \otimes  \Phi ( x_4 )  >_{TFT_2} \cr 
&& = ( \hat\eta^{ 12} \hat\eta^{34} + \hat\eta^{13} \hat\eta^{ 24} + \hat\eta^{ 14} \hat\eta^{23} )
     (  \Phi  ( x_1) \otimes \Phi (x_2 )  \otimes \Phi (x_3 )  \otimes \Phi ( x_4 )  )  \cr 
&& = { 1 \over 2 } 2 G ( x_1 , x_2 ) { 1\over 2  }  2 G ( x_3 , x_4 )  
 +  { 1 \over 2 } 2 G ( x_1 , x_2 ) { 1\over 2  } 2 G ( x_3 , x_4 ) + { 1 \over 2 } 2 G ( x_1 , x_2 ) { 1\over 2  } 2 G ( x_3 , x_4 )
\cr  && ~ 
\eea 
The $ { 1 \over 2 } $ for each $ \hat\eta^{ij} $ comes from multiplying the two factors of ${1 \over \sqrt { 2 }}$ appearing in (\ref{Foundational}). 
The factor of $2$ comes from adding the $\hat\eta(\Phi^+(x_1),\Phi^{-}(x_2))=G(x_1,x_2)$ 
and $\hat\eta(\Phi^-(x_1),\Phi^{+}(x_2))=G(x_1,x_2)$ - where  $\Phi^+(x)$ and $\Phi^-(x)$ live in $V_+ $ and $V_-$. 
This is in perfect agreement with the usual free field computation 
\bea 
&& < \varphi ( x_1 ) \varphi ( x_2) \varphi ( x_3  ) \varphi ( x_4 ) >_{CFT_4}  \cr 
&& = G ( x_1 , x_2 ) G ( x_3 , x_4 ) + G ( x_1 , x_3 ) G ( x_2 , x_4 ) 
+ G ( x_1 , x_4 ) G ( x_2 , x_3 ) 
\eea

In general, to construct the map $\eta:V^{\otimes\, 2k }\to \mC$ we sum over ${(2k )!\over 2^k  k!}$ pairings which can be parametrized by choosing 
$ i_1 < i_2 \cdots < i_k  $ and $ i_1 < j_1 , i_2 < j_2 , \cdots , i_k < j_k $. 
The pairings are 
\bea 
( i_1 , j_1 ) \cdots  ( i_k , j_k )  
\eea
These are determined  by permutations $ \sigma $  with $n$ cycles of length $2$, which form a  conjugacy class 
denoted as $  [ 2^k  ]$ of $S_{2 k } $, so we may 
write $ \{  i^{   \sigma }_{ l  }  ,  j^{  \sigma }_l   \}  $, which is a set  of $ k  $ pairs  $ ( i , j )$  
uniquely determined by $ \sigma $.  
Here we start with the state 
\bea\label{tensprodPHI} 
\Phi ( x_1 ) \otimes \cdots \otimes \Phi ( x_{2 k } )   \in V^{ \otimes 2 k  } 
\eea
and then act with the $SO(4,2)$ invariant map
\bea \label{thebasicmap}
\sum_{ \sigma \in [ 2^k  ]} 
\left ( \prod_{ l =1}^k \hat\eta^{ i_l^{ \sigma }  j_l^{ \sigma} }     \right)
\eea
Thus 
\bea 
&& < \Phi ( x_1 ) \cdots \Phi ( x_{2 k  } ) >_{ TFT_2}  =
\sum_{ \sigma \in [ 2^k ]} 
\left ( \prod_{ l =1}^k  \hat\eta^{ i_l^{ \sigma }  j_l^{ \sigma} }     \right) \Phi ( x_1 ) \otimes \cdots \otimes \Phi ( x_{2 k } )  \cr 
&& = \sum_{ \sigma \in [2^k ] } \prod_{ l=1}^k { 1 \over 2 } 2 G ( x_{ i_l^{ \sigma}  } , x_{j_l^{ \sigma } }   )  \cr 
&& =  \sum_{ \sigma \in [2^k ] } \prod_{ l =1}^k  G ( x_{ i_l^{ \sigma}  } , x_{j_l^{ \sigma } }   ) 
\eea
The $ { 1 \over 2 } $ comes from the normalization factors and the two from the $ \hat\eta ( \Phi^+ , \Phi^- ) $ and 
$ \hat\eta ( \Phi^- , \Phi^+ )$. This is the correct CFT$_4$ correlator. 

Consider next correlators of descendents of $\varphi (x)$. 
Our TFT$_2$ proposal is to insert the derivatives of the basic $ \Phi $ from (\ref{Foundational})
\bea
   {\partial\over\partial x^\mu}\Phi (x) &=&
   {1\over\sqrt{2}}\left( iP_\mu e^{iP\cdot x}v^+ 
                      + 2 x'_\alpha {\partial x^{\prime \alpha} \over\partial x^\mu} \rho (e^{iP\cdot x'}v^+)
                     -i {\partial x^{\prime \alpha} \over\partial x^\mu} (x')^2 \rho (P_\alpha e^{iP\cdot x'}v^+)\right)\cr
 &=& {1\over\sqrt{2}}\left( iP_\mu e^{iP\cdot x}v^+ 
                      + 2 x'_\alpha I_\mu^\alpha (x')^2 \rho (e^{iP\cdot x'}v^+)
                     -i (x')^4 I^\alpha_\mu \rho (P_\alpha e^{iP\cdot x'}v^+)\right)
\eea
where $I_\mu^\rho$ appearing in
\bea
  {\partial x^{\prime\rho}\over\partial x^\mu}={1\over x^2}\left[\delta^\rho_\mu -{2x^\rho x_\mu\over x^2}\right]
                                              ={I_\mu^\rho\over x^2}=I^\rho_\mu (x')^2
\eea
is the local Lorentz transformation for an inversion.

In CFT$_4$, correlators of descendents follow by taking the appropriate derivatives of the Green's functions.
The TFT$_2$ applies  the map (\ref{thebasicmap}) after the derivatives of $\Phi (x)$ have been taken.   
The equality of the TFT$_2$ and CFT$_4$ computations follows because taking derivatives of (\ref{tensprodPHI}) and then applying 
(\ref{thebasicmap}) is the same as computing  the pairings   and then doing the derivatives. 
These commute because things like  $  x_1^{\mu_1} \cdots x_1^{ \mu_s} P_{ \mu_1} \cdots P_{ \mu_s } v^+ $ 
as live in $ V_+ \otimes \mC [ x_1  ] $, the tensor product of the $SO(4,2)$ irrep with the function space $\mC[x_1]$. 
The different orders of taking derivatives commute because $\hat\eta$ acts on the $V_{+}\otimes V_-$ whereas the derivatives 
act on the $ \mC [ x_1 ] \otimes \mC [ x_2 ]$ and operators acting on different tensor factors commute.  

As an example consider 
\bea
&& \hat\eta\left( \partial_\mu\Phi (y),\partial_\nu\Phi (x)\right)=\cr
&&{1\over 2}\hat\eta\left( iP_\mu e^{iP\cdot y}v^+, 2x_\rho'I_\nu^\rho (x')^2\rho (e^{iP\cdot x'}v^+)\right)
+{1\over 2}\hat\eta\left( iP_\mu e^{iP\cdot y}v^+, (x')^4 I_\nu^\rho \rho (iP_\rho e^{iP\cdot x'}v^+)\right)\cr
&&+{1\over 2}\hat\eta\left( 2y_\rho'I_\mu^\rho (y')^2\rho (e^{iP\cdot y'}v^+),iP_\nu e^{iP\cdot x}v^+\right)
+{1\over 2}\hat\eta\left((y')^4 I_\mu^\rho \rho (iP_\rho e^{iP\cdot y'}v^+),iP_\nu e^{iP\cdot x}v^+\right)\cr 
&& ~~ 
\label{terms2compute}
\eea
Lets discuss the evaluation of the first term on the RHS.
Use the $so(4,2)$ invariance of $\hat\eta$ to move the $P_\mu$ from the first slot to the second slot.
After evaluating the $\rho$ map in the second slot, we have to evaluate $P_\mu$ acting on $e^{-iK\cdot x'}v^-$.
This can be done using the $so(4,2)$ algebra, as explained in section \ref{invpair}.
Evaluating all four terms in this way, it is now simple to find
\bea
\hat\eta\left( \partial_\mu\Phi (y),\partial_\nu\Phi (x)\right)&&=
   {1\over 2}{\partial\over\partial y^\mu}{\partial\over\partial x^{\nu}}
{(x')^2\over 1-2y\cdot x'+y^2 (x')^2}
+   {1\over 2}{\partial\over\partial y^\mu}{\partial\over\partial x^{\nu}}
{(y')^2\over 1-2y'\cdot x+x^2 (y')^2}\cr
&&={\partial\over\partial x^\nu}{\partial\over\partial y^\mu} {1\over |x-y|^2}
\eea
which confirms the argument given earlier that taking  derivatives with respect to  the spacetime coordinates 
commutes with the evaluation of $ \hat  \eta $, so that the CFT$_4$ correlator for scalars with derivatives 
is reproduced by  the TFT$_2$ construction. 

To complete the discussion, consider correlators involving powers of the elementary field. 
\bea 
< \varphi^{n_1}  ( x_1 ) \cdots \varphi^{ n_k }  ( x_k ) >_{CFT_4} 
\eea
The starting point to get this on the TFT$_2$ side is 
\bea\label{startcomposite} 
\Phi^{ n_1} ( x_1 )  \otimes \Phi^{ n_2} (x_2)  \cdots \otimes  \Phi^{ n_k}  ( x_k ) \in Sym ( V^{ \otimes n_1} ) \otimes \cdots Sym ( V^{ \otimes n_k } ) 
\eea
We need $ n_1 + n_2 + \cdots + n_k \equiv 2 M  $ to be even for a non-zero correlator. 
To this we apply a sum of products of $ \hat\eta $'s, schematically written as
\bea\label{schemsumprodeta}  
\sum \prod \hat\eta^{ ij } 
\eea
The sum is over all pairings of $ 2M$ objects, avoiding cases where $ ( i , j ) $ belong to 
the same subset of $ n_1 $ or $n_2$ etc.  integers.
To write a more explicit version of (\ref{schemsumprodeta})  note that 
the pairings we are summing over correspond  to permutations in  the conjugacy class $ [2^M]$ 
in the symmetric group $ S_{ 2M } $, but are  not in the subgroup $ S_{ n_1 } \times S_{ n_2 } \times \cdots 
\times S_{n_k  } \equiv S_{ \vec n } $. Hence the invariant map is 
\bea\label{exctsumprodeta}  
\sum_{ \substack { \sigma \in [2^M] \in S_{ 2M } \\ 
\sigma \notin S_{ \vec n } } }  \prod_{  l =1}^M  \hat\eta^{i^\sigma_l  j^\sigma_l } 
\eea

 For each $ \hat\eta $ factor there is a $ { 1 \over 2 } \times 2 G ( x_i , x_j ) $ 
as before. In this way we again see that the TFT$_2$ correctly computes the CFT$_4$ correlator 
\bea 
&& < \varphi^{n_1}  ( x_1 ) \cdots \varphi^{ n_k }  ( x_k  ) >_{CFT_4}  = \sum \prod  G ( x_i , x_j ) \cr 
&&  = \sum \prod \hat\eta^{ ij } \Phi^{ n_1} ( x_1 )  \otimes \Phi^{ n_2} (x_2)  \cdots \otimes  \Phi^{ n_k }  ( x_k )\cr 
&& = < \Phi^{ n_1} ( x_1 )  \cdots \Phi^{ n_k } ( x_k  )   >_{ TFT_2} 
\eea
One can also consider applying derivatives in each of the $ x_1 , \cdots , x_k  $. 
Again the TFT$_2$ proposal is to apply the derivatives to the state (\ref{startcomposite}) which gives, schematically 
\bea 
\sum \prod \hat\eta^{ ij }  ( \partial_{ x_i}   \Phi ( x_i ) ,  \partial_{ x_j } \Phi ( x_j  ) )  
\eea
which is equal to 
\bea 
\sum \prod \partial_{ x_i} \partial_{ x_j } \hat\eta^{ ij }  (   \Phi ( x_i ) ,  \Phi ( x_j  ) ) 
\eea
by using the argument given above.

\section{Non-degeneracy  of the Invariant   Pairing }\label{sec:nondegen} 

For $ V = V_+ \oplus V_-$,  we  have described an invariant  pairing $\hat\eta : V \otimes V \rightarrow \mC $. 
\bea 
   \hat\eta ( v_1 , v_2 ) = 0 
\eea
if $ v_1 , v_2 $ are both in $ V_+ $ or both in $V_-$.
We can explicitly describe the pairing using $ SU(2) \times SU(2)$ and $ SL(2 )$ sub-algebras.

Using the states 
\bea 
&& v^+_{ l , p_L , p_R } = (J_+^R )^{p_{R}} (J_+^L)^{p_{L}} (H_+)^l v^+ \cr 
&& v^-_{ l' , p_L' , p_R'} = (J_-^R )^{p_{R}'}(J_-^L )^{p_{L}'}(H_-)^{l'} v^-
\eea
we have 
\bea 
&& \hat\eta ( v^+_{ l , p_L , p_R } , v^-_{ l' , p_L' , p_R'} ) = \hat\eta (  v^-_{ l' , p_L' , p_R'} ,  v^+_{ l , p_L , p_R } )   = 
f_{l,p_L }^{su(2)}f_{l,p_R }^{su(2)} f_{d,l }^{sl(2)}  \delta_{p_{L},p_{ L}'}\delta_{p_{R},p_{R}'}\delta_{l,l'} 
\eea
The $f$ factors are given earlier in Section \ref{fsection}. 

There is an inverse $\tilde{\hat{\eta}}$ which obeys 
\bea 
&& \hat\eta ( v_a , v_b )   \equiv \hat\eta_{ ab } \cr 
&& \tilde{\hat{\eta}}( v_b , v_c )  \equiv  \tilde{\hat{\eta}}^{bc} \cr  
&& \hat\eta_{ab} \tilde{\hat{\eta}}^{ bc } = \delta_a^c 
\eea 
In the $ ( SL(2), SU(2) \times SU(2))$ basis, we can describe $\tilde{\hat{\eta}}$ explicitly. 
\bea 
&& \tilde{\hat{\eta}}( v^+_{ l , p_L ,  p_R  } , v^+_{ l' , p'_L   ,  p'_R  }  )  =   0 \cr 
&& \tilde{\hat{\eta}}( v^-_{ l , p_L   ,  p_R } , v^-_{ l' , p'_L   ,  p'_R  }  )  =   0 \cr 
&& \tilde{\hat{\eta}}( v^+_{ l , p_L   ,  p_R }      ,    v^-_{ l' , p'_L   ,  p'_R}  )        )   =
   \tilde{\hat{\eta}}( v^-_{ l , p_L   ,  p_R} , v^+_{ l' , p'_L   ,  p'_R  }  )  =         (  f_{l,p_L }^{su(2)} )^{-1} ( f_{l,p_R }^{su(2)} )^{-1} 
  (  f_{d,l }^{sl(2)}  )^{-1}  \delta_{p_{L},p_{ L}'}\delta_{p_{R},p_{R}'}\delta_{l,l'}                                     \cr 
&& 
\eea

For the space $ \cW$  given in (\ref{statespace})
we define $\eta$ 
\bea 
  \eta  ( v_1 , v_2 ) = 0 
\eea
if $v_1,v_2$ belong to $\Sym(V^{\otimes n_1})$ and $ \Sym ( V^{ \otimes n_2} ) $ 
for $n_1 \ne n_2$.  Define 
\bea 
e_{a_1} \circ \cdots \circ e_{a_n} = { 1 \over n! } \sum_{ \tau \in S_n } e_{ a_{ \tau ( 1) } } \otimes \cdots \otimes  e_{ a_{ \tau ( n ) }  } 
= P_{sym} e_{ a_1} \otimes \cdots \otimes \circ e_{a_n } 
\eea
The projector $P_{sym} = { 1 \over n! } \sum_{ \sigma \in S_n } \sigma $ is the projector for 
the symmetric part.  

For fixed $n$, define 
\bea 
\eta_{a_1 , \ldots , a_n ; b_1 , \cdots , b_n  } \equiv \eta ( e_{a_1} \circ  e_{a_2 }   \cdots \circ e_{ a_n } ,  e_{b_1} \circ \cdots \circ  e_{b_n} ) 
\eea 
by  the equation 
\bea 
\eta_{a_1 , \ldots , a_n ; b_1 , \ldots , b_n  } 
= \sum_{\sigma \in S_n} \prod_{ i=1}^n \hat\eta_{ a_i , b_{ \sigma ( i) } } 
\eea

There is an inverse, $ \tilde \eta $, which obeys 
\bea 
\eta_{AB} \tilde \eta^{ B C } = \delta_A^C 
\eea
where these capital indices run over the states in $\cW$. 
Choose a basis running over the degree $n$ of the symmetric tensor product $ Sym ( V^{ \otimes n } ) $, 
and for each $n$, we run over states $v_{ a_1} \circ \cdots v_{a_n} $. 
We define 
\bea 
\tilde \eta ( v_1 , v_2  )  =  0  
\eea
for $ v_1 \in Sym ( V^{ \otimes n_1} ) \subset \cW , v_2 \in Sym (  V^{ \otimes n_2}  )  \subset \cW $ with $ n_1 \ne n_2$.  
And for $v_1 , v_2$ in the same symmetric power, 
\bea 
\tilde \eta ( v_{b_1}  \circ \cdots \circ  v_{ b_n } , 
 v_{c_1} \circ \cdots  \circ v_{c_n} ) \equiv  \tilde \eta^{ b_1    \cdots  b_n   , c_1   \cdots    c_n } 
  = { 1 \over n! } \sum_{\sigma  } \tilde{\hat{\eta}}^{ b_1 ; c_{ \sigma (1)}} \cdots \tilde{\hat{\eta}}^{ b_n , c_{ \sigma (n)} }
\eea
With these definitions both $\eta,\tilde{\hat{\eta}}$ are block-diagonal, with blocks labeled 
by the degree of the symmetric tensors. 

The non-trivial part of the check of the inverse property, in each block, involves showing 
\bea 
\eta_{a_1,\cdots ,a_n;b_1\cdots b_n}\tilde{{\eta}}^{ b_1 , \cdots b_n  ; c_1 , \cdots , c_n } 
= { 1 \over n! } \sum_{ \sigma \in S_n } \delta_{ a_1 }^{  c_{ \sigma ( 1) }  } \cdots \delta_{ a_n }^{ c_{ \sigma (n) } }
\eea
This expresses the fact the LHS is non-zero only when the symmetric part of
$ e_{ a_1} \otimes e_{ a_2} \cdots  \otimes e_{ a_n } $ is identical to the symmetric part of 
$ e_{ c_1} \otimes \cdots  \otimes e_{ c_n} $. The proof is a simple calculation. 
\bea 
&& \eta_{ a_1 , \cdots , a_n  ;  b_1 \cdots b_n  }  \tilde \eta^{ b_1 , \cdots b_n  ; c_1 , \cdots , c_n }  
= { 1 \over n! } \sum_{ \alpha  \in S_n } \prod_{ i =1}^{ n } \hat\eta_{ a_{\alpha ( i ) }  ,  b_i }     ~~ 
{ 1 \over n! } \sum_{  \beta \in S_n } 
\prod_{ i =1}^{ n } \tilde{\hat{\eta}}^{ b_i , c_{ \beta  ( i ) } }   \cr 
&&  = { 1 \over (n!)^2 } \sum_{ \alpha , \beta } \prod_{ i =1}^{ n } \hat\eta_{ a_{\alpha ( i ) }  ,  b_i } 
\tilde{\hat{\eta}}^{ b_i , c_{ \beta  ( i ) } } 
= { 1 \over (n!)^2 } \sum_{ \alpha , \beta } \prod_{ i =1}^{ n } \delta_{ a_{ \alpha (i)} }^{ c_{ \beta (i) } } \cr  
&& = { 1 \over (n!)^2 } \sum_{ \alpha , \beta  } \prod_{ i =1}^{ n } 
\delta_{ a_{ \beta^{-1} \alpha (i)} }^{ c_{ i  } } = { 1 \over n! } \sum_{ \alpha } \prod_{ i =1}^{ n } \delta_{ a_{ \alpha (i)} }^{ c_i } 
\eea
In the last step we use the invariance of the sum over $S_n$, under group multiplication. 
The final answer on the right is the matrix element of the  identity operator on symmetric tensors. 

There is a physical way to understand the non-degeneracy we have just described.
Let us define the vector space $V_{T}^+ $ spanned by states of the form 
\bea 
P_{\mu_1 } \cdots P_{\mu_s }  v^+  
\eea
This corresponds to local operators $ \partial_{ \mu_1 } \cdots \partial_{ \mu_s } \varphi $ in CFT.  
Similarly we have the dual representation $V_T^-$  spanned by 
\bea 
K_{\mu_1 } \cdots K_{\mu_s }  v^- 
\eea
The map $\hat \eta$ can be  defined to act on  vector space $V_{T}^+  \otimes V_T^-  $. 
Using the $SO(4,2)$ algebra as well as the $SO(4,2)$ invariance of $\eta$, it is straight forward to demonstrate that
\bea
   \eta (P_\mu P_\mu v^+ ,K_\alpha K_\beta v^- )=0
\eea
This shows that $ P_\mu P_\mu v^+$ is a degenerate state for the invariant pairing. 
It is also a degenerate state for the inner product $ g $ defined in terms of $\eta$ in Section \ref{sec:etarogee}.   
This is a purely representation theoretic fact related to the possibility of imposing the 
equation of motion $\partial_{\mu}\partial_{\mu}\phi = 0$ in a way consistent with $SO(4,2)$. 
The $P_\mu P_\mu v^+$ state, together with all of its descendants, are null. 
We can quotient  $V_{T}^+ $ by the null states to get $V_+$: the space $V_+$ is the vector space made from states 
obtained by acting with symmetric traceless products of $ P$'s, i.e products of the form 
$ C^I_{ \mu_1 \cdots \mu_p } P_{\mu_1} \cdots P_{\mu_p}$. This corresponds to 
the  derivatives of the elementary scalar $ \varphi$, with the equation of motion imposed.
The 2-point function $ F ( x_1 , x_2' )$ derived from the path integral 
 obeys the property $ \partial_{ x_1^{ \mu} } \partial_{ x_1^{ \mu} }  F =0$, 
related to the fact that equations of motion are satisfied by the quantum field, inside correlators, 
as operator equations. From the representation theory point of view, when we work with  
 $V_+$ the equation of motion has already been imposed. 
We then take $ V = V_+ \oplus V_-$ to get a self-dual object which has a non-degenerate
$SO(4,2)$  pairing. In this section we considered  
$\cW = \bigoplus_{n=0}^{\infty}Sym (V^{\otimes n})$
and demonstrated that the  non-degenerate bilinear pairing on $V$ extends to this space. 
From the physical point of view, this is not surprising. 
We know that once we have accounted for the equations of  motion on the basic field $\varphi$, the construction of composite fields has no further source of null states, than the ones that come from the equations of motion on each  elementary field. It is nevertheless useful to exhibit this directly as a fact in representation theory, since 
  the non-degeneracy is a crucial ingredient in the $SO(4,2)$-invariant TFT$_2$.

\section{ Three-point functions and OPE  }\label{sec:3ptOPE} 

In Section \ref{sec:corrs} we have described a state space  $ \cW $ (with discrete basis $e_A$) and shown how to construct a multi-linear map from 
the tensor product $ \cW^{ \otimes k } $
to the complex numbers $ \mC$ using tensor products of the basic
invariant map $ \hat \eta$ according to Wick combinatorics. Insertions 
of  states in $ \cW $ corresponding to composite  scalar field operators, labeled by 
positions $ x_1 , x_2 , \cdots , x_k $, gives the $k$-point functions of free 4D scalar field  theory. 
In particular  there are  $ C_{ABC}$ which give the $3$-point function. 
In TFT$_2$ we can form $ C_{AB}^{~~~C}   = C_{ABD} \tilde \eta^{ DC } $ 
which gives a product structure to $ \cW $, i.e. a map  $ \cW \otimes \cW \rightarrow \cW$. 

Given any state $ e_{ A } \otimes e_B $ in $ \cW \otimes \cW $ we have 
a state $ C_{AB}^{ ~~~D}  e_D $ in $ \cW $. This state has the property that the 
$3$-point function of $ e_A \otimes e_B $ with any $e_C $ is just the 
two-point function  $ \eta ( C_{AB}^{ ~~~ D}  e_D , e_C )$. Indeed 
\bea 
\eta ( C_{AB}^{ ~~~D}  e_D , e_C ) = C_{AB}^{ ~~~D }  \eta ( e_D , e_C ) = C_{ABE } \tilde \eta^{ E D} \eta_{ D C } = C_{ABE} \delta^E_C = C_{ABC} 
\eea

This is the TFT2 expression of a familiar construction in CFT, the operator product expansion (OPE). 
For any two local operators at spacetime positions $x_1$ and $x_2$, the OPE expresses their product as a sum of local operators at $x$ where $x$ 
may be $x_1$, $x_2$ or even the midpoint ${x_1+x_2\over 2}$, according the convention adopted.
In conformal field theories this expansion is a convergent expansion \cite{Pappadopulo:2012jk}
and it provides a powerful approach for understanding the correlation functions in the theory. 
The three point function for operators located at positions  $x_1$, $x_2$ and $x_3$ is reduced to the computation 
of a sum of two point functions, after the OPE is used to take the product of two of the operators in the correlator.

To make this link between product in TFT$_2$ and operator product in CFT$_4$ explicit, recall 
that  the basic field $ \varphi ( x ) $  corresponds to the following state
\bea 
\Phi ( x ) =  \Phi^+ ( x )  + \Phi^- ( x ) = e^{ i x \cdot P } v^+ +  ( x')^2  e^{ -i x'\cdot K } v^- 
\eea
in the state space $ \cW$ of the TFT$_2$. 
The composite field $\varphi^2(x)$ of QFT corresponds to the tensor product $\Phi(x)\otimes\Phi(x)\in Sym(V^{\otimes 2})$. 
More generally any local operator corresponds to an expansion of the form 
\bea 
\cO_1 ( x ) = \cO_1^A (x) e_A  
\eea
for $e_A $ running over a discrete basis  in $ \cW$. 
Given the way the  $ C_{ABC } $ are constructed from tensor products of $ \hat \eta $ (equations (\ref{thebasicmap}), (\ref{exctsumprodeta}))
we see that we may write the operator product of $\cO_1 (x_1)$ and $\cO_2 (x_2)$ in the TFT$_2$ language 
as a sum of the form $(1+\hat \eta + \hat \eta \otimes \hat\eta + \cdots) $ acting on 
\bea 
  \cO_1^{A} ( x_1 )  \cO_2^{B } (x_2) e_{A} \otimes e_{ B } 
\eea
We are organizing the product by the number of Wick contractions between the first two operators.
The first term in the sum corresponds to the $3$-point function $\langle\cO_1 (x_1)\cO_2(x_2)\cO_3(x_3)\rangle$ 
for an $\cO_3(x_3)$ that receives only contribution from terms having no Wick contractions between the first two operators. 
This is followed  by terms with one or more Wick contractions.  Hence the first term is the ordinary tensor 
product, while the subsequent terms involve the application of the invariant map $ \hat \eta $ which acts 
on $V_{ { \pm } }  $ and $ V_{ \mp } $ factors from $ \cO_1 $ and $ \cO_2$. The symmetric role 
played by positive and negative energy representations in the TFT$_2$ construction of CFT$_4$ naturally leads to 
the correct form of the OPE. 

Let us make the discussion even more concrete by taking $ \cO_1 $ and $ \cO_2$ to correspond 
to the quadratic composite field $ \varphi^2 $. Here we will have 
\bea 
   &&   (1+\hat\eta^{13 } + \hat \eta^{ 14 } + \hat \eta^{ 23} + \hat \eta^{ 24}   + \hat\eta^{ 13 } \otimes \hat \eta^{ 24} +   \hat\eta^{ 13 } \otimes \hat \eta^{24}) \bigl( ( \Phi (x_1) \otimes \Phi ( x_1) ) \otimes  ( \Phi (x_2)  \otimes \Phi (x_2 ) ) \bigr )  \cr 
&& = 
    (  \Phi  (x_1) \otimes \Phi ( x_1 ) ) \otimes  ( \Phi ( x_2 ) \otimes \Phi ( x_2 ) ) +{4  \over (x_1-x_2)^2}\Phi(x_1)\otimes\Phi (x_2)+{2 \over (x_1-x_2)^4}  \nonumber
\eea 
In this expression we have not chosen whether to expand the RHS around $x_1$ or $x_2 $ or the  mid-point. 
That is a subsequent choice that can be made and the above state in $ \cW $ expanded accordingly.

\section{ CFT crossing and associativity }\label{sec:crossass} 

The OPE is associative. Since the OPE can be used to construct correlation functions, associativity of the OPE 
implies relations among the CFT correlators, namely crossing symmetry.
It is known that crossing symmetry implies strict constraints on operator dimensions and OPE coefficients 
(or equivalently, on the $C_{AB}^{ ~~ C }$ of TFT$_2$). In this section we give a combinatoric proof 
of crossing in the TFT$_2$ framework. The reader will expect this to work since  the TFT$_2$ framework 
has already been shown to reproduce correlators of the free CFT$_4$.  
It is however useful to give an explicit proof without appealing to the path integral of the free CFT$_4$. 
It allows us to see that it continues to work for generalizations where $SO(4,2)$ is replaced 
by any $G$ and $V$ by any space with a unique invariant in $ V \otimes V$.

The correlator is constructed by applying contraction maps to 
$ Sym ( V^{ \otimes n_1 } ) \otimes  Sym ( V^{ \otimes n_2 } ) \otimes  Sym ( V^{ \otimes n_3 } ) \otimes  Sym ( V^{ \otimes n_4 } ) $. 
Define the sets 
\bea 
&& S_1  = \{ 1, 2, \cdots , n_1 \} \cr 
&& S_2 = \{ n_1 +1 , \cdots , n_1  + n_2 \} \cr 
&& S_3 = \{ n_1 + n_2 +1 , \cdots , n_1 + n_2  + n_3 \} \cr 
&& S_4 = \{ n_1 + n_2 + n_3 +1 , \cdots , n_1 + n_2 + n_3 + n_4 \} 
\eea 
To compute the correlator, use the contraction maps $\prod \hat\eta^{(ij)}$. 
Each $ \hat\eta^{(ij)} $ acts on the $V \otimes V$ where the first $V$ is located in the $i$'th slot 
and the second in the $j$'th slot of 
\bea 
V^{ \otimes (  n_1 + n_2 + n_3 + n_4  ) } 
\eea
There are ${n_1 + n_2 + n_3 + n_4 \over 2}$ of these contractions in 
the product, and no pair involves two elements from the same subset $S_a$. 

We can parametrize the sum over contractions by decomposing the sets $S_a$ as 
\bea 
&& S_1 = S_{12} \cup S_{13} \cup S_{14} \cr 
&& S_2 = S_{21} \cup S_{23} \cup S_{24} \cr 
&& S_3 = S_{31} \cup S_{32} \cup S_{34} \cr 
&& S_4 = S_{41} \cup S_{42} \cup S_{43} 
\label{decomps}
\eea
For a pair of sets  $U , V$ of same cardinality $ |U| = |V| $, we define the { \it contractor}  
\bea 
   C^{U,V}=\sum_{\sigma\in Sym (V)}\prod_{k=1}^{|U|}\hat\eta^{(i_k\sigma (j_k))} 
\eea
where $ Sym ( V) $ is the symmetric group of all permutations of the set $V$. 
So the correlator is computed by applying the map 
\bea 
  \sum_{S_{ij}} C^{S_{12},S_{21}} C^{S_{13},S_{31}} C^{S_{14},S_{41}} C^{S_{23},S_{32}} C^{S_{24},S_{42}} C^{S_{34},S_{43}} 
  \label{FllCrrltr}
\eea
The sum runs over all possible decompositions (\ref{decomps}).

When we compute the correlator by using the OPE in the $(12)(34)$ channel, we first choose subsets $S_{12},S_{21}$ and $S_{34},S_{43}$ 
and then do the corresponding contractions. 
Then we do the contractions between $(S_1\setminus S_{12})\cup (S_2\setminus S_{21})$ and $(S_{3}\setminus S_{34})\cup (S_{4}\setminus S_{43})$. 
So we are applying the map 
\bea 
\sum_{S_{12},S_{21},S_{34},S_{43}}C^{S_{12},S_{21}}C^{S_{34},S_{43}}C^{(S_1\setminus S_{12})\cup(S_2\setminus S_{21}), 
  (S_{3}\setminus S_{34})\cup(S_{4}\setminus S_{43})} 
  \label{1234channel}
\eea
We will show that 
\bea 
&& C^{(S_1\setminus S_{12})\cup (S_2\setminus S_{21}),(S_3\setminus S_{34})\cup (S_4\setminus S_{43})}\cr 
&& = \sum_{S_{13},S_{31},S_{14},S_{41},S_{23},S_{32},S_{24},S_{42}} C^{S_{13},S_{31}} C^{S_{14},S_{41}} C^{S_{23},S_{32}} C^{S_{24},S_{42}}
    \label{equalityC}
\eea
The correlator computed using the OPE in the $(12)(34)$ channel (\ref{1234channel}) can be obtained from the original expression for the
correlator (\ref{FllCrrltr}) by simply reordering sums. 
To see why (\ref{equalityC}) is true, we consider 
 (\ref{1234channel}) and find that  $C^{S_{12},S_{21}}C^{S_{34},S_{43}}$ includes a sum over 
$Sym(S_{21})\times Sym (S_{43})$ and 
$C^{(S_1\setminus S_{12})\cup(S_2\setminus S_{21}),(S_{3}\setminus S_{34})\cup(S_{4}\setminus S_{43})}$ includes a sum over
$Sym (S_2\setminus S_{21})\times Sym (S_4\setminus S_{43}))$.
The sum in (\ref{1234channel}) is a sum over the cosets $S_2/(Sym(S_{21})\times Sym (S_2\setminus S_{21}))$ and
$S_3/(Sym (S_{43})\times Sym (S_4\setminus S_{43}))$.
Putting these sums together, we reconstruct the sum in (\ref{FllCrrltr}). 

Although we have written $C^{S,T}$ in terms of sums over permutations above, we can also write it without permutations but using lists  and give another  proof of \ref{equalityC}.  
Let $L^S$ be a list constructed from the set $S$. 
A set does not know about any ordering. 
Let $|S|$ be the cardinality of $S$. 
Then $ |S|! $ is the number of lists constructed from $S$, with list size $|S|$. 
Let $L_0^S$ be a fixed list, with some fixed chosen ordering of the elements. 
We can write 
\bea 
  C^{S,T} = \sum_{L^T} \delta (L_0^S,L^T) \delta (|S|,|T|) 
\eea
By definition, the delta function on the lists pairs the first from $L_0^S$ with the first from $L^T$, second with second etc. 
In what follows the delta on the cardinalities is often suppressed. 
Let us define $U=(S_1 \setminus S_{12})\cup (S_2\setminus S_{21})$ and $V=(S_3\setminus S_{34})\cup (S_4\setminus S_{43})$.
We can write the LHS of (\ref{equalityC}) as
\bea
  C^{U,V}=\sum_{L^V}\delta (L_0^U,L^V)
\eea
Now a pairing $\delta (L_0^U,L^V)$ determines 

\begin{figure}[ht]%
\begin{center}
\includegraphics[width=0.5\columnwidth]{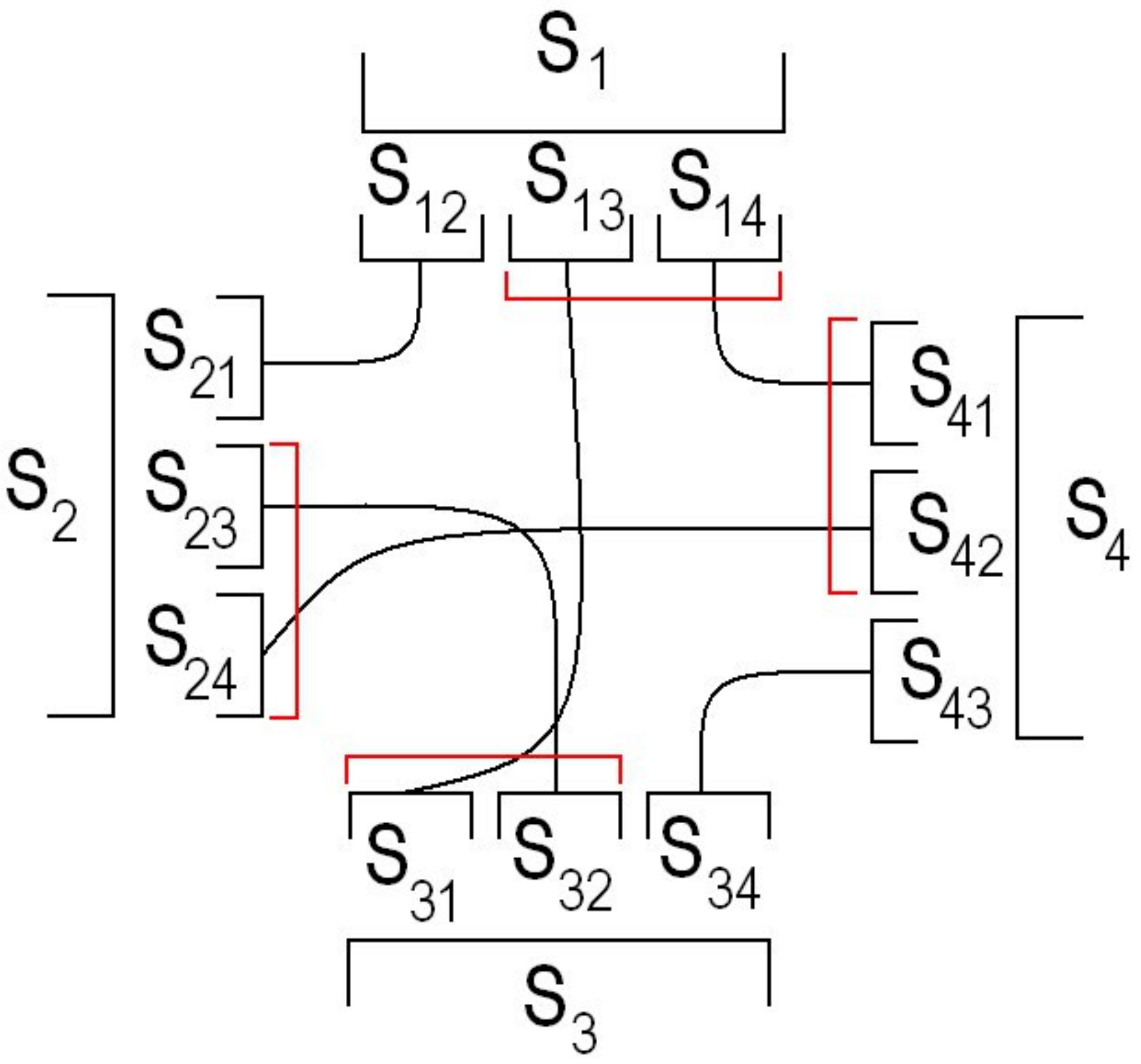}%
\caption{ crossing argument with $V $   }%
\label{fig:crossing-Varg}%
\end{center}
\end{figure}

\begin{itemize} 

\item a subset $S_{13}$ of $(S_1 \setminus S_{12})$ and a subset $S_{31}$ of  $S_3 \setminus S_{34}$ which are contracted with each other. 

\item  the order in $L_0^U$ picks an ordered list $L_0^{S_{13}}$ from $S_{13}$.
       The sum over $L^V$ includes a sum over all ordered lists $L^{S_{31}}$ in $S_{31}$. 

\item Similarly $S_{14}$, $S_{41}$ are determined as in Figure \ref{fig:crossing-Varg}. 
      $L_0^{S_{14}}$ and $L^{S_{41}}$ are also determined. 

\item Likewise $ S_{23} , S_{32} $ and $ L_0^{ S_{23} } , L^{S_{32}} $. 

\item Likewise $ S_{24} , S_{42 } $ and $ L_0^{ S_{24} } , L^{ S_{42}}$.

\end{itemize} 

We conclude that
\bea 
C^{U,V} && = \sum_{S_{13},S_{31},S_{14},S_{41}} \sum_{S_{23},S_{32},S_{34},S_{42}} \sum_{L^{S_{31}},L^{S_{41}},L^{S_{32}},L^{S_{43}}}\cr 
        &&\delta (L_0^{S_{13}},L^{S_{31}})\delta (L_0^{S_{14}},L^{S_{41}})\delta (L_0^{S_{23}},L^{32})\delta (L_0^{S_{24}},L^{S_{42}})\cr 
        && = \sum_{S_{13},S_{31},S_{14},S_{41},S_{23},S_{32},S_{24},S_{42}}C^{S_{13},S_{31}}C^{S_{14},S_{41}}C^{S_{23},S_{32}}C^{S_{24},S_{42}}
\eea
which is the desired identity (\ref{equalityC}). See Figure \ref{fig:crossing-Varg} for a graphical representation of the discussion. 

Computing the correlator by using the OPE in the $(14)(23)$ channel, we first choose subsets $S_{14},S_{41}$ and $S_{32},S_{23}$ 
and then do the corresponding contractions. 
Then we do the contractions between $(S_1\setminus S_{14})\cup (S_2\setminus S_{23})$ and $(S_{3}\setminus S_{32})\cup (S_{4}\setminus S_{41})$. 
For this channel we are applying the map 
\bea 
\sum_{S_{14},S_{41},S_{32},S_{23}}C^{S_{14},S_{41}}C^{S_{32},S_{23}}C^{(S_1\setminus S_{14})\cup(S_2\setminus S_{23}), 
  (S_{3}\setminus S_{32})\cup(S_{4}\setminus S_{41})}
  \label{1423channel} 
\eea
with
\bea 
&& C^{(S_1\setminus S_{14})\cup (S_2\setminus S_{23}),(S_{3}\setminus S_{34})\cup (S_{4}\setminus S_{43})}\cr 
&& =\sum_{S_{13},S_{31},S_{34},S_{43},S_{21},S_{12},S_{24},S_{42}} C^{S_{13},S_{31}} C^{S_{12},S_{21}}C^{ S_{43},S_{34}}C^{S_{24},S_{42}}
\label{1423channelOPE}
\eea
The equality between (\ref{1234channel}) and (\ref{1423channel}) follows simply by swapping the orders of summation.

\section{Counting of Primaries}\label{sec:counting} 

A rather basic question about free scalar field theory is to enumerate the $SO(4,2)$ irreducible representations appearing 
among the composite fields made out of $n=2,3, \cdots$ fundamental fields. 
For example, we need these multiplicities to compute the spectrum of primary operators in the CFT$_4$.
This question amounts to decomposing, into irreducible representations, the tensor product $Sym(V_+^{\otimes n})$,
where $V_+=D_{[1,0,0]}$ in the notation of \cite{Dolan05}. 
The three integer labels in $D_{[d,j_L,j_R]}$ are the dimension and two Lorentz spins.

We have seen simple examples of TFT$_2$ in Section \ref{sec:tft2eqs},  arising 
from 1-variable Gaussian integration, as well as from matrix integration. 
In this section we will introduce a simple TFT$_2$ that organizes the counting of primaries. 
Our results give a formula for the multiplicities of irreducible representations of $SO(4,2)$ in 
the tensor product $V_+^{\otimes  n}$.

Some key results that are needed to reproduce the results of this section have been collected in Appendix \ref{sec:countappendix}.

\subsection{Results on tensor products}\label{tprods}

It is known that \cite{Heidenreich:1980xi}
\bea\label{V+tensV+} 
   D_{[100]} \otimes D_{[100]} = \cA_{ [200] } + \sum_{ k_1 =1} D_{ [ k_1 + 2 , {k_1 \over 2 } , {k_1 \over 2 } ] }
\eea
where $\cA_{d,j_L,j_R}$ is another class of irreducible representation of $SO(4,2)$, see \cite{Dolan05} for more details.  
For $n=3$, we have, using results in \cite{Dolan05}, or alternatively by manipulating characters, 
\bea\label{cubictensprod}  
&& D_{[100]}^{\otimes 3 } = \sum_{ k_1 =0} ^{ \infty } \sum_{ k_2 = 0}^{\infty } 
 \cA_{ [k_1 + k_2 + 3 , { k_1 + k_2 \over 2 } , { k_1 + k_2 \over 2 } ]} \cr 
 && + \sum_{ k_1 = 1 }^{ \infty } \sum_{ k_2 = 1 }^{\infty} 
    \cA_{ [ k_1 + k_2 + 3 , { k_1- 1 \over 2 }  \otimes { k_2 - 1 \over 2 }  , { k_1 + k_2 \over 2 } ] } + 
   \cA_{ [ k_1 + k_2 + 3 , { k_1 + k_2 \over 2 } ,  { k_1- 1 \over 2 }  \otimes { k_2 - 1 \over 2 }   ]} 
\eea
For $n=4$, we get 
\bea 
&& D_{[100]}^{\otimes 4 } = \sum_{ k_1 , k_2 , k_3 =0 }^{\infty} 
\cA_{ [ k_1 + k_2 + k_3 + 4 , { k_1 + k_2 \over 2 }  \otimes { k_3 \over 2 } , { k_1 + k_2 \over 2 } \otimes { k_3 \over 2 } ] } 
+ \sum_{ k_1 , k_2  =1}^{\infty} \sum_{ k_3 =0}^{\infty} 
 \cA_{ [ k_1 + k_2 + k_3 + 4 , { k_1 + k_2 \over 2 } \otimes { k_3 \over 2 }  , { k_1 -1 \over 2 } \otimes { k_2 -1 \over 2 } \otimes { k_3 \over 2 } ]} \cr 
 && + \cA_{ [ k_1 + k_2 + k_3 + 4  , { k_1 -1 \over 2 } \otimes { k_2 -1 \over 2 } \otimes { k_3 \over 2 }  , { k_1 + k_2 \over 2 } \otimes { k_3 \over 2 } ]}
\eea
This is an easy application of the previously derived formula for $V_+^{\otimes 3}$ along with equation (4.7) of \cite{Dolan05}.
For general $n$, we have 
\bea\label{gendecomp} 
&& D_{[100]}^{\otimes (n+1) }  = \sum_{ k_1 , \cdots , k_{n} = 0 }^{ \infty} 
\cA_{ [ n+1 + k_1 + \cdots + k_n , { k_1 + k_2 \over 2 } \otimes { k_3 \over 2 } \cdots \otimes  { k_n \over 2 } , { k_1 + k_2 \over 2 } \otimes { k_3 \over 2 } \cdots \otimes {k_n \over 2 }  ] } \cr 
&& +  \sum_{ k_1, k_2 =1}^{ \infty} \sum_{ k_3 , \cdots , k_n =0}^{\infty} \cA_{ [ n+1 + k_1 + \cdots + k_n , { k_1-1\over 2 } \otimes { k_2 -1\over 2} \otimes 
 { k_3 \over 2 } \otimes  \cdots  \otimes  { k_n \over 2 } , { k_1 + k_2 \over 2 } \otimes { k_3 \over 2 } \otimes  \cdots \otimes { k_n \over 2 }  ]} \cr 
 && 
  +    \cA_{ [ n+1 + k_1 + \cdots + k_n , { k_1 + k_2 \over 2 } \otimes { k_3 \over 2 } \otimes  \cdots \otimes { k_n \over 2 } 
  ,{ k_1-1\over 2 }  \otimes { k_2 -1\over 2}  \otimes  { k_3 \over 2 } \otimes \cdots  \otimes  { k_n \over 2 } ]}
\eea

\subsection{TFT$_2$ for counting primaries} 

Consider an algebra of polynomials in variables $X_{j}$ where $j\in\{0,1/2,1,\cdots\}$. 
The structure constants are given by the fusion rules of $SU(2)$
\bea \label{TFTProd}
 X_{ j_1 } X_{j_2} && = N_{j_1j_2}^{j_3} X_{j_3} \cr
 && = \sum_{ \substack{ j = |j_1 - j_2 | \\ \Delta j = 1 }   }^{ j_1 + j_2 } X_j 
\eea
This is the large $k$ limit of the fusion rule algebra in the $SU(2)$ WZW model. 
Call this algebra $\cA_{su}$. Introduce the pairing
\bea 
< X_{j_1} , X_{j_2} >  = \delta_{ j_1 j_2 } 
\eea
Given the above structure constants and the pairing, we have associativity equations such as 
\bea 
\sum_{ j } N_{j_1 j_2}^{j} N_{ j j_3}^{j_4}  =\sum_{ j }  N_{j_1 j_3}^{j} N_{ j j_2 }^{j_4} 
\eea
Now, take a second copy of this algebra, generated by $Y_j$.
Introduce the algebra of power series in $s$ with rational coefficients, called $\mQ [s]$. 
A pairing on $\mQ [s]$ is given by
\bea 
   < s^{k_1} , s^{k_2} > = \delta_{ k_1 k_2 } 
\eea
In summary, we have an algebra $\cA_{su}\otimes\cA_{su}\otimes\mQ [s]$ which is associative and 
has a non-degenerate pairing. 
Thus, we have defined a TFT$_2$ of the kind we described in Section \ref{sec:tft2eqs}. 
Let us call this the $ SU(2) \times SU(2)$-fusion-TFT$_2$. Computations of the $SU(2)$ fusion multiplicities can be 
easily programmed in Mathematica. 

We will now explain how this TFT$_2$ can be used to compute the multiplicities of the primaries in the CFT$_4$.
The character $\chi_{V_+}(s,x,y)$ is obtained by computing $tr_{V_+} X $ with $X=s^\Delta x^{J_3^L} y^{J_3^R}$.
In computing this trace, the null state associated with the equation of motion for the free massless boson,
together with all of its descendents, are subtracted. 
The character of $V^{\otimes n}$ is then $\chi_{V_+}(s^n,x^n,y^n)$.
Since the fundamental field of our CFT is a boson, taking a product of $n$ copies gives $Sym(V_+^{\otimes n})$.
The character $\chi_{Sym(V_+^{\otimes n})}(s,x,y)$ can be computed as the trace of a  symmetrizer acting
on $V^{\otimes n}$ which leads to
\bea
   \chi_{Sym(V_+^{\otimes n})}(s,x,y)&&={1\over n!}\sum_{\sigma\in S_n} tr_{V_+^{\otimes n}} (X^{\otimes n}\sigma)\cr 
                                     &&={1\over n!}\sum_{\sigma\in S_n} \prod_{i=1}^n\left(\chi_{V_+}(s^i,x^i,y^i)\right)^{C_i(\sigma)}
   \label{Asschur}
\eea
In the last line, $C_i(\sigma)$ is the number of cycles of length $i$ in the permutation $ \sigma $.  
Using the formula (\ref{withoutP}) from the Appendix \ref{sec:countappendix}, we can express
 the characters $\chi_{V_+}(s^i,x^i,y^i)$ in terms of $\chi_{j_L}(x)\chi_{j_R}(y)$, which can be replaced 
by $ X_{j_L} Y_{j_R}$. The result is then simplified using the product (\ref{TFTProd}) and written as a sum of characters of $SO(4,2)$ irreducible representations.
We will illustrate the procedure for $n=3$ in the next subsection.

Finally, the characters useful for understanding the tensor product decomposition are 
\bea 
&&\cA_{[\Delta,j_L,j_R]}(s,x,y)=s^{\Delta}\chi_{j_L} (x)\chi_{j_R}(y)P(s,x,y)\cr 
&&D_{[\Delta,j_L,j_R]}(s,x,y)=s^{j_L+j_R+2}\left(\chi_{j_L} (x)\chi_{j_R}(y)-s\chi_{j_L-1/2}(x)\chi_{j_R-1/2}(y)\right)P(s,x,y)\cr 
&&\chi_{V_+}(s^i,x^i,y^i)=D_{[1,0,0]}(s,x,y)=s(1-s^2)P(s,x,y)
\label{charresults}
\eea
where
\bea
&&P(s,x,y)={1\over (1-sx^{1/2}y^{1/2})(1-sx^{1/2}y^{-1/2 })(1-sx^{-1/2}y^{1/2})(1-sx^{-1/2}y^{-1/2})}\cr 
&&=\sum_{p,q=0}^{\infty } s^{2p+q}\chi_{{q\over 2}}(x)\chi_{{q\over 2}}(y)  
\eea

\subsection{ Symmetrized tensor products and primaries in free field theory }  

In this section we will focus on the case $n=3$.
A simple application of (\ref{Asschur}) gives
\bea
\chi_{Sym(V^{\otimes 3})}={1\over 6}\left((\chi_{V_+}(s,x,y))^3 +3\chi_{V_+}(s,x,y)\chi_{V_+}(s^2,x^2,y^2)
                                          +2\chi_{V_+}(s^3,x^3,y^3)\right)
\eea
The first term above can be simplified using (\ref{cubictensprod}).
For the second and third terms we use the formula (\ref{withoutP}) in Appendix \ref{usingQ} for $\chi_{V_+}(s^n,x^n,y^n)$.
In view of (\ref{charresults}), to identify the sum of characters we obtain we need to have an overall factor of $P(s,x,y)$ appearing.
One way to achieve this is to introduce the inverse of $P(s,x,y)$
\bea
 Q = ( 1 + s^4 ) - s ( 1 + s^2 ) X_{ 1 \over 2 } Y_{ 1 \over 2 }  + s^2 ( X_1 + Y_1 ),\qquad P(s,x,y)Q=1 
\eea
An overall factor of $P(s,x,y)$ can now be arranged by including a factor of $Q$. 
For more details see Appendix \ref{usingQ}.
With this motivation, consider the following state of the TFT  
\bea\label{chingen} 
   {\boldsymbol \chi }_n = Q \chi_n
\eea
where
\bea
&& { \chi}_n = s^n \sum_{ q =0}^{ \infty } s^{nq} \sum_{ l_1 = 0 }^{ \infty } \left ( X_{ nq/2 - nl_1 } - \sum_{l_1=0}^{ \infty } X_{nq/2 - nl_1  -1 } \right ) 
\sum_{ l_2 = 0 }^{ \infty } \left ( Y_{ nq/2 - nl_2 } - \sum_{l_2=0}^{ \infty } Y_{nq/2 - nl_2  -1 } \right ) \nonumber
\eea
This state is obtained by replacing $SU(2)$ characters $\chi_{j_L}(x)\chi_{j_R}(y)\to X_{j_L} Y_{j_R}$ in $\chi_{V_+}(s^n,x^n,y^n)$. 
It is straight forward to compute ${\boldsymbol \chi}_n$ in Mathematica, with the result
\bea 
   \label{firstform}
   {\boldsymbol \chi }_n  =\sum_{k,k_1,k_2=0}^{\infty}s^{n+k}X_{k_1}Y_{k_2}C_{n,k,k_1,k_2}    
\eea
The coefficients of $C_{n,k,k_1,k_2}$ give the multiplicities of $SO(4,2)$ irreducible representations in $V^{\otimes n}$.
By computing these numbers explicitly we could for example, derive the results of section \ref{tprods}.

We now specialize to the case $n=3$.
For fixed $k$, at least one of $k_1$ or $k_2$ in (\ref{firstform}) is equal to ${k\over 2}$. 
Also, ${\boldsymbol \chi}_n$ is symmetric under exchange of $X$ and $Y$. 
With these properties in mind, it is useful to introduce the operation $\boldsymbol {S}$ defined by
\bea 
    \boldsymbol{S} X_a Y_b = (X_a Y_b + X_b Y_a) - \delta_{ab} X_a Y_b  
\eea
$\boldsymbol{S}$ acts on  $\cA_{su}\otimes\cA_{su}\otimes \mQ [s]$.
In terms of $\boldsymbol{S}$ we can write
\bea\label{sym3ans}
   \chi_{Sym^3(V)}(s,x,y)=(Ps^3 \boldsymbol{S})\left(\sum_{q =0}^{\infty}\sum_{l=0}^{\infty}s^q X_{q/2}Y_{q/2-l} 
   \left(\lfloor {q\over 6}\rfloor - \lfloor {l\over 3}\rfloor + A(\mu,\nu)\right)\right) 
\eea
where $\mu$ is the residue of $q$ modulo $6$, $\nu$ is the residue of $l$ modulo $6$ and
\bea 
A (\mu,\nu) && =\delta(\mu) \left (\delta(\nu)-\delta(\nu-1)-\delta(\nu-5)\right) 
   +\delta(\mu-2)\left(\delta(\nu)+\delta(\nu-4)-\delta(\nu-5)\right)\cr 
   &&+\delta(\mu-3)(\delta(\nu)+\delta(\nu-3))+\delta(\mu-4)\left(\delta(\nu)+\delta(\nu-4)+\delta(\nu-2)\right)\cr 
&&+\delta(\mu-5)\left(\delta(\nu)+\delta(\nu-3)+\delta(\nu-1)+\delta(\nu-4)\right) 
\eea
where we have used the Kronecker delta $\delta(x)$ which is 1 if $x=0$ and zero otherwise.

\section{ Discussion and Future directions  }\label{sec:discuss} 

\subsection{ Genus zero  TFT$_2$ :  Equations in search of Axioms  }\label{sec:eqsrax}

We have given a set of genus zero equations which form part of a  truncation of 
the usual finite dimensional TFT$_2$ equations. 
These include a product, co-product, unit, co-unit, associativity and 
crossing. This construction is realized in ordinary Gaussian integrals, 
and also as an $SO(4,2)$-covariant TFT$_2$ which reproduces the correlators and 
OPE of 4D CFT. It is a genus zero system because of the infinite dimensionality 
of the state space. A torus amplitude would diverge. 

An obvious open problem is to find the axiomatic framework which accommodates
this truncated TFT$_2$ system of equations and which therefore admits solutions in terms of infinite dimensional state spaces. 
The usual definitions of TFT$_2$ and Frobenius algebras 
force the state spaces to be finite dimensional and naturally include amplitudes for all genera.

Infinite dimensionality of the state spaces $ \cW $ implies some accompanying subtleties. 
We avoided these issues by focusing attention on a discrete basis, which can be chosen 
to have nice properties, such as diagonalising the CFT$_4$ inner product. 
We gave the equations in terms of arrays of numbers $C_{ABC},\eta_{AB},\tilde \eta^{AB}$ 
with indices labeling  this basis set. Going beyond this somewhat basis dependent 
approach, requires being more careful about specifying what sorts of linear combinations 
we have in mind when we talk of the vector space $\cW$. As a first step let us consider defining as the vector space  $ \cW$ of 
finite linear combinations of the specified basis set. Since the basis set has an infinite number of elements, this is an 
infinite dimensional space.  For $ \cW \otimes \cW $, we also 
consider finite linear combinations $ { \mathcal F }^{AB} e_A \otimes e_B $. 
Acting on these $\eta$ is obviously well defined. For generic infinite sums, it would not be well defined. 
Similarly we can work with $ \cF^{ A_1 \cdots A_n } e_{A_1} \otimes e_{A_2} \cdots \otimes e_{A_n } $, with 
$\cF$ having only finitely many nonzero entries, as our definition of $ \cW^{ \otimes n }$. 
Then the map from $ \cW^{\otimes n } \rightarrow \mC $ is well defined. 

Now consider the amplitude for vacuum going to two circles. 
Using  the discrete data $\tilde{{\eta}}^{AB}$ we can build 
\bea 
  \tilde\eta^{AB} e_A \otimes e_B 
\eea
But this is an infinite linear combination. 
So we are not getting a map from $\mC \rightarrow \cW \otimes \cW$, as we might have expected. 
The infinite dimensionality has thus entered in a crucial way here. 
Indeed this is expected. If  the amplitude for vacuum to 2 circles exists, and 2 circles to vacuum exists, 
then if we stick with a category theory framework , we can compose 
these two and get a torus, which looks like it cannot be anything but infinite
(hence ill-defined) in the case of infinite dimensional state spaces. 

So we conclude that  we cannot  make sense of the amplitude for the  vacuum going to 2 circles 
in terms of a map $\mC \rightarrow \cW \otimes \cW$. However, we found 
good uses of $ \tilde \eta^{AB}  $ in our system of equations, e.g. 
in relating crossing to associativity. How can we modify the axiomatic framework 
so as to associate some sort of algebraic data to the picture of 
vacuum going to 2 circles ? 

One sensible possibility would be  to associate specified finite dimensional subspaces  of $ \cW $ to 
one of the two  outgoing circles corresponding to $\tilde\eta^{AB} $. 
We could think of this as placing a {\it subspace defect} on one of the circles.
The effect is to restrict one of the indices of $\tilde \eta^{AB}$, to run over a 
specified subset of the basis vectors in $\cW $. If we do that to one of the outgoing circles, 
the torus amplitude is no longer infinite, but depends on the dimension of the chosen subspace.  
In general, for an amplitude $C_{A_1 , A_2 , \cdots A_n }^{ B_1, \cdots , B_m }$
corresponding to $ n $ incoming circles and $ m$ outgoing circles, we would restrict $m-1$ of 
the outgoing circles with subspace defects.

There are generalizations of TFT$_2$  as a  functor from the geometrical category of cobordisms to 
a more general target category which replaces the category of vector spaces (e.g. chain complexes) \cite{segal-moore}. 
The target category we need here would
be the one where an object is not a vector space, but rather a vector space along with 
some class of its subspaces. In our applications the vector space would be $\cW$
and we would also allow its subspaces. So a circle can be associated to $\cW $ or 
to some subspace thereof. 

Subspace defects can be considered even in TFTs with finite dimensional state spaces. 
They may have interesting combinatoric applications. For the case of 2D TFTs based on lattice 
constructions with a gauge group $G$, natural observables to consider, with some similarity to subspace defects, 
include those where the holonomy of a specified boundary is in a subgroup $ H\subset G $. Amplitudes including 
such observables have applications in counting Feynman graphs and BPS states of quiver theories 
\cite{feyncount,quivcalc}.  

We believe there is an interesting story with the space of finite linear combinations, with 
the inclusion of subspace defects. Even if that were solved, there would remain 
another layer of subtleties involved  in giving a complete axiomatic 
framework for CFT$_4$,  based on the TFT$_2$ picture developed here. 
This is because states of interest in computing correlators, namely the  $ \Phi (x)$ (eq. \ref{Foundational}) used to 
construct CFT$_4$ correlators, are actually infinite linear combinations! 
Careful treatment of these issues gets into the subject of topological vector spaces. 
This can be a subtle subject, e.g. how to define tensor products, see for example \cite{Wiki-top}. 
% But having an inner product usually helps. We can perhaps  talk about $ g ( \Phi (x)  , \Phi (x) ) $ here. 
% ( REFS). 
The right definition of ``tensor product'' in our case, would have to 
take into account the fact that if we insert $\Phi (x) $  at one circle and again the same 
$ \Phi ( x ) $ at another circle ( with the same $x$ ), then the amplitude is not well-defined. 
As pragmatic physicists, we can  be content with a description  of the TFT$_2$
in terms of genus-restricted version of the usual TFT$_2$ equations, and this is the point 
of view we have taken in most of the paper. However,  solving the problem of finding the right axiomatic framework
for the current TFT$_2$ formulation of CFT$_4$ would give an interesting 
mathematical perspective on AdS/CFT.  

\subsection{ Correlators and Clebsch-Gordan coefficients }

We have written 
\bea 
\Phi ( x ) = \Phi^{+} ( x ) + \Phi^{-} ( x ) = e^{ i P \cdot x } v^+ +  ( x^{\prime} )^2 e^{ - i K \cdot x^{\prime}  } v^- 
\eea
Correlators of the type $\langle  \Phi ( x_1 ) \Phi ( x_2 ) \Phi^2 ( x_3) \rangle $ involve Wick contractions 
between the operator at $ x_1$ and $x_2$ with the operator $ x_3$. This is an invariant in $ V^+_{ [1,0,0] } \otimes V^+_{ [1,0,0] } \otimes V^{-}_{ [2,0,0] } $. 
Equivalently they are  $ SO(4,2)$-covariant homomorphisms from $ V^+_{ [1,0,0] } \otimes V^+_{ [1,0,0] }  \rightarrow V_{ [2,0,0]}^+$. 
This means that the correlators 
\bea 
{ 1 \over ( x_1 - x_2 )^2 ( x_1 - x_3)^2 } 
\eea
 can be used to generate the matrix elements of this homomorphism, i.e. the Clebsch-Gordan coefficients, 
which are often of interest as special functions. As far as we are aware, this QFT connection has  not been used to develop explicit formulae 
for  $ SO(4,2)$ Clebsch-Gordan coefficients in the discrete basis. This would be an interesting exercise. 

Similar remarks apply to correlators such as 
$ \langle \Phi^{ n_1} ( x_1) \Phi^{ n_2} ( x_2 ) \cdots \Phi^{ n_k } ( x_k ) \Phi^{ n_1+ n_2 + \cdots + n_k } ( x_{ k+1} ) \rangle $
which will generate the Clebsch-Gordan coefficients for coupling $ V^+_{[n_1,0,0] } \otimes V^{+}_{ [n_2, 0 ,0 ] } \otimes \cdots \otimes V_{ [n_k , 0 ,0 ] }^+ \rightarrow V^+{ [n_1 + n_2 + \cdots + n_k , 0 ,0 ] } $.  By inserting more general primary fields we can get Clebsch-Gordan coefficients 
involving more general irreps of $ SO(4,2)$.

\subsection{Further examples of CFT$_4$/TFT$_2$ }\label{sec:furtherexs}  

Let us recap, from section \ref{sec:modelGaussian},   that for  the case of the Gaussian integration model, we have a state space 
\bea 
\cW = \cW_0 \oplus \cW_1 \oplus \cdots = \bigoplus_{ n = 0 }^{ \infty} \cW_n 
\eea
At each level we have a 1-dimensional space, isomorphic to $ \mC$. 
$ \cW_n $ corresponds to $ \phi^n $. The product is given by Gaussian integration.
For the free scalar field in four dimensions, we had $ \cW_n  = Sym ( V^{ \otimes n } ) $, 
where $ V = V_+ \oplus V_-$. 
For Gaussian Hermitian Matrix model, let us work at large $N$, we have 
$ \cW_0  = \mC $. At level $n$, we have a vector space of dimension 
$p(n)$, the number of partitions of $N$. The basis vectors can 
be taken to be the multi-traces 
\bea 
( \tr \Phi )^{p_1}  ( \tr  \Phi^2 )^{ p_2} \cdots  ( \tr \Phi^n )^{ p_n }  
\eea
The numbers $ p_i $ for $ i \in \{ 1, \cdots , n \} $ determine a 
partition of $n$,  i.e. $ n = \sum_i i p_i $. 
We may write 
\bea
\cW_n  = \mC^{ p(n)} 
\eea
The product  is obtained from matrix integration. If can be written neatly  in terms of 
permutations in $S_n$. We won't make that explicit here.

For the 4D CFT of a free  hermitian matrix field $ \Phi$, in the large $N$ limit, 
we have for $ \cW_n$,  a direct sum over partitions $p$. For each $p$
\bea 
\Sym^{ p_1}  ( V ) \otimes \Sym^{ p_2 }  ( Cyc^{2} ( V ) ) \otimes \Sym^{ p_n }   ( Cyc^{ n  } ( V )  ) 
= \otimes_{ i=1}^n \Sym^{ p_i }   ( Cyc^{ i  } ( V ) )     
\eea
The $ Cyc^i ( V )  $ is defined as the projection of $ V^{ \otimes i } $ to the 
subspace invariant under cyclic permutations of the $i$ factors.  $ \Sym^{ p_i } ( W )$ is 
the subspace of $ W^{ \otimes p_i } $ invariant under symmetric group $S_{p_i}$. 

If we want to do this at finite $N$, we can use the technology from \cite{BHR2}.  
The primaries in a scalar field theory 
are obtained from $\oplus_{ n=0}^{ \infty } Sym V_+^{ \otimes n } $,  
where $V_+ = D_{[1,0,0] } $. For the matrix scalar theory, we need 
to consider $ ( V_+ \otimes V_N \otimes \bar V_N) ^{\otimes n } $
and project to invariants of $U(timesN)$ and $S_n$. We can use the following decompositions 
\bea 
&&  V_+^{\otimes n } = \oplus_{ \Lambda , \L_1 } V_{\L}^{SO(4,2)} \otimes V_{ \L_1}^{ S_n } \otimes V_{ \L , \L_1} \cr 
&& V_N^{ \otimes n } = \oplus_R V_R^{ U(N)}   \otimes V_R^{ S_n } \cr 
&&   \bar V_N^{ \otimes n } = \oplus_S \bar V_S^{ U(N)}   \otimes \bar V_S^{ S_n } \cr 
&& 
\eea
In the first line we are decomposing $ V_+^{ \otimes n } $ in terms of
irreps of $ SO(4,2)$ and the commuting $S_n$. The vector space $ V_{ \L , \L_1 } $ is the 
space of multiplicities for given irreps of $ SO(4,2) $ and $ S_n$.  
$U(N)$ invariance forces $R = S$. $S_n$ invariants 
requires that $ V_R \otimes V_R \otimes V_{ \Lambda_1} $ contains the trivial irrep of $S_n$.
The multiplicity $ C ( R , R , \Lambda ) $ of this trivial irrep. 
appears in the construction of the operators.

Let us outline some problems related to the above models and 
obvious, but interesting, generalizations :

\begin{itemize}

\item For the free scalar or hermitian matrix CFT$_4$, write the products $ C_{AB}^C $  and corresponding 
associativity equations explicitly. 

\item For  the extremal correlators of  the complex one-matrix model, the description of the 
state space will follow much the same steps as above for the matrix scalar field. This is relevant to
the half-BPS sector in AdS/CFT. 
The explicit formulae for the invariant maps leading to correlators will need to 
take into account that the correlators involve insertion of multiple holomorphic operators 
and a single anti-holomorphic operators.

\item  Develop the CFT$_4$/TFT$_2$  of free Maxwell Gauge fields, including  TFT$_2$ construction of the 
correlators of $E$ and $B$-fields. Extend this to free non-abelian gauge fields and  free fermionic fields.
These steps would provide the foundation for approaching  perturbative $ \cN=4$ super-Yang Mills theory
as a TFT$_2$. We give more discussion of going beyond free fields in Section \ref{sec:beyondfree}. 

\item Conversely, the TFT$_2$ construction we have developed with $ SO(4,2)$ can be  modified 
by replacing $ SO(4,2)$ with a compact group $G$. Any finite dimensional  irreducible representation $V$  of $G$ 
which contains the trivial irrep in the tensor product decomposition $ V \otimes V $, will allow  the definition of 
an invariant map $ \hat \eta : V \otimes V \rightarrow \mC $.  Then we can define the state space 
\bea 
\cW = \bigoplus_{ n = 0}^{\infty} Sym ( V^{ \otimes n } ) 
\eea
and the amplitudes $C_{A_1 , \cdots , A_k } $ can be defined using tensor products of the elementary $ \hat \eta$. 
This will give $G$-invariant TFT$_2$'s. Do these constructions  have a realization in terms of a path integral ?

\end{itemize} 

\subsection{Generalized free fields and the stress tensor condition}\label{sec:genfree} 

Our reformulation of the free scalar field CFT$_4$ in terms of TFT$_2$ has started from the representation $V^+=V_{\Delta =1,j_L=0,j_R=0}$ 
and the dual $V^-=(V^+)^*$, has constructed $V=V^+\oplus V^-$, and then built the state space $\cW=\bigoplus_{n=0}^{\infty} Sym(V^{\otimes n})$.
There is a basic $SO(4,2)$ invariant map $\hat \eta: V \otimes  V \rightarrow \mC$ which leads to the 2-point function. 
Multilinear $SO(4,2)$ invariant maps corresponding to the cobordism between $k$ circles and the vacuum in TFT$_2$, are constructed from 
$\cW^{\otimes k}\rightarrow\mC$, using tensor products of the basic invariant $\hat \eta$. 
These include, for the case $k=2$, a bilinear invariant pairing $\eta:\cW\otimes\cW\rightarrow\mC$ with $\eta_{AB}$ in some convenient 
discrete basis $e_A$ of $\cW$.  For the amplitude for 3-circles to vacuum, we have $C_{ABC}$.

Given this reformulation, it is very natural to ask if all $SO(4,2)$ invariant TFT$_2$'s lead to some CFT$_4$.
A key result of this section is to construct examples of $SO(4,2)$ invariant TFT$_2$'s that do not correspond to a CFT$_4$.
Our discussion was motivated by recent activity aimed at solving the constraints imposed by 
crossing symmetry \cite{Rattazzi:2008pe,Caracciolo:2009bx,Rattazzi:2010gj}.
By understanding the crossing symmetry constraints one may hope to understand and completely characterize the space of all possible CFTs.
In this section we will consider the problem of solving the constraints imposed by crossing, in the framework of our novel approach to CFT$_4$. 

The construction of new solutions might begin by replacing $V^+\oplus V^-$ above with $V = V_{\Delta,j_L,j_R}\oplus  V_{\Delta,j_L,j_R}^*$. 
Since any discrete series lowest weight irrep (having positive energy) 
has a dual highest weight irrep (having negative energy) such that $ V_{\Delta,j_L,j_R}\otimes V_{\Delta,j_L,j_R}^*$ contains a unique copy of the identity 
representation, we know that we will have a unique $ \hat \eta_{ab}$ as matrix elements of $ V \otimes V \rightarrow \mC$.  
Aside from this replacement, the construction of $ \cW , \eta , C $ proceeds as before. 
This is a {\it generalized free field construction}, which, as we will explain, gives an $SO(4,2)$ covariant 
TFT$_2$. Further conditions can be imposed, such as unitarity and existence of a stress tensor, 
which can be reasonably considered necessary for these to be proper  CFT$_4$. 

To describe $ \hat \eta_{ab}$ explicitly, we need a simple generalization of the analysis of section \ref{sec:basic}. 
For simplicity we will discuss the case with $j_L,j_R=0$. 
The generalization to $j_L,j_R\ne 0$ is straight forward.
Start from
\bea
   \Phi_d (x) = {1\over\sqrt{2}}\left( e^{iP\cdot x}v^+ + (x')^{2 d} \rho (e^{iP\cdot x'}v^+) \right)
\eea
where now we have
\bea
   D v^+ = d v^+,\qquad  D v^- = -dv^-\qquad M_{\mu\nu}v^+ = M_{\mu\nu}v^- = 0
\eea
Again consider the tensor
\bea\label{Tdefined} 
   T_{p_1 p_2\cdots p_n,q_1 q_2\cdots q_n}=\eta\left( P_{p_1}P_{p_2}\cdots P_{p_n} v^+,K_{q_1}K_{q_2}\cdots K_{q_n} v^-\right)
\eea
Using nothing but the $so(4,2)$ algebra and the $so(4,2)$ invariance of the pairing $\eta$, we find
\bea\label{Tequals} 
T_{p_1\cdots p_n, q_1\cdots q_n}\prod_{a=1}^n y^{\prime q_a}x^{p_a}
=2^n n! \sum_{l=0}^{\big[{n\over 2}\big]}(-1)^l (x\cdot y')^{n-2l}|x|^{2l}|y'|^{2l}{(n-l-1+d)!n!\over (n-2l)!(d-1)!2^{2l}l!}
\eea
Thus,
\bea
   F(x,y')&&\equiv \eta\left( e^{-ix\cdot P'}v^+,e^{iy'\cdot K'}v^-\right)\cr
&&=\sum_{n=0}^\infty T_{p_1\cdots p_n, q_1\cdots q_n}\prod_{a=1}^n y^{\prime q_a}x^{p_a}{1\over (n!)^2}\cr
&&=\sum_{n=0}^\infty 2^n \sum_{l=0}^{\big[{n\over 2}\big]}(-1)^l (x\cdot y')^{n-2l}|x|^{2l}|y'|^{2l}{(n-l-1-d)!\over (d-1)!(n-2l)!2^{2l}l!}\cr
&&={1\over (1-2x\cdot y' +x^2 y^{\prime 2})^d}
\eea
and
\bea
   \eta\left(\Phi_d(x),\Phi_d(y)\right)={(y')^{2d}\over (1-2x\cdot y' +x^2 y^{\prime 2})^d}={1\over |x-y|^{2d}}
\eea
The $C_{ABC}$ are constructed from tensor products of the elementary $ \eta$'s as before. 
With these choices, it is now clear that the argument demonstrating crossing from section \ref{sec:crossass} goes through without 
modification, so that we have indeed obtained a new TFT$_2$ and hence a new solution to the crossing constraints.

We will want to place additional constraints on this solution. 
For example we can ask for a unitary TFT$_2$. We have explained how 
to define an inner product $ g : V \otimes V \rightarrow \mC$, which is 
sesquilinear and hermitian. 
This extends to an inner product on $ \cW$ which coincides with the CFT inner product. 
If we require that this inner product on the TFT$_2$ state space is non-degenerate with strictly positive 
eigenvalues, then we have the usual restriction of unitarity in CFT$_4$, which leads to constraints on dimensions of fields. 
For example, for a scalar (where $ j_L , j_R = 0 $) we have $\Delta\ge 1$. 

Another natural constraint is to require that the theory contains a stress tensor.  For the free scalar
field which has $\Delta =1$, $Sym(V^{ \otimes 2 } )$ contains a local stress tensor, for which $V_{\Delta,j_L,j_R}=V_{4,1,1} $.  
 
An interesting case to consider is where the basic free field is itself the representation $V_{4,1,1}$. 
Now $V$ has the right $SO(4,2)$ transformation properties to be a stress tensor.
There is a $T_{\mu\nu}$ which is treated as an elementary field in this generalized free field construction. 
We want to know whether we can make sense of 
\bea 
&&  P_{\mu}  = \int  T_{\mu 0} (x) d^4 x  \cr 
 && [ P_{\mu},\cO] = \partial_{\mu} \cO 
\eea
in our TFT$_2$ approach to this generalized free field system. 
To define $P_{\mu}$ along the lines of the first equation, we would need to define some sum over states in the representation $V=V^+\oplus V^-$.  
This would give $P_{\mu}$ as a state in $\cW$. 
So $P_{\mu}=P_{\mu}^A e_A$ where $e_A$ is a set of basis vectors for $ \cW$. 
There is a corresponding operator $(\widehat P_{\mu})_B^C \equiv P_{\mu}^A C_{AB}^{~~~C} $ in the endomorphisms of $\cW$. 
The state $\cO$ is a general state in $\cW$, i.e. $\cO=\cO^A e_A$ and there is a corresponding operator $(\widehat \cO)_B^C\equiv\cO^A C_{AB}^{ ~~~C} $.  
The LHS can be written as 
\bea 
 ( \widehat P_{\mu} )_A^C  (\widehat \cO )_C^D - ( \hat \cO )_A^C ( \widehat P_{\mu} )_C^D   
\eea
The RHS of the second equation is another well defined state in $\cW$, so we can consider 
$(\widehat{\partial_{\mu}\cO})_B^C=(\partial_{\mu}\cO)^A C_{AB}^C$.   
The question is whether there is a good definition of $P_{\mu}$, i.e. whether the description as a sum above can be made precise enough 
to prove that the equality 
\bea 
 (\widehat P_{\mu})_A^C (\widehat\cO)_C^D-(\widehat\cO)_A^C(\widehat P_{\mu})_C^D=(\widehat {\partial_{\mu}\cO})_A^D 
\label{toprove}
\eea
holds for arbitrary $\cO$. 

Since we do not have a Lagrangian for this prospective new generalized free field CFT, we have to formulate the requirement 
of existence of a good stress tensor along the algebraic lines outlined here. 
Although this has been instructive as an exercise in using the new TFT language, the above construction does not work, as we now explain.
The argument we employ appeared recently in the context of CFT$_d$ in \cite{Dymarsky:2014zja}.
Our structure constants $C_{ABC}$ are symmetric.
By looking at the dimensions of fields, it is clear that the energy momentum tensor does not appear in the product of the field with itself.
Equivalently, the field will not appear in the product of the energy momentum tensor with the field which contradicts (\ref{toprove}). 
This contradiction rules out generalized free fields in CFT$_d$ unless they have a dimension $\Delta = {d-2\over 2}$.
Although the generalized free field theories are not CFT's, our discussion makes it clear what they are: they are $SO(d,2)$ covariant TFT$_2$'s.

\subsection{Beyond free fields   }\label{sec:beyondfree} 

We highlight here some  key ingredients in the connection between CFT$_4$ and TFT$_2$ we have developed, 
which  will continue to hold beyond the set-up of free fields that we have described explicitly here. 

\begin{itemize} 

\item The description of the state space in terms of discrete series representations of $ SO(4,2)$. 
The description of the two-point function of operators in terms of an invariant pairing between positive and negative 
energy representations. This is necessitated simply by  the fact that there is no
$ SO(4,2)$  invariant in the tensor product of two positive energy representations.

\item  The relation of the bilinear  invariant pairing $\eta_{ AB } $  to a sesquilinear  positive definite  inner product, via a map 
$ \rho $ between positive and negative energy representations. 

\item The non-degeneracy of the pairing, i.e existence of  $ \tilde \eta^{AB}$, 
 and its relation to non-degeneracy of the Zamolodchikov metric.

\item The OPE and $ C_{AB}^{~~C} = C_{ABD } \tilde\eta^{ DC} $. 

\item Relation between crossing and associativity. 

\end{itemize}

One element which needs to be worked out in a case by case basis, is 
the explicit construction of the  correlators in terms of representation theoretic data
as we have done for the free field case. Here we have developed the construction by expressing 
the usual Wick contraction rules of free field theory in terms of applications of 
a tensor product of elementary invariant pairings. 

A very interesting problem is to develop this beyond free fields to perturbative 
CFTs, e.g. $\cN =4$ supersymmetric Yang Mills theory. Elucidating the role of regularization and renormalization 
in this TFT$_2$ framework will be a fascinating problem. There is good reason to believe  that this problem 
will have an elegant solution. This expectation  comes from the fact that one-loop $ \cN =4$ SYM has been formulated 
using  representation theory of $ PSU(2,2|4)$  via the formalism of the one-loop dilatation operator \cite{Beisert:2003tq,Beisert-1loop} 
and this approach is also known to generalize to higher loops\cite{Beisert:2005fw}. 
The key elements of the link to TFT$_2$ have been aspects of
$SO(4,2)$ representation theory. In the case of $ \cN=4$ SYM we would expect a $ PSU(2,2|4)$-invariant  TFT$_2$. 

Perturbative CFT$_4$ correlators are constructed by inserting local operators in the path integral 
and expanding the interaction terms, which are interaction vertices integrated over space-time. 
The computation involves two steps : the first being the calculation of  an integrand by doing Wick contractions involving 
the inserted  operators as well as the interaction vertices. These are free field Wick contractions, which 
have been interpreted in terms of TFT$_2$. The next step is the computation of space-time integrals. A key question 
is whether this step can be incorporated in the framework of $SO(4,2)$ invariant TFT$_2$, 
where a correlator is calculated by evaluating $SO(4,2)$ equivariant maps between specified 
states at the boundaries.  Very interestingly
some recent mathematical developments starting from a completely different perspective have arrived at an interpretation of 
 integrals in perturbative QFT suggestive of the type of TFT$_2$ considered here \cite{FL07,Lib1407}.
 The motivations of these papers stem from the idea  of applying 
quarternionic analysis to four dimensional quantum field  theory, much in the way 
that complex analysis is a powerful tool in two dimensional CFT.  
 Conformal integrals have been written in terms of $SO(4,2)$-equivariant maps $ \rho : V_+ \otimes V_+ \rightarrow V_+ \otimes V_+$ - precisely the structure needed for $SO(4,2)$ invariant TFT2.  
 With a view to an $SO(4,2)$-invariant-TFT$_2$ interpretation for general CFTs
beyond the ones accessible as perturbations from free fields, it is also worth noting that these same conformal integrals have played a key role in the analysis of conformal blocks \cite{SimmDuff1204}. These recent developments provide good grounds to believe that the TFT$_2$ perspective developed here has the potential to go beyond free fields. This is a fascinating topic for future research.

\bigskip 

\begin{center} 
{ \bf Acknowledgements}
\end{center} 
SR is supported by STFC Grant ST/J000469/1, String Theory, Gauge Theory, and Duality.
RdMK is supported by the South African Research Chairs Initiative of the Department of Science and Technology and the National Research Foundation.
This project was initiated while both authors were visiting Durham University.
RdMK would like to thank Collingwood College, Durham for their support and especially the IAS Durham for a fellowship which made the visit possible. SR thanks Grey College, Durham for a Grey Fellowship and the Department 
of Mathematical Sciences, Durham for hospitality. RdMK and SR also thank the Perimeter Institute for hospitality 
while part of this work was done. Research at the Perimeter Institute is supported by the Government of Canada through
Industry Canada and by the Province of Ontario through the Ministry of Research and
Innovation. We thank Chong Sun Chu, David Garner,   Hally Ingram,  Dimitri Polyakov, Douglas Smith, Gabriele Travaglini, Rodolfo Russo for discussions.

\vskip3cm

\begin{appendix} 

\section{$sl(2)$ and $ su(2)\times su(2)$ subalgebras of $so(4,2)$}\label{sec:subgroupapp}

In this Appendix we will describe the $sl(2)$ and $ su(2)\times su(2)$ subalgebras of $so(4,2)$ that were used in section \ref{fsection}
to construct the invariant pairing $\hat\eta$.
In terms of the generators of section \ref{radialalgebra}, introduce
\bea
   K \equiv K_1 + iK_2 \qquad P \equiv P_1 -iP_2 
\eea
which obey
\bea
  [K ,P ]=-4D +4iM_{21},\qquad [M_{21} ,K ]=-iK ,\qquad [M_{21} ,P ]=iP 
\eea
Consequently
\bea 
&& H = D  -  i M_{21}   \cr 
&& H_+  = { P \over 2 } \cr 
&& H_-  = { K \over 2 } 
\eea
obey the standard $SL(2, R )$ relations 
\bea 
[  H_+  , H_-  ] =  H_3 \qquad [ H_3 ,  H_+  ] = 2 H_+  \qquad [ H_3 , H_-  ] = - 2 H_-  
\eea

Consider now the $SU(2)\times SU(2)$ subgroup.
The generators of the left and right $SU(2)$ subgroups are
\bea\label{JMs} 
&& J_3^{L} = -i (M_{12} + M_{34})\cr 
&& J_{+}^L = -{1 \over 2}\left( (M_{13} - M_{24}) + i(M_{14} + M_{23})\right)\cr 
&& J_-^L   =  {1 \over 2}\left(( M_{13} - M_{24}) - i(M_{14} + M_{23})\right)\cr 
&& J_3^{R} = -i(M_{12} - M_{34})\cr 
&& J_+^R   = {1\over 2}\left(-(M_{13} + M_{24}) + i(M_{14} - M_{23})\right) \cr 
&& J_-^R   = {1\over 2}\left(( M_{13} + M_{24}) + i(M_{14} - M_{23})\right) 
\eea
One way to understand this formulae is to realize 
$M_{ij} = x_i { \partial \over \partial x_j }  - x_j { \partial \over \partial x_i } $ 
and then convert to complex variables 
\bea 
&& z_1 = x_1 + i x_2  \cr 
&& z_2 = x_3 + i x_4 \cr 
&& \bar z_2 = x_{3} - i x_4 \cr 
&& \bar z_1 = x_1 - i x_2 
\eea
These complex coordinates have $(J_3^L,J_3^R )$ charges
\bea 
z_1  \rightarrow  ( 1,1 ) ~~~ z_2  \rightarrow  ( 1,-1) ~~~  \bar z_2  \rightarrow  ( -1, 1 ) ~~ \bar z_1 \rightarrow  ( -1 , -1 ) 
\eea

It is straightforward to find
\bea 
 && [ J_3^L , P_1 -i P_2 ] = -( P_1 - i P_2 ) \cr  
 && [ J_3^L , K_1 +i K_2 ] =  ( K_1 + i K_2 ) \cr 
 && [ J_3^R , P_1 -i P_2 ] = -( P_1 - i P_2 ) \cr 
 && [ J_3^R , K_1 +i K_2 ] =  ( K_1 + i K_2 ) 
\eea 
Thus the state $ (P)^l v^+$ is a lowest weight state with $( J_3^L , J_3^R ) = ( -l , -l ) $
and the state $ ( K)^{l'} v^- $ has $ ( J_3^L , J_3^R ) = ( l' , l' ) $.

\section{ Useful formulae for deriving counting of primaries } \label{sec:countappendix}

The results given in the Appendix are needed to carry out the counting arguments given in section \ref{sec:counting}.
In the first section we derive a formula for the character $\chi_{{k\over 2}}(x^n)$ which naturally appears when we
consider the character of $V_+^{\otimes n}$. 
In the second section of this Appendix, we derive a number of identities useful for decomposing the character of $V_+^{\otimes n}$
into a sum of characters of $SO(4,2)$ irreps.

\subsection{Character evaluated at $x^n$}

The character for spin $j$ is
\bea
\chi_{j}(x) &&= x^j + x^{j-1} + \cdots + x^{-j+1} + x^{-j}\nonumber\\
             &&={x^{j+{1\over 2}}-x^{-j-{1\over 2}}\over x^{1\over 2}-x^{-{1\over 2}}}  
\eea
Thus, we find
\bea
 \chi_{{k\over 2}}(x^n) &&= x^{nk\over 2} + x^{n(k-2)\over 2} + \cdots + x^{-{n(k-2)\over 2}} + x^{-{nk\over 2}}\nonumber\\
         &&= \left({x^{nk\over 2} + x^{n(k-2)\over 2} + \cdots + x^{-{n(k-2)\over 2}} + x^{-{nk\over 2}}\over x^{1\over 2}-x^{-{1\over 2}}}\right)
              (x^{1\over 2}-x^{-{1\over 2}})\nn\\
         &&=\sum_{l=0,1,...}^{\lfloor k/2\rfloor}\chi_{{{kn\over 2}-nl}}(x)-\sum_{l=0,1,...}^{\lfloor (k-1)/2\rfloor}
            \chi_{{{kn\over 2}-nl-1}}(x)\label{charwewant}
\eea
To obtain the last line, multiply the numerator out and collect terms.

\subsection{Inverse of $P$ in terms of characters and reduction to $SU(2)$ fusion multiplicities } \label{usingQ}
Using the formulas for $\chi_{q\over 2}(x^n)$ derived in the first section of this Appendix, we find
\bea
&&P(s^n,x^n,y^n)=\sum_{p,q=0}^\infty s^{2np+nq}\chi_{q\over 2}(x^n)\chi_{q\over 2}(y^n)\nn\\
&&={1\over 1-s^{2n}}\sum_{q=0}^\infty s^{nq}\chi_{q\over 2}(x^n)\chi_{q\over 2}(y^n)\nn\\
&&={1\over 1-s^{2n}}\sum_{q=0}^\infty s^{nq}
\left[\sum_{l=0,1,...}^{\lfloor q/2\rfloor}\chi_{{{qn\over 2}-nl}}(x)-\sum_{l=0,1,...}^{\lfloor (q-1)/2\rfloor}
            \chi_{{{qn\over 2}-nl-1}}(x)\right]\nn\\
&&\quad\times\left[\sum_{l=0,1,...}^{\lfloor q/2\rfloor}\chi_{{{qn\over 2}-nl}}(y)-\sum_{l=0,1,...}^{\lfloor (q-1)/2\rfloor}
            \chi_{{{qn\over 2}-nl-1}}(y)\right]
\eea
Thus,
\bea
  \chi_{V_+}(s^n,x^n,y^n)&&=P(s^n,x^n,y^n)s^n (1-s^{2n})\nn\cr
&&=s^{n}\sum_{q=0}^\infty s^{nq}
\left[\sum_{l=0,1,...}^{\lfloor q/2\rfloor}\chi_{{{qn\over 2}-nl}}(x)-\sum_{l=0,1,...}^{\lfloor (q-1)/2\rfloor}
            \chi_{{{qn\over 2}-nl-1}}(x)\right]\nn\\
&&\quad\times\left[\sum_{l=0,1,...}^{\lfloor q/2\rfloor}\chi_{{{qn\over 2}-nl}}(y)-\sum_{l=0,1,...}^{\lfloor (q-1)/2\rfloor}
            \chi_{{{qn\over 2}-nl-1}}(y)\right]
\label{withoutP}
\eea
It is useful to consider the special case
\bea
  \chi_{V_+}(s,x,y)&&=P(s,x,y) s(1-s^2)\cr
                   &&=s(1-s^2)\sum_{p,q=0}^\infty s^{2p+q}\chi_{q\over 2}(x)\chi_{q\over 2}(y)\cr
                   &&=\sum_{q=0}^\infty s^{1+q}\chi_{q\over 2}(x)\chi_{q\over 2}(y)
\label{nisonehere}
\eea
which contains useful information about the states that appear in $V_+$.

We also need an identity which rewrites $\chi_{V_+}(s^n,x^n,y^n)$ as $SU(2)$ characters multiplied by $P(s,x,y)$; these
can very easily be translated into ${\cal A}_{[\cdot,\cdot,\cdot]}$s. 
Towards this end, note that
\bea
  1&&=P(s,x,y)(1-sx^{1/2}y^{1/2})(1-sx^{1/2}y^{-1/2})(1-sx^{-1/2}y^{1/2})(1-sx^{-1/2}y^{-1/2})\nn\\
   &&=P(s,x,y)\big[1+s^4-s(1+s^2)\chi_{1\over 2}(x)\chi_{1\over 2}(y)+s^2(\chi_1(x)+\chi_1(y))\big]
\eea

With  these formulae in hand, we can write any of the characters in terms of 
sums of $\chi_{j_L}(x)\chi_{j_R}(y)P$, the coefficients being determined by 
computing some $SU(2)$ fusion multiplicities.  These fusion multiplicity computations
are programmed in Mathematica to yield the results we have given above. 

\end{appendix}

\end{document}